\documentclass[apj]{emulateapj}

\usepackage{color}
\usepackage{rotating}

\begin{document}

% for emulateapj
\title{Testing Models of Intrinsic Brightness Variations
       in Type Ia Supernovae, \\
       and their Impact on Measuring Cosmological Parameters }

% preprint  title
%\title{Testing Models of Intrinsic Brightness Variations \\
%       in Type Ia Supernovae and their Impact on \\
%       Measuring Cosmological Parameters }

\email{kessler@kicp.uchicago.edu}
\submitted{accepted by ApJ}

% ===========================================================
%  author list for 
%  ``Testing Models of Intrinsic Brightness Variations
%
% ===========================================================
% institution nick-names

\newcommand{\NUMUCASTRO}{1}        % U.C astro
\newcommand{\NUMKICP}{2}           % U.Chicago KICP
\newcommand{\NUMLPNHE}{3}          % LPNHE, Paris
\newcommand{\NUMFNAL}{4}           % Fermilab
\newcommand{\NUMAPO}{5}            % Apache Point Observatory
\newcommand{\NUMWAYNE}{6}          % Wayne State (Detroit)
\newcommand{\NUMRUTGERS}{7}        % Rutgers University
\newcommand{\NUMUPENN}{8}          % U Penn
\newcommand{\NUMPENNSTATEAA}{9}    % Penn State U, Astro Dept.
\newcommand{\NUMPENNSTATECOS}{10}  % Penn State U, Cosmo dept.

% ====================================================
% names

\author{
Richard~Kessler,\altaffilmark{\NUMUCASTRO,\NUMKICP}
Julien~Guy,\altaffilmark{\NUMLPNHE}
John~Marriner,\altaffilmark{\NUMFNAL}
Marc~Betoule,\altaffilmark{\NUMLPNHE}
Jon~Brinkmann,\altaffilmark{\NUMAPO}
David~Cinabro,\altaffilmark{\NUMWAYNE}
Patrick~El-Hage,\altaffilmark{\NUMLPNHE}
Joshua~A.~Frieman,\altaffilmark{\NUMUCASTRO,\NUMKICP,\NUMFNAL}
Saurabh~Jha,\altaffilmark{\NUMRUTGERS}
Jennifer~Mosher,\altaffilmark{\NUMUPENN}
and
Donald~P.~Schneider\altaffilmark{\NUMPENNSTATEAA,\NUMPENNSTATECOS}
} % end authorlist

% ===============================================
%       INSTITUTIONS
% ===============================================

\altaffiltext{\NUMUCASTRO}{
Department of Astronomy and Astrophysics,
University of Chicago, 5640 South Ellis Avenue, Chicago, IL 60637, USA
}

\altaffiltext{\NUMKICP}{
Kavli Institute for Cosmological Physics, 
University of Chicago, 5640 South Ellis Avenue Chicago, IL 60637, USA
}

\altaffiltext{\NUMLPNHE}{
Laboratoire de Physique Nucl\'eaire et des Hautes Energies, 
UPMC Univ. Paris 6, UPD Univ. Paris 7, CNRS IN2P3, 
4 place Jussieu F-75005,  Paris, France
}

\altaffiltext{\NUMFNAL}{
Center for Particle Astrophysics, 
  Fermi National Accelerator Laboratory, 
  P.O. Box 500, Batavia, IL 60510, USA
}

\altaffiltext{\NUMAPO}{
  Apache Point Observatory, P.O. Box 59, Sunspot, NM 88349, USA
}

\altaffiltext{\NUMWAYNE}{
Department of Physics,
Wayne State University, Detroit, MI 48202, USA
}

\altaffiltext{\NUMRUTGERS}{
Department of Physics and Astronomy, 
Rutgers University, 136 Frelinghuysen Road, Piscataway, NJ 08854, USA
}

\altaffiltext{\NUMUPENN}{
Department of Physics and Astronomy,
University of Pennsylvania, 203 South 33rd Street, 
Philadelphia, PA  19104, USA
}

\altaffiltext{\NUMPENNSTATEAA}{
 Department of Astronomy and Astrophysics, 
   The Pennsylvania State University, University Park, PA 16802, USA
}

\altaffiltext{\NUMPENNSTATECOS}{
  Institute for Gravitation and the Cosmos, 
  The Pennsylvania State University, University Park, PA 16802, USA
}

% =========== END ==============

\newcommand{\xScale}{1.10}   % 1.1 for emulateapj
\newcommand{\xxScale}{1.15}  % 1.15 for emulateapj

\newcommand{\SNANA}{{\tt SNANA}}
\newcommand{\SDSS}{SDSS--II}
\newcommand{\SNLS}{SNLS3}
\newcommand{\unc}{uncertainty}
\newcommand{\uncs}{uncertainties}
\newcommand{\Uncs}{Uncertainties}
\newcommand{\SALTII}{{\sc salt-ii}}
\newcommand{\SIFTO}{{\sc sifto}}

\newcommand{\SALTtomu}{{\tt SALT2mu}}

\newcommand{\Trestobs}{T_{\rm rest,obs}}
\newcommand{\Trest}{T_{\rm rest}}
\newcommand{\Tobs}{T_{\rm obs}}

\newcommand{\Lrest}{\lambda_{\rm rest}}
\newcommand{\Lobs}{\lambda_{\rm obs}}

\newcommand{\specy}{spectroscopically}
\newcommand{\spec}{spectroscopic}
\newcommand{\Specy}{Spectroscopically}
\newcommand{\Spec}{Spectroscopic}
\newcommand{\eff}{efficiency}
\newcommand{\effspec}{\epsilon_{\rm spec}}
\newcommand{\ineff}{inefficiency}
\newcommand{\obss}{observations}
\newcommand{\obs}{observation}
\newcommand{\intonly}{intrinsic-only}

\newcommand{\LCDM}{\Lambda{\rm CDM}}
\newcommand{\OM}{\Omega_{\rm M}}
\newcommand{\OL}{\Omega_{\Lambda}}
\newcommand{\DMU}{\Delta\mu}
\newcommand{\MUfit}{\mu_{\rm fit}}
\newcommand{\MUcalc}{\mu_{\rm calc}}

\newcommand{\SYMDELmu}{\Delta_{\mu}}
\newcommand{\SYMDELz}{\Delta_{z}}
\newcommand{\SYMDELc}{\Delta_{c}}
\newcommand{\DIFmu}{(\mu_{\rm fit}-\MUcalc)/(1+z)}
\newcommand{\DIFz}{(Z_{\rm phot}-Z_{\rm spec})/(1+z)}
\newcommand{\DIFc}{(B-V-\cc)/(1+z)}

\newcommand{\SIMDELmB}{\Delta_{m_B}^{\rm SIM}}
\newcommand{\SIMDELc}{\Delta_{c}^{\rm SIM}}

\newcommand{\RATIO}{R_{\rm MC/Data}}

\newcommand{\sigCOHvalue}{0.13}
\newcommand{\sigFUNONEvalue}{0.06}
\newcommand{\sigFUNTWOvalue}{0.045}
\newcommand{\lamFUNvalue}{1000}
\newcommand{\sigCOH}{\sigma_{\rm COH}}
\newcommand{\signode}{\sigma_{\rm node}}
\newcommand{\sigV}{\sigma_{5500}}
\newcommand{\sigint} {\sigma_{\rm int}}
\newcommand{\siginti} {\sigma_{{\rm int},i}}
\newcommand{\sigstati}{\sigma_{{\rm stat},i}} 
\newcommand{\FUNLAM}{\exp[-(\lambda_{\rm node}-5500)/3000]}

\newcommand{\sigDATA}{\sigma_{\rm DATA}}
\newcommand{\sigSIM}{\sigma_{\rm SIM}}
\newcommand{\sigZPT}{\sigma_{\rm ZPT}}
\newcommand{\sigOFF}{\sigma_0}
\newcommand{\ZPTpe}{{\rm ZPT}_{\rm pe}}
\newcommand{\sigHOST}{\sigma_{\rm HOST}}

\newcommand{\sigFARUV}{0.59}  % for Chotard extrap to farUV

\newcommand{\C}{\Sigma}  % intrinsic covariance matrix

\newcommand{\sigintmB}{\sigma_{m_B}^{\rm int}}
\newcommand{\sigintc}{\sigma_{c}^{\rm int}}
\newcommand{\sigintx}{\sigma_{x1}^{\rm int}}

\newcommand{\sigmB}{\sigma_{m_B}}
\newcommand{\sigc}{\sigma_{c}}
\newcommand{\sigx}{\sigma_{x1}}

\newcommand{\mBtrue}{m_B({\rm true})}
\newcommand{\mBfit}{m_B({\rm fit})}
\newcommand{\ctrue}{c({\rm true})}
\newcommand{\cfit}{c({\rm fit})}

\newcommand{\UU}{U^{\prime}}
\newcommand{\rhoUUU}{\rho_{\UU,U}}

\newcommand{\WIS}{$m_B$-only}  % Wavelength-Independent Scatter
\newcommand{\alphaSIM}{\alpha_{\rm SIM}}
\newcommand{\betaSIM}{\beta_{\rm SIM}}

\newcommand{\Ndof}{N_{\rm dof}}

\newcommand{\photoz}{photo-$z$}
\newcommand{\chisqr}{\chi^2_{r}}
\newcommand{\Pfit}{P_{\rm fit}}

\newcommand{\Nignit}{N_{\rm ignit}}
\newcommand{\Kamin}{{\rm Ka}_{\rm min}}
\newcommand{\cc}{c^{\prime}}

\newcommand{\sigMINUS}{\sigma_{-}}
\newcommand{\sigPLUS}{\sigma_{+}}
\newcommand{\xbar}{\bar{x}_1}
\newcommand{\cbar}{\bar{c}}

\newcommand{\MED}{M_{\Delta}}
\newcommand{\sigMED}{\sigma_{M}}
\newcommand{\sigMEDplus}{\sigma_{M}^{+}}
\newcommand{\sigMEDminus}{\sigma_{M}^{-}}
\newcommand{\sigHalf}{\sigma_{(N/2)}}

\newcommand{\FKRW}{F_{\rm KRW09}}
\newcommand{\wwwSNANA}{\tt http://www.sdss.org/supernova/SNANA.html}

\newcommand{\dwsyst}{\delta w_{\rm syst}}

\newcommand{\bfit}{\beta_{\rm fit}}
\newcommand{\mubiasTrue}{\Delta\mu_{\rm Malm}({\rm true})}
\newcommand{\mubiasEval}{\Delta\mu_{\rm Malm}(\bfit)}

% --------------

% Numbers/Results for intrinsic magSmear paper 

% Statistics. 
\newcommand{\NSDSS}{251}    % Global comment 
\newcommand{\NSDSSforHub}{251}    % Hubble anal 
\newcommand{\NSDSSforColor}{245}    % B-V-c anal 
\newcommand{\NSDSSforPhotoz}{208}    % Photoz anal 
\newcommand{\MCDATAratioSDSS}{6}    % Exact= 5.96 

\newcommand{\NSNLS}{191}    % Global comment 
\newcommand{\NSNLSforHub}{191}    % Hubble anal 
\newcommand{\NSNLSforColor}{120}    % B-V-c anal 
\newcommand{\NSNLSforPhotoz}{176}    % Photoz anal 
\newcommand{\MCDATAratioSNLS}{6}    % Exact= 6.06 

% w Bias. 
\newcommand{\wBias}{-0.01, -0.00, 0.01, 0.02}    % w-bias 
\newcommand{\wBiasFUN}{-0.01}    % w-bias, FUN-MIX 
\newcommand{\wBiasGten}{-0.00}    % w-bias, G10 
\newcommand{\wBiasCzero}{0.01}    % w-bias, C11_0 
\newcommand{\wBiasCone}{0.02}    % w-bias, C11_1 
\newcommand{\MCDATAratiowbias}{20}    % MC/data ratio for w-bias 
\newcommand{\MAXMUBIAS}{0.05}    % max MU-bias variation over redshift 

% ===========================
\begin{abstract}
For \specy\ confirmed Type Ia supernovae
we evaluate models of intrinsic brightness variations 
with detailed data/Monte Carlo comparisons
of the dispersion in the following quantities:
Hubble-diagram scatter, 
color difference ($B-V-c$) between the true $B-V$ 
color and the fitted color ($c$) from the \SALTII\
light curve model, and photometric redshift residual.
The data sample includes
\NSDSS\ $ugriz$ light curves from the three-season 
Sloan Digital Sky Survey-II and 
\NSNLS\ $griz$  light curves from the 
Supernova Legacy Survey 3 year data release.
We find that the simplest model of a wavelength-independent 
(coherent) scatter is not adequate, 
and that to describe the data the intrinsic-scatter model 
must have wavelength-dependent variations resulting in a 
$\sim 0.02$~mag scatter in $B-V-c$.
Relatively weak constraints are obtained on the nature of 
intrinsic scatter because a variety of different models 
can reasonably describe this photometric data sample.
We use Monte Carlo simulations to examine the standard approach 
of adding a coherent-scatter term in quadrature to the
distance-modulus \unc\ in order to bring the reduced $\chi^2$ 
to unity when fitting a Hubble diagram.
If the light curve fits include model \uncs\ with the correct 
wavelength dependence of the scatter,
we find that this approach is valid and that the bias
on the dark energy equation of state parameter $w$ is 
much smaller ($\sim 0.001$) than current systematic \uncs.
However, incorrect model \uncs\ can lead to a significant 
bias on the distance moduli,
with up to $\sim 0.05$~mag redshift-dependent variation. 
This bias is roughly reduced in half after applying a 
Malmquist bias correction.
For the recent SNLS3 cosmology results 
% \citep{Sullivan11}
we estimate that this effect introduces an additional
systematic \unc\ on $w$ of  $\sim 0.02$, well below
the total \unc. 
This \unc\ depends on the choice of viable scatter models and
the choice of supernova (SN) samples,
and thus this small $w$-\unc\ is not guaranteed in future
cosmology results.
For example, the $w$-\unc\ for SDSS+SNLS (dropping the nearby SNe) 
increases to $\sim 0.04$.

\vspace{.5cm}
\end{abstract}

% \large  % for editing only

% ##################################
 \section{Introduction}
 \label{sec:intro}
% ##################################

For more than a decade Type Ia supernovae (SN~Ia)
have been used as standardizable candles to
measure luminosity distances.  These distances,
along with the associated redshifts, have been
used to measure properties of dark energy
\citep{Riess98,Saul99,Riess04,Astier06,WV07,Freedman09,K09cosmo,Conley11}.
The uncorrected variation in the SN~Ia peak brightness
is $\sim 1$ mag, and this variation 
is reduced to $\sim 0.1$~mag after empirical corrections
based on the measured stretch \citep{Phillips_93}
and color \citep{Riess_96,Tripp_97}.
This 0.1~mag intrinsic scatter increases the scatter 
in the Hubble diagram
well beyond what is expected from the
distance modulus \uncs,
and the resulting cosmology fits have reduced chi-squared 
($\chisqr \equiv \chi^2/\Ndof$) significantly larger than unity.
To obtain $\chisqr = 1$, all SNIa-cosmology analyses to date 
have introduced an ad hoc intrinsic-scatter term
($\sigint \sim 0.1$~mag) that is added in quadrature
to the measured distance modulus \uncs.

This procedure of adding a constant ad hoc scatter term 
would be correct if the unknown source of intrinsic variation 
is independent of redshift, and if it is fully coherent
such that the variation is the same for all wavelengths 
and passbands. 
% -------
\citet[hereafter K10]{K10photoz}
found evidence 
contradicting a coherent variation in a study comparing 
the \photoz\ precision between data and 
Monte Carlo simulations (MC).
Using the coherent-scatter model in the MC
underestimated the fitted \photoz\ precision observed 
in the data, while simulating a model using color variations
gave better agreement.
\citet[hereafter G10]{Guy10} 
examined residuals from the \SALTII\ training
and showed that the variation about the best-fit
spectral model is wavelength dependent;
this wavelength-dependent \unc\ is included in the
light curve fitting model.
\citet[hereafter M11]{JM11}
presented a more formal
treatment of $\sigint$ based on an intrinsic scatter
covariance matrix that depends on 
a coherent term, a stretch term, and
a color term.\footnote{Within the \SALTII\ model framework,
these three parameters are $m_B$, $x_1$ and $c$.}
They compared Sloan Digital Sky Survey-II (\SDSS) 
SN~Ia data and simulations
within this framework and suggest
that the coherent and color terms are significant,
while the stretch term is negligible.

The importance of understanding the nature of intrinsic
scatter is tied to understanding systematic \uncs\
in cosmology analyses using SNe~Ia. If this scatter is
truly random, as suggested by explosion models showing
brightness variation with viewing angle 
\citep[hereafter KRW09]{KRW09}, 
then there is no intrinsic bias
and the \unc\ will decrease with increasing sample size.
Even in this optimistic scenario a wavelength-dependent
scatter results in a redshift-dependent dispersion because 
with broadband filters different rest-frame wavelengths
are probed as a function of redshift.
This variation must be properly accounted for,
except in the hypothetically ideal 
scenario of measuring high-quality spectra to determine
synthetic magnitudes with the same rest-frame
passbands at all redshifts.
%
%unless a sufficiently broad range of passbands
%can be used to probe the same rest-frame wavelengths
%at all redshifts.
%
If this scatter depends on more subtle physics related to
the explosion mechanism and the host-galaxy environment,
there could be additional redshift-dependent effects
not yet detected with current data samples, 
but that become apparent in future surveys with
much larger samples.

Significant effort to reduce this scatter has been attempted
using near infrared (NIR) photometry and \spec\ features.
\cite{Mandel_11} report that optical+NIR photometry result in
a Hubble scatter that is $\sim 30$\% smaller compared to 
using only optical data.
A decade of effort on \spec\ correlations can be summarized
with results from three groups. 
\cite{B11spec} examined spectra from the CfA Supernova Program
and used spectral features to reduce the
Hubble scatter by at most 10\%, 
albeit with only $2\sigma$ significance.
From the Berkeley SNIa Program,
\cite{Silverman12} examined 108 high quality SNe~Ia with 
a spectrum taken within 5 days of maximum brightness 
and found similar results.
The best Hubble scatter reduction was obtained by
\citet{Bailey09} using very high quality
spectra from the Supernova Factory \citep{SNF2002}.
Using spectra within 2.5 days of peak brightness,
they scanned every possible flux-ratio in $\sim 40$~{\AA} 
bins and found a minimum Hubble scatter using the ratio
$F(642~{\rm nm})/F(443~{\rm nm})$; 
the resulting scatter is about 25\% lower compared to 
the traditional photometric analysis with the 
\SALTII\ light curve model.

To realize significant reductions ($\sim 30$\%) 
in the Hubble scatter requires optical photometry combined 
with either rest-frame NIR photometry or
very high quality spectra near the epoch of peak brightness.
Both of these supplemental data samples are difficult to obtain
at low redshift, and it is not yet clear what resources could 
be allocated to
obtain large data samples for higher redshift SNe that
are needed to construct a cosmologically interesting 
Hubble diagram.
Given the unlikely prospects for significantly reducing
the Hubble scatter, we take a different approach here and 
explore models to describe the scatter in more detail. 
Such models of intrinsic scatter can be used to evaluate 
and constrain systematic \uncs\ from assuming an incorrect model,
and possibly lead to a better understanding of the 
underlying wavelength dependence of SN brightness variations.

In this work we demonstrate a method for evaluating 
models of intrinsic scatter by computing three
scatter-dependent dispersion variables and making 
the following data/MC comparisons:
(1) the traditional Hubble-diagram residual;
(2) $B-V-c$, where $B-V$ is the true rest-frame color
and $c$ is the fitted color parameter
from the \SALTII\ model; and
(3) the \photoz\ residual.
In the hypothetical limit of observing SNe~Ia with
infinite photon statistics and no intrinsic scatter,
the distribution for each variable would be a 
Dirac-$\delta$ function.
Simulations that include fluctuations from photon statistics,
but no intrinsic-scatter model,
underestimate the measured dispersion in these variables.
A viable model of intrinsic scatter must predict 
the dispersion for each variable and for multiple data sets.
These three variables do not constitute an exhaustive
list of photometric observables; for example one could
examine other rest-frame colors ($U-B$, $V-R$),
correlations among colors, 
and the dependence of the scatter on redshift, stretch, and color.
With our current statistics and signal to noise we limit 
this initial study to the three variables described above,
but larger and higher-quality SN samples from current and 
future surveys should enable a more thorough study.

Three classes of wavelength-dependent intrinsic-scatter 
models are investigated. 
First we try purely phenomenological functions of rest-frame 
wavelength with parameters tuned to match observations.
The second class is based on measurements from data.
The third class uses theoretical explosion models (KRW09)
to perturb the \SALTII\ spectral model.

This work is part of the SDSS+SNLS joint analysis,
and the data sets used here include \NSDSS\
\specy\ confirmed SNe~Ia from the 3 year \SDSS\ sample
\citep{Frieman07},  
and another \NSNLS\ \specy\ confirmed SNe~Ia from the 
Supernova Legacy Survey \citep[{\SNLS};][]{Conley11}.
All simulations and light curve fitting are done
with the publicly available \SNANA\ 
package\footnote{\wwwSNANA}
\citep[version {\tt v10\_07}]{SNANA09} 
and the \SALTII\ light curve model (G10).

The outline of the paper is as follows.
The data samples are described in Section~\ref{sec:data}
and the simulation and intrinsic-scatter models are described in 
Section~\ref{sec:sim}.
The determination of each dispersion variable is in 
Section~\ref{sec:anal},
and the resulting data/MC comparisons are in
Section~\ref{sec:results} along with some systematics tests.
Finally, in Section~\ref{sec:results_Hubble} we investigate the 
potential Hubble diagram bias from using an incorrect 
model of intrinsic scatter.

% ##################################
 \section{The \SDSS\ and SNLS Data Samples}
 \label{sec:data}

% ##################################

We use two SN~Ia data samples that are well calibrated
with $\sim 1$\% photometric
precision, and that span complementary
redshift ranges. The lower redshift SNe ($z<0.4$)
are from the full three-season \SDSS\ sample \citep{Frieman07}, 
and the higher redshift SNe ($z<1$) are from the 
publicly available 3 year \SNLS\ sample \citep{Conley11}.
Below we give a brief description of these samples.

The \SDSS\ Supernova Survey used the SDSS camera \citep{Gunn98} 
on the SDSS 2.5 m telescope \citep{SDSS_telescope,York00} 
at the Apache Point Observatory to search for SNe in the Fall 
seasons (September 1 through November 30) of 2005--2007.
This survey scanned a region (designated stripe~82) 
centered on the celestial equator in the 
Southern Galactic hemisphere that is 
2.5$^{\circ}$ wide and runs between right ascensions of
20$^{\rm h}$ and 4$^{\rm h}$, 
covering a total area of~300~deg$^2$ with a typical cadence 
of every four nights per region. 
Images were obtained in five broad passbands,
$ugriz$ \citep{Fukugita96}, with 55~s exposures and 
processed through the PHOTO photometric pipeline \citep{Lupton01}. 
Within 24~hr of collecting the data, the images were 
searched for SN candidates that were selected for \spec\ \obss\
in a program involving about a dozen telescopes.
The \SDSS\ Supernova Survey discovered and \specy\ confirmed
a total of $\sim 500$ Type Ia SNe. 
Details of the SDSS-II SN Survey are given in 
\citet{Frieman07} and \citet{Sako08},
and the procedures for \spec\ identification and redshift determinations
are described in \citet{Zheng08}.

The SN photometry for \SDSS\ is based on 
Scene Model Photometry (SMP) described in \citet{Holtz08}. 
The basic approach of SMP is to simultaneously model the ensemble
of survey images covering an SN location as a time-varying point
source (the SN) and sky background plus 
time-independent galaxy background and nearby calibration stars, 
all convolved with a time-varying 
point-spread function (PSF).
The fitted parameters are SN position,
SN flux for each epoch and passband,
and the host-galaxy intensity distribution in each passband.
The galaxy model for each passband is a $20 \times 20$ grid 
(with a grid scale set by the CCD pixel scale, 
$0.4\arcsec \times 0.4\arcsec$) 
in sky coordinates, and each of the $400 \times 5 = 2000$  
galaxy intensities is an independent fit parameter.
As there is no pixel re-sampling or image convolution, 
the procedure yields correct statistical error estimates.

The SNLS was a 5 year survey covering four 1~deg$^2$ fields
using the MegaCam  imager on the 
3.6 m Canada-France-Hawaii Telescope (CFHT). 
Images were taken in four bands similar to those used
by the SDSS: $g_M, r_M, i_M, z_M$, where 
the subscript $M$ denotes the MegaCam system.
The SNLS exposures were $\sim 1$~hr in order
to discover SNe at redshifts up to $z\sim 1$.
The SNLS images were processed in a fashion similar to the 
\SDSS\ so that \spec\ observations could be used to confirm the 
identities and determine the redshifts of the SN candidates.
Additional information about the SNLS can be found in 
\cite{Astier06} and references within.

The \SNLS\ SN photometry is based on a simultaneous fit of the 
SN flux and position, a residual sky background per image, 
and a galaxy intensity map. Images are resampled to the same 
reference pixel grid prior to the fit. The SN+galaxy image model 
is PSF-matched to the resampled images. Only sky noise is included 
in the photometric uncertainties 
(host galaxy and source noise are negligible for most SNe). 
Because resampling introduces pixel correlations, the \uncs\ 
ignoring correlations are scaled such that the reduced
$\chi^2$ is one when assuming a constant SN flux per night.

To ensure good quality fits to the light curves, the following
selection criteria are applied to both the \SDSS\ and \SNLS\
data samples,
\begin{itemize}
  \item At least one epoch before the epoch of peak brightness
        in the $B$ band (defined as $\Trest=0$)
  \item At least one epoch with $\Trest > 10$ days.
  \item At least three filters with an \obs\ that has a
        signal-to-noise ratio (S/N) above 8
  \item At least five \obss\ in the fitted epoch range 
        $-12 < \Trest < +25$~days.
        The maximum $\Trest$ is set by the range for one
	of the models (KRW09) of intrinsic brightness variation.
 \item  Color excess from Milky Way Galactic extinction 
         \citep{Schlegel98} is $E(B-V)<0.2$.
  \item After fitting each light curve to the \SALTII\ 
        model (Section~\ref{subsec:anal_SALT2}),
        we require the SNIa fit probability to be $\Pfit > 0.02$, 
        where $\Pfit$ is computed from the fit-$\chi^2$ 
        and the number of degrees of freedom. 
	After all other requirements, this cut removes 
	14 (3) events from the \SDSS\ (\SNLS)
	samples.
\end{itemize}
The sample statistics after these requirements are given in
Section~\ref{subsec:stats}.

% ##################################
 \section{Simulations}
 \label{sec:sim}
% ##################################

We use the \SNANA\ MC code \citep{SNANA09} 
to generate realistic 
SN~Ia light curves that are analyzed in exactly the 
same manner as the data.
The MC is used to make detailed comparisons 
with the data using different models of intrinsic
SN~Ia brightness variations.
All simulations are based on a standard
$\LCDM$ cosmology with $w=-1$, $\OM=0.3$, $\OL=0.7$.
Details of the simulation are described in \citet{SNANA09} 
and in Section~6 of \citet[hereafter K09]{K09cosmo}; 
here we give a brief overview.

Simulations are generated using the \SALTII\ model (G10)
that is based on a time sequence of rest-frame spectra.
The spectral model is explained in more detail in
Section~\ref{sec:anal} within the context of light curve fitting.
Observer-frame magnitudes are computed by redshifting
the rest-frame spectrum for each epoch, 
reddening the spectra from Galactic extinction \citep{Schlegel98}
using $R_V=3.1$,
and summing the flux in the appropriate 
filter-response curves.

To account for non-photometric conditions and varying 
time intervals between observations due to bad weather,
actual observing conditions are used for both the \SDSS\
and SNLS surveys.
For each simulated observation, the noise is determined from
the measured PSF,\footnote{The PSF is
described by a double-Gaussian function.}
Poisson noise from the source, and sky background.
Noise from the host-galaxy background is included for the 
\SDSS\ simulations where it has a small effect at low redshifts.
Host-galaxy noise for the higher redshift SNLS sample is
negligible, and thus not simulated.
Additional details of the simulation of noise are given in 
Section~\ref{subsec:sim_noise}.
The simulated flux in CCD counts is based on a mag-to-flux 
zeropoint, and a random fluctuation drawn from the noise estimate.

The parent distributions of the \SALTII\ stretch ($x_1$) 
and color ($c$) are well described by an asymmetric Gaussian 
that is a function of three parameters,
\begin{eqnarray}
   e^{ [-(x_1 - \xbar)^2/2\sigMINUS^2] } & ~~~~ & x_1 < \xbar \\
   e^{ [-(x_1 - \xbar)^2/2\sigPLUS^2]  } & ~~~~ & x_1 > \xbar 
\end{eqnarray}
and a similar function with $x_1 \to c$.
The parameters for each distribution are shown in 
Table~\ref{tb:unfold_x1c}. 
After accounting for Malmquist bias,
we find that the higher-redshift \SNLS\ sample is slightly 
brighter and bluer compared to the \SDSS\ sample. 
This difference is expected from previous results 
showing that younger star-former galaxies 
host brighter/bluer SNe~Ia than older passive galaxies 
\citep{Sullivan06,Lamp_2010,Smith2012}. 
The younger star-forming galaxies are more abundant
at higher redshifts, thus qualitatively explaining
the brightness difference between the two surveys.
While the redshift-dependent variation in the stretch 
population is well established,
the variation in the color population has been reported
only in \citet{Smith2012} where they show that the SN~Ia color
population is the same for passive and moderately 
star-forming galaxies, but different in {\em highly} 
star-forming galaxies.  
Previous studies comparing passive and 
{\em all} star-forming galaxies found no color variation,
and should not be considered inconsistent with the
results of \citet{Smith2012}.
We show in the systematics analysis 
(Section~\ref{subsec:results_syst}) that the simulated intrinsic scatter 
is rather insensitive to the parameters describing the parent 
populations in Table~\ref{tb:unfold_x1c},
and therefore it does not matter if these parameters are 
the same or slightly different for each survey.

\begin{table}[h!]
\caption{
  Asymmetric Gaussian Parameters to Describe the
  Parent Distribution of $x_1$ and $c$.
  } % end caption
\begin{center}
\begin{tabular}{l | cccc  }
\tableline\tableline
          &  Value at    &              &             & Generation \\
          &  Peak Prob   & $\sigMINUS$  &  $\sigPLUS$ & Range      \\
   \hline\hline
   $x_1$ ({\SDSS})  &  0.5    &  1.4   & 0.7   & $-5,+3$     \\
   $x_1$ ({\SNLS})  &  0.5    &  1.3   & 0.7   & $-5,+3$     \\
   $c$  ({\SDSS})   &  0.0    &  0.08  & 0.13  & $-0.4,+0.6$ \\
   $c$  ({\SNLS})   & $-0.04$ &  0.06  & 0.14  & $-0.4,+0.6$ \\
\tableline  % ------------------------------------------------------
\end{tabular}
\end{center}
  \label{tb:unfold_x1c}
\end{table}

\newcommand{\MiPeak}{M_i}
\newcommand{\effCOR}{C_{\epsilon}(\MiPeak)}

The simulation includes a detailed treatment of the search \eff,
including \spec\ selection effects. For the \SDSS, 
the search-pipeline \eff\ has been measured separately 
for each $g,r,i$ filter using fake SNe inserted into the images
\citep{Dilday08}.
The \spec\ selection \eff\ ($\effspec$) has been estimated 
from matching data/MC distributions for redshift and for the 
fitted observer-frame magnitudes at the epoch of peak brightness. 
$\effspec$ is adequately described as a function of peak $r$-band
magnitude and the peak color $g-r$.
These \eff\ functions are available in tabular 
form.\footnote{See {\tt \$SNDATA\_ROOT/models/searcheff} in \SNANA\ download.}
For the {\SNLS}, $\effspec$ has been evaluated as a
function of peak $i_M$-band magnitude ($\MiPeak$) in
Figure~9 of \cite{Perrett10}.
For the \SNANA\ simulation we parameterize this function as
\begin{equation}
  \effspec^{\rm \SNLS} = 
       \left\{
        0.5 + \frac{1}{\pi}
        \tan^{-1} \left[ \frac{24.3 - \MiPeak}{0.2} \right] 
      \right\}
        \times \effCOR
  \label{eq:effspec_SNLS}
\end{equation}
where 
$\effCOR=1$ for $\MiPeak<23$ and 
$\effCOR = \exp[(23-\MiPeak)/0.63]$ for $\MiPeak>23$.
The function in parentheses is a first-order estimate
and $\effCOR$ is a correction obtained from a fit to the 
data/MC ratio as a function of $\MiPeak$.

\begin{figure}[h!]
\centering
\epsscale{\xScale}  % 1.1 for emulateapj 
\plotone{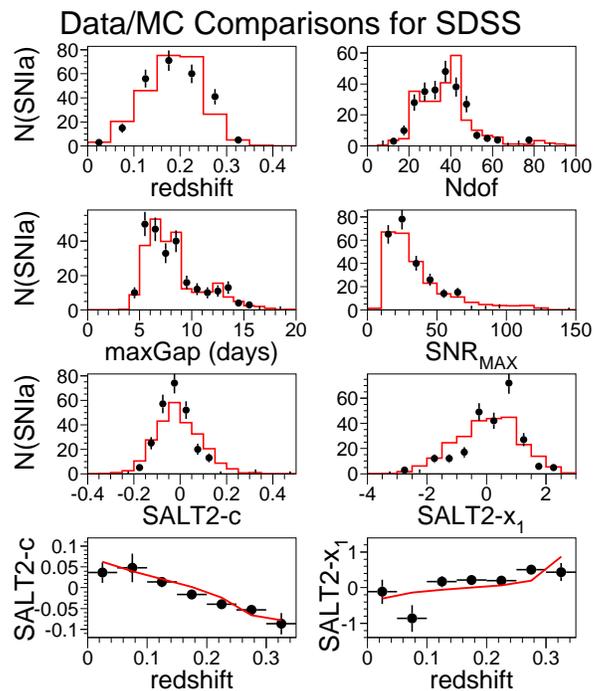}
  \caption{
    Comparison of distributions for \SDSS\ data (dots) and 
    MC (histogram), where each MC distribution is scaled to 
    have the same sample size as the data. 
    The distributions are 
    redshift, 
    number of degrees of freedom in the \SALTII\ light curve fit, 
    maximum rest-frame time difference (gap) between {\obss},
    maximum S/N, 
    fitted \SALTII\ color ($c$) and stretch parameter ($x_1$).
    The bottom two panels show the mean fitted 
    \SALTII\ color ($c$) and shape parameter ($x_1$) vs. redshift. 
  }
  \label{fig:ov1_SDSS}
\end{figure}

\begin{figure}[h!]
\centering
\epsscale{\xScale}   % 1.1 for emulateapj 
\plotone{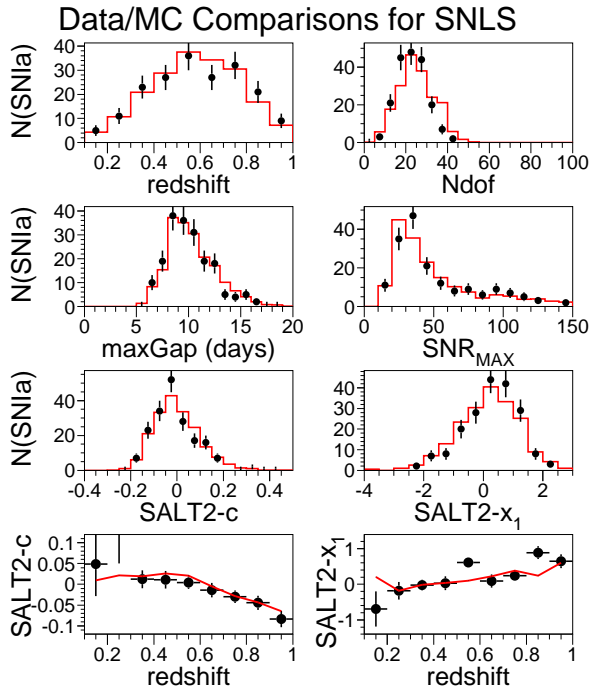}
  \caption{
    Same as Figure~\ref{fig:ov1_SDSS}, except for the \SNLS\ sample.
  }
  \label{fig:ov1_SNLS}
\end{figure}

For this analysis we generate MC samples 
with sizes corresponding to 
%% \MCDATAratioSDSS\ 
six times the data statistics.
The quality of the simulation for each sample is 
illustrated with several data/MC comparisons in 
Figures~\ref{fig:ov1_SDSS} and \ref{fig:ov1_SNLS};
the overall agreement is good.

% ##################################
 \subsection{Simulation of Noise}
 \label{subsec:sim_noise}
% ##################################

Since the three scatter-dependent variables 
(Hubble scatter, $B-V-c$  \photoz\ precision)
are sensitive to the flux \uncs,
it is important to accurately simulate these \uncs. 
The simulation strategy is to first calculate the \uncs\ 
from a model based on measurements of the 
sky level and PSF. 
To accurately check the model,
the true \unc\ for each \obs\ in the data\footnote{
True data \uncs\ are from the photometric pipeline codes 
describe in Section~\ref{sec:data}} % end footnote
is compared to the calculated model \unc.
Discrepancies between the true and calculated \uncs\
are corrected by fitting for ad-hoc parameters.
The simulated \unc\ model in photoelectrons 
($\sigSIM$) is given by
\begin{eqnarray}
  \sigSIM^2 = F & + & (A\cdot b) 
                  + (qF)^2 
                  +  (\sigOFF \cdot 10^{0.4\cdot\ZPTpe})^2 
        \nonumber \\
                & + &  \sigHOST^2
  \label{eq:uncModel}
\end{eqnarray}
where $F$ is the flux, 
$A = [2\pi\int {\rm PSF}^2(r,\theta) r dr]^{-1}$ 
is the noise-equivalent area,
$b$ is the effective sky level including dark current 
and readout noise, and 
$q$ and $\sigOFF$ are fitted ad-hoc parameters.
$\ZPTpe$ is defined such that the number of CCD photoelectrons
for a point source of magnitude $m$ is given by $10^{-0.4(m-\ZPTpe)}$;
thus the $\sigOFF$ term 
%corresponds to a fixed magnitude that 
is independent of the PSF, sky level and host-galaxy.
The quantity $\sigHOST$ is simulated for the \SDSS\ sample 
using a library of galaxies that have a \spec\ redshift 
and a well measured profile consistent with either 
exponential or de Vaucouleurs.

\begin{figure}[ht!]
\centering
\epsscale{\xScale}  % 1.1 for emulateapj 
\plotone{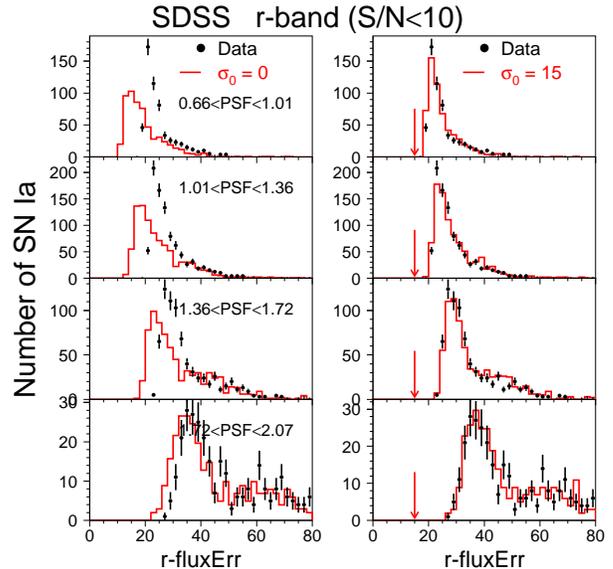}
  \caption{    
    Left panels show the $r$-band flux-\unc\ distribution
    for the data (black dots) and for the calculation
    (histogram), $\sigSIM = \sqrt{F+Ab}$ with $\sigOFF=0$ and $q=0$.
    Each descending plot is for a different PSF range (FWHM, arcsec)
    as indicated on the plot.
    Right panels show the $\sigSIM$ calculation including the best-fit
    $\sigOFF$ value labeled on the plot and shown by the arrow.
    The flux unit shown is that used for light curve fits
    (not photoelectrons).
  }
  \label{fig:SDSS_noisefit_r}
\end{figure}

To check the \unc\ calculation, $\sigSIM$ is computed
for each epoch in the data and compared to the measured
\unc\ $\sigDATA$.
The left panels in Figure~\ref{fig:SDSS_noisefit_r} show that
the first two terms, $F+Ab$, are not adequate to reproduce
the \obss.
The right panels in Figure~\ref{fig:SDSS_noisefit_r} 
show that the fitted $\sigOFF$ term
results in good agreement over a wide range of PSF values.
A separate $\sigOFF$ value is evaluated for each filter
and for each sample (Table~\ref{tb:sigOFF}).
The quadratic term $q$ is sensitive
to large flux values with ${\rm S/N} \sim 10^2$.
The value of $q$ is obtained from minimizing
$\chi^2 = \sum_s[(\sigSIM/\sigDATA)_s -1]^2$,
where the sum ($s$) is over $\log_{10}({\rm S/N})$ bins;
$q\simeq 0.01$ for the SDSS bands,  and 
$q\sim 0.001$ for the SNLS bands.
For SNLS, it is difficult to interpret this low value of $q$ because 
\uncs\ on the SN flux (Poisson noise and flat-fielding noise) 
only arise via the normalization of errors based on the
intra-night flux scatter.

Finally, note that the terms $F + (A\cdot b) + \sigHOST^2$ 
are determined from \obss\ and first principles, 
while  $q$ and $\sigOFF$ are empirically determined parameters.
The $q$ term corresponds to a zeropoint \unc.
The $\sigOFF$ term is not understood, although this term works 
surprisingly well for both  the \SDSS\ and \SNLS\ surveys
even though the respective photometry codes are 
independent.

\begin{table}[ht]
\caption{
  Fitted $\sigOFF$ values for \SDSS\ and \SNLS.
  } % end caption
\begin{center}
\begin{tabular}{c | c c}
\tableline\tableline
            &  \multicolumn{2}{c}{$10^{11}\times \sigOFF$ for} \\
   Filter   &  \SDSS\  &  \SNLS\   \\
       \hline\hline
% --------------------------------------------------
   $u$      &  28  &  ---   \\
   $g$      &  10  &  0.22  \\
   $r$      &  15  &  0.30  \\
   $i$      &  23  &  1.08  \\
   $z$      &  63  &  1.62 \\
\tableline  % ------------------------------------
\end{tabular}
\end{center}
  \label{tb:sigOFF}
\end{table}

% ##################################
 \bigskip
 \subsection{Intrinsic-scatter Models}
 \label{subsec:scatter_models}
% ##################################

\begin{figure}[hb!]
\centering
\epsscale{\xScale}  % 1.1 for emulateapj 
\plotone{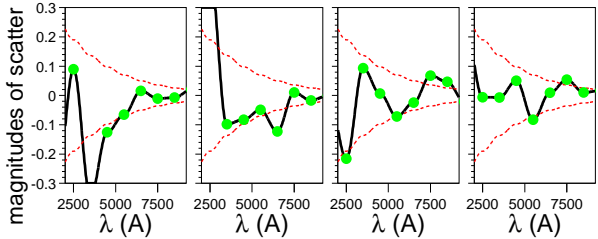}
  \caption{
    Illustration of FUN-COLOR for four simulated SNe.
    Black lines show the $\Lrest$-dependent variation.
    Solid dots show the Gaussian-random variations at
    the nodes, 
    and dashed lines show $\pm 1~\sigma$ vs. $\Lrest$.
  }
  \label{fig:FUN-COLOR}
\end{figure}

The intrinsic scatter models are 
summarized in Table~\ref{tb:models}.
These models are defined as wavelength-dependent perturbations 
to the \SALTII\ spectral model, and these perturbations average
to zero so that the underlying \SALTII\ model is not changed.
All models are independent of redshift,
and only the explosion models from KRW09 depend on epoch.
We begin with the phenomenological functions
(see ``FUN'' prefix) with parameters arbitrarily
chosen to increase the scatter.
The coherent model (FUN-COH) assigns a 
single magnitude shift for all wavelengths;
for each SN this shift is given by a 
Gaussian-random number with $\sigCOH = \sigCOHvalue$~mag.
The other two FUN functions are designed to probe a wider 
variety of wavelength-dependent scatter
with a coherence length of a few hundred {\AA}.
First a sequence of nodes is defined at $\lamFUNvalue$~{\AA} 
intervals in the rest frame. An independent
Gaussian random scatter is selected at each node with 
$\signode = \sigV\FUNLAM$ so that there is more scatter 
at bluer wavelengths.
The variation is the same at all epochs.
A continuous function of wavelength is constructed by 
connecting the node values with sine functions so that 
the derivative is zero at each node.
FUN-COLOR is defined with $\sigV = \sigFUNONEvalue$
and is shown in Figure~\ref{fig:FUN-COLOR} for a few
simulated SN.
FUN-MIX is defined with $\sigV = \sigFUNTWOvalue$ along with
a coherent term $\sigCOH = 0.09$~mag.

\begin{table}[ht]
\caption{
  Summary of Intrinsic-scatter Models.
  $\Lrest$ refers to rest-frame wavelength.
  } % end caption
\begin{center}
\begin{tabular}{ll}
\tableline\tableline
   Model  &              \\
   Name   &  Description \\
       \hline\hline
   NONE         & Poisson noise only \\  \hline
% ---------------------
   FUN-COH      &  Coherent variation at all $\Lrest$  \\
                & ($\sigCOH = \sigCOHvalue$~mag) \\
% ----
   FUN-COLOR   &  $\signode = \sigFUNONEvalue\FUNLAM$ \\
               &  (Figure~\ref{fig:FUN-COLOR}) \\
% ---
   FUN-MIX     & $\signode = \sigFUNTWOvalue\FUNLAM$ \\
               &  and $\sigCOH=0.09$ \\
% -------------
  \hline
% -------------
   G10      &  $\sigCOH = 0.09$~mag and $\signode$  \\
            &   from solid curve in Figure~8 of G10 \\
% --
   C11\_0   &  $UBVRI$ correlation matrix. $\rhoUUU = 0$. \\
   C11\_1   &  Same for $UBVRI$, $\rhoUUU = +1$. \\
   C11\_2   &  Same for $UBVRI$, $\rhoUUU = -1$. \\
% --------------
  \hline
% ---------
   \multicolumn{2}{l}{(KRW09 2D explosion models\tablenotemark{a})}   \\
  iso3\_dc1  & $\Nignit=~60$, $\Kamin = 1$     \\
  iso6\_dc1  & $\Nignit=100$, $\Kamin = 1$     \\
  iso6\_dc2  & $\Nignit=100$, $\Kamin = 250$   \\
  iso6\_dc3  & $\Nignit=100$, $\Kamin = 750$   \\
  iso6\_dc4  & $\Nignit=100$, $\Kamin = 1500$  \\
  iso6\_dc5  & $\Nignit=100$, $\Kamin = 2250$  \\
  iso8\_dc3  & $\Nignit=150$, $\Kamin = 750$   \\
  asym1\_dc3\tablenotemark{b}
             & $\Nignit=120$, $\Kamin = 750$  \\
\tableline  % ------------------------------------
\end{tabular}
  \tablenotetext{1}{``iso'' = isotropic  and ``dc'' = detonation criteria.}
  \tablenotetext{2}{This mis-labeled asym model is really an isotropic model.}
\end{center}
  \label{tb:models}
\end{table}

The next two models (G10 and C11)  
are based on measurements from data combined with
assumptions needed to create a model that is a continuous
function of wavelength.
The G10 error model was obtained as part of the \SALTII\ training
process in which they minimized the likelihood of light curve
amplitude residuals using a parametric function of central
rest-frame wavelength, assuming uncorrelated residuals 
in different passbands.
The resulting wavelength-dependent function (Figure~8 of G10) 
has approximate values of 0.07, 0.03, 0.02, 0.03, 0.06~mag 
at the $U,B,V,R,I$ central wavelengths, respectively.
This function is not intended to represent a
wavelength dependent scatter,
but rather it is a model of independent
broadband scatter as a function of central wavelength.
To translate this broadband model into a wavelength model,
independent random scatter values ($\signode$) are selected
every 800~{\AA}, and these node values are connected with 
the same sine-interpolation that is used for the 
phenomenological ``FUN'' functions. 
Since this procedure reduces the resulting broadband scatter, 
$\signode$ is multiplied by $1+(\Lrest-2157)/9259$
so that the simulated $UBVRI$ broadband scatter matches the G10 function.
In addition to a wavelength-dependent function, 
the G10 model includes a coherent term, $\sigCOH=0.09$~mag.

The model of 
\citet[hereafter C11]{Chotard_thesis,Chotard_2011}
is based on a covariance scatter matrix among
the $UBVRI$ filter passbands, and is derived from
an analysis of spectral correlations using
high quality spectra from the Supernova Factory \citep{SNF2002}.
The broadband covariance model is translated into a
wavelength-dependent model as follows.
First, the model is extrapolated to wavelengths below the 
$U$ band (3600~\AA) by defining an ad-hoc $\UU$ filter
with central wavelength $\bar\Lobs = 2500$~\AA.
The G10 scatter value of $\signode = \sigFARUV$~mag is used
for $\UU$,
and we model three different assumptions for the
reduced correlation between $\UU$ and $U$:
$\rhoUUU = 0$ (incoherent,C11\_0), 
$\rhoUUU = +1$ (C11\_1), and 
$\rhoUUU = -1$ (C11\_2).
For each simulated SN, 
six random magnitude shifts are selected according 
to the C11 correlation matrix in upper half of
Table~\ref{tb:C11matrix};
these shifts are assigned to the central 
${\UU}UBVRI$ wavelengths.
A continuous function of wavelength is obtained by
interpolating these six points with a sine function,
similar to the FUN-COLOR interpolation in 
Figure~\ref{fig:FUN-COLOR}. 
Finally, the scatter function is multiplied by 1.3
to compensate for the fact that the wavelength interpolation
reduces the broadband covariances.
The correlation matrix realized by the simulation is
shown in bottom half of Table~\ref{tb:C11matrix}
for the C11\_1 model.
The realized correlation matrix is slightly different than the
input model because the input model is described by
broadband covariances, while the simulated model depends on 
wavelength. In principle a more finely tuned spectral model 
in the simulation would result in the exact C11 covariances, 
but we believe that the simple and approximate model used here 
is adequate, especially in light of the large and unknown 
\uncs\ on the covariances.

\begin{table}[hb!]
\caption{
  Reduced Correlation Matrix from the C11\_1 Model\tablenotemark{a}
  and Realized from the Simulation
  } % end caption
\begin{center}
\begin{tabular}{l | cccccc  }
\tableline\tableline
        &         &        &       &         &         &        \\
        &  $\UU$  &  $U$   &  $B$  &   $V$   &   $R$   &   $I$  \\
   \hline\hline
        &  \multicolumn{6}{c}{(From C11)}   \\
  $\UU$ & $+1.00$ & $+1.00$ & $-0.12$ & $-0.77$ & $-0.91$ & $-0.22$ \\ 
  $U$   & $+1.00$ & $+1.00$ & $-0.12$ & $-0.77$ & $-0.91$ & $-0.22$ \\
  $B$   & $-0.12$ & $-0.12$ & $+1.00$ & $+0.57$ & $-0.24$ & $-0.89$ \\
  $V$   & $-0.77$ & $-0.77$ & $+0.57$ & $+1.00$ & $+0.53$ & $-0.40$ \\
  $R$   & $-0.91$ & $-0.91$ & $-0.24$ & $+0.53$ & $+1.00$ & $+0.49$ \\
  $I$   & $-0.22$ & $-0.22$ & $-0.89$ & $-0.40$ & $+0.49$ & $+1.00$ \\
  Diag\tablenotemark{b}
      & $\sigFARUV$ &  0.06   &   0.04  &   0.05  &   0.04  &   0.08  \\ \hline
% -----
        &  \multicolumn{6}{c}{(Realized in simulation)}   \\
  $\UU$ & $+1.00$ & $+0.99$ & $+0.12$ & $-0.78$ & $-0.96$ & $-0.28$ \\
  $U$   & $+0.99$ & $+1.00$ & $+0.20$ & $-0.73$ & $-0.96$ & $-0.34$ \\
  $B$   & $+0.12$ & $+0.20$ & $+1.00$ & $+0.40$ & $-0.27$ & $-0.93$ \\
  $V$   & $-0.78$ & $-0.73$ & $+0.40$ & $+1.00$ & $+0.72$ & $-0.34$ \\
  $R$   & $-0.96$ & $-0.96$ & $-0.27$ & $+0.72$ & $+1.00$ & $+0.36$ \\
  $I$   & $-0.28$ & $-0.34$ & $-0.93$ & $-0.34$ & $+0.36$ & $+1.00$ \\
  Diag\tablenotemark{a}
        & $\sigFARUV$ & 0.073   &  0.036  &   0.051 &   0.042 &  0.083  \\
\tableline  % ------------------------------------
\end{tabular}
  \tablenotetext{1}{See Table 14.2 of \citet[PhD thesis]{Chotard_thesis}.}
  \tablenotetext{2}{Diag = $\sqrt{{\rm COV}_{ii}}$ with $i= \UU,U,B,V,R,I$
           are from the PhD thesis and differ slightly from those 
           given in \citet{Chotard_2011}.      }
\end{center}
  \label{tb:C11matrix}
\end{table}

The final class of brightness variations is based on 
2D explosion models with random
ignition points (KRW09), 
followed by radiative transfer calculations using the 
{\tt SEDONA} program \citep{SEDONA06}.
Isotropic models are obtained from ignition points that are 
randomly placed throughout the white dwarf (WD),
while asymmetric models are obtained from ignition points
within a cone whose apex is at the center of the WD.
Both the isotropic and asymmetric models result in 
explosion asymmetries and a viewing angle dependence
that contributes significantly to the intrinsic scatter.
The width-luminosity relation is related to the number
of ignition points ($\Nignit$) because $\Nignit$ affects the 
amount of pre-expansion before detonation, 
and hence the amount of $^{56}{\rm Ni}$ produced in the explosion. 
In a recent study by \citet[hereafter B11]{B11KRW09},
detailed comparisons between data and the KRW09 models 
were made. They conclude that the KRW09 models with
the best spectroscopic agreement also have the best
photometric agreement, and they identified a subset
of eight models with the best agreement to data.
Here we use these same eight models; they are shown in 
Table~\ref{tb:models} along with a few parameters
describing the number of ignition points and the
detonation criteria.  All of these models have an
isotropic distribution of ignition points, and B11
note that radial fluctuations in isotropic models 
can lead to significant viewing angle asymmetries.

We initially used these KRW09 models in the \SNANA\ simulation 
to generate light curves corresponding to the \SDSS\ and \SNLS. 
While the data/MC comparisons are visually impressive, 
the simulated light curves are not adequate for this study
because the \SALTII\ light curve fits are in general rather poor.
This trend of poor light curve fits was also noted in B11.

\begin{figure}[ht!]
\centering
\epsscale{\xScale}  % 1.1 for emulateapj 
\plotone{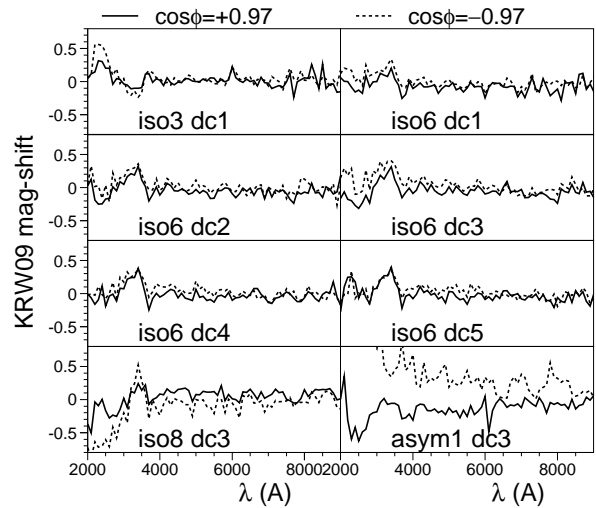}
  \caption{
    Magnitude shift applied to the \SALTII\ spectral model
    vs. wavelength for the KRW09 models.
    The solid and dashed  curves reflect different
    viewing angles as indicated in the legend above the plots.
    The label on each panel indicates the specific explosion model
    as defined in Table~\ref{tb:models}.
  }
  \label{fig:KRW09_magshift}
\end{figure}

Instead of attempting an absolute prediction with the KRW09 models,
we have instead used these models as a perturbation on the
\SALTII\ model. In short, the \SALTII\ model describes the
stretch and color relations, while the KRW09 models are used
to describe the intrinsic scatter. The spectral flux ($F$)
is given by
\begin{equation}
   F = F_{\rm SALT2} \times 
   \frac{\FKRW({\rm random}~\cos\phi)}{\FKRW(\cos\phi=0)}
\end{equation}
where $\phi$ is the viewing angle.
The corresponding mag-shifts are illustrated in
Figure~\ref{fig:KRW09_magshift} as a function of wavelength
for a two extreme viewing angles.

% ##################################
 \section{Analysis}
 \label{sec:anal}
% ##################################

Here we describe the determination of the three
scatter-dependent quantities used to evaluate
models of intrinsic brightness variations.
All analyses are based on light curve fits
using the \SALTII\ model.

% ----------------------------------------
%  \bigskip
  \subsection{Review of \SALTII\ Model}
  \label{subsec:anal_SALT2}
% ----------------------------------------

The \SALTII\ SN~Ia model flux is a function of 
wavelength ($\lambda$) and time ($t$) in the rest-frame, 
\begin{equation}
   F(t,\lambda)   =  x_0\left[
      M_0(t,\lambda) + x_1 M_1(t,\lambda)
      \right]
     \times \exp[c\cdot {\rm CL}(\lambda)]
   \label{eq:Frest}
\end{equation}
where the spectral sequences ($M_0$ and $M_1$)
and color law (${\rm CL}(\lambda)$) 
are derived from the training in G10.
Synthetic photometry in the observer frame is obtained
by redshifting Eq.~\ref{eq:Frest} and multiplying by
the filter response and Galactic transmission.
The overall scale ($x_0$), stretch ($x_1$), color ($c$)
and time of peak brightness ($t_0$)
are determined for each SN in a light curve fit that 
minimizes a $\chi^2$ based on the difference between the 
data and synthetic photometry.
Eq.~\ref{eq:Frest} is valid for $2000 < \lambda < 9200$~{\AA},
and the model is valid for observer-frame filters that satisfy
\begin{equation}
  2800 < \bar\Lobs/(1+z) < 7000~~{\rm \AA},
  \label{eq:Lobs}
\end{equation}
where $\bar\Lobs$ is the central wavelength of the filter.

An effective $B$-band magnitude is defined to be
$m_B = -2.5\log_{10}(x_0) + 10.635$;
this is the observed magnitude through an idealized
filter that corresponds to the $B$ band in the rest-frame
of the SN. The fitted distance modulus is given by
\begin{equation}
   \MUfit = m_B - M + \alpha x_1 - \beta c
   \label{eq:muSALT2}
\end{equation}
where $\alpha$, $\beta$ and $M$ are determined
from a global fit to all of the SNe using the
``{\SALTtomu}'' program  described in M11
and below in Section~\ref{sec:results_Hubble}.

% ----------------------------------------
  \subsection{Hubble Scatter}
  \label{subsec:anal_hubble}
% ----------------------------------------

The well known Hubble scatter is defined as the dispersion on $\DMU$,
the difference between the fitted (measured) distance modulus 
and the distance modulus calculated from the best-fit 
cosmological parameters.
To simplify the analysis here we do not fit
for the $\alpha$ and $\beta$ parameters,
nor do we fit for the best-fit cosmological parameters.
Instead we compute the dispersion of
\begin{equation}
    \DMU \equiv \MUfit({\rm Eq.}\ref{eq:muSALT2}) - \MUcalc(z,\OM,\OL,w) 
    \label{eq:DMU}
\end{equation}
where $\alpha=0.11$, $\beta=3.2$, $M=-19.36$ 
($H_0=70$~km s$^{-1}$ Mpc$^{-1}$ ), 
and $\MUcalc$ is the
calculated distance modulus assuming a 
$\LCDM$ cosmology with $w=-1$, $\OM=0.3$, $\OL=0.7$.
Although the fitted $\alpha$ and $\beta$ may have given
slightly different $\DMU$ values,
the resulting bias is more than an order of magnitude smaller
than the dispersion, and hence the impact of this approximation
is negligible.

% ----------------------------------------
  \subsection{Color Precision}
  \label{subsec:anal_color}
  \newcommand{\Mstar}{M^{\star}}
% ----------------------------------------

The color precision test compares the fitted \SALTII\ color
($c$) to the true $B-V$ rest-frame color at the epoch of
peak brightness. Since the fitted color is really a color
excess, $c=E(B-V)$, 
and the color also depends slightly on the stretch,
the true
$B-V$ color does not exactly correspond to $c$. 
A numerical examination of the model shows that with no 
intrinsic scatter,
\begin{equation}
   B-V = \cc = (1.016\times c) + (x_1/1250) + 0.0232
   \label{eq:cprime}
\end{equation}
and therefore we examine the dispersion on $B-V-\cc$.
The dispersion on $c-\cc$ is $\sim 0.001$,
more than an order of magnitude smaller than the 
dispersion on $B-V-\cc$,
and thus this correction has little effect.
The evaluation of $\cc$ is from simply plugging the
fitted $c$ and $x_1$ values into Eq.~\ref{eq:cprime}.
The naive rest-frame magnitudes $\Mstar_B$ and $\Mstar_V$
are obtained from Eq.~\ref{eq:Frest} using the 
best-fit parameters ($c$, $x_0$, $x_1$) and using the
$B$ and $V$ filter-transmission functions.
However, these naive magnitudes are not necessarily
the true values if there are intrinsic color variations.
To obtain a better approximation for the
magnitudes we fit only the two nearest 
observer-frame bands that bracket the $B$ or $V$ band in 
wavelength.

The details of the fitting procedure are as follows.
First a normal fit is done using all filters to determine
the fit parameters ($t_0$,$c$,$x_0$, $x_1$).
For each rest-frame band one additional fit is performed
using only the two nearest observer-frame bands and holding $
t_0$ and $x_1$ fixed from the normal fit.
The floated color ($c$) and distance ($x_0$) parameters provide 
the flexibility to fit both observer-frame bands regardless
of how much intrinsic color variation exists.
The two-band fit parameters are 
$c^B, x_0^B$ for the $B$ band, and 
$c^V, x_0^V$ for the $V$ band.
After finishing both two-band fits the 
$B-V$ color is computed as
\begin{eqnarray}
  B-V & = & {\Mstar}(T_B,~t_0,~x_1=0,~c^B,~x_0^B) 
                   \nonumber \\
      & - & {\Mstar}(T_V,~t_0,~x_1=0,~c^V,~x_0^V)
\end{eqnarray}
where $\Mstar$ is the magnitude computed from Eq.~\ref{eq:Frest}
using filter-transmission functions $T_{B,V}$, and with $x_1=0$
so that all $B-V$ colors correspond to an SN~Ia with 
the same stretch. 

This fitting procedure was tested on an \SNLS\ simulation
in which the maximum S/N was artificially set to 1000 for 
every SN regardless of redshift. The rms on $B-V-\cc$
is 0.002 mag, an order of magnitude smaller
than the observed dispersion.

% ----------------------------------------
  \subsection{Photo-$z$ Precision}
  \label{subsec:anal_photoz}
% ----------------------------------------

The \photoz\ precision is based on the difference between
the SN redshift determined from broad band photometry and 
the more precise \spec\ redshift.
The basic \photoz\ method is to extend the usual methods of 
fitting light curves to include the redshift as a 
fifth fit parameter.
Particular attention is needed to estimate initial parameter
values near those corresponding to the global minimum-$\chi^2$,
and to iteratively determine which filters satisfy Eq.~\ref{eq:Lobs}.
Details of the \photoz\ fitting process are given in K10.

There are two modifications in our \photoz\ fitting
procedure compared to K10.
The first change is that we use the known \spec\
redshift as the initial estimate in order to reduce catastrophic
outliers. The fitting task has thus been changed to find a local
\photoz\ minimum near the true redshift instead of searching
the entire redshift range for a global minimum.
The second change is related to estimating the initial 
parameter $x_0$ for each color value along the coarse-grid 
search in color.
In K10, $x_0$ at each grid point was calculated using the current
color, \photoz\ and a reference cosmology. Here we analytically
minimize for $x_0$, making the fitted \photoz\ less sensitive
to the absolute brightness. 
To check that the fitted \photoz\ depends only on the
SN colors we have applied this method to simulations
with no intrinsic scatter and with the coherent scatter model
(see COH entry in Table~\ref{tb:models});
the \photoz\ precision is the same in both cases.

% ------------------------------------
  \subsection{Statistics Summary}
  \label{subsec:stats}
% ------------------------------------

After applying the selection requirements in Section~\ref{sec:data},
along with the light curve fitting requirements for each
dispersion variable, the number of SNe~Ia for each sample 
and for each dispersion variable
is shown in Table~\ref{tb:NSN}

The smaller \photoz\ samples arise from a light curve 
fitting requirement. For each successive fit iteration,
observer-frame filters are added or dropped based
on which filters satisfy the \SALTII\ wavelength range
in Eq.~\ref{eq:Lobs} with $z=$~\photoz.
If any filter fails this wavelength requirement after 
the last fit iteration, the SN is rejected;
this requirement avoids fitting to wavelength regions
in which the \SALTII\ model may be poorly defined.

The smaller \SNLS\ sample for the $B-V-\cc$ analysis is
due to SNe~Ia at $z>0.7$; for these objects the observer-frame 
$i$ and $z$ bands no longer bound the rest-frame $V$ band.

\begin{table}[hb]
\caption{
  Number of SNe~Ia After Selection Requirements.
  } % end caption
\begin{center}
\begin{tabular}{l | ccc  }
\tableline\tableline
                &  \multicolumn{3}{c}{Number of SNe~Ia for:}   \\
   Survey       &  Hubble Resid  &  $B-V-\cc$    & Photo-$z$   \\
\tableline\tableline
   \SDSS\   &  \NSDSSforHub\  &  \NSDSSforColor\ & \NSDSSforPhotoz\  \\
   \SNLS\   &  \NSNLSforHub\  &  \NSNLSforColor\ & \NSNLSforPhotoz\  \\
\tableline  % ------------------------------------
\end{tabular}
\end{center}
  \label{tb:NSN}
\end{table}

% ##################################
 \bigskip\bigskip
 \subsection{Quantifying the Dispersions}
 \label{subsec:quant}
% ##################################

\begin{figure}[h!]
\centering
\epsscale{\xxScale}  % 1.15 for emulateapj 
\plotone{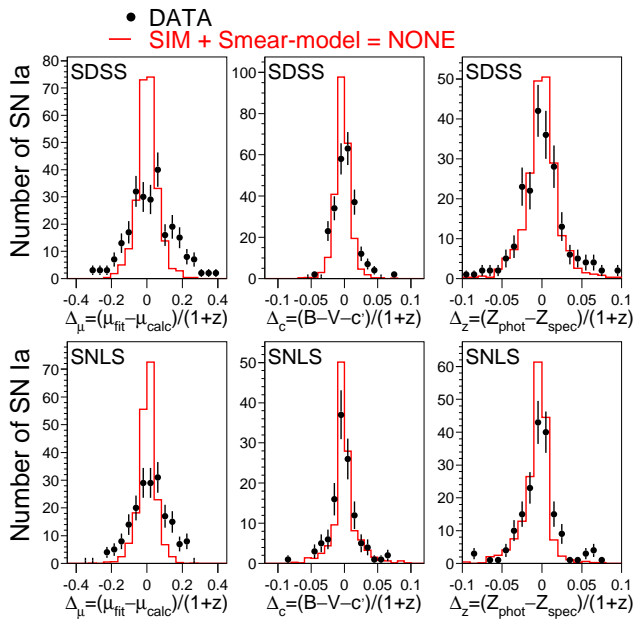}
  \caption{    
    Comparison of $\SYMDELmu$, $\SYMDELc$ and $\SYMDELz$ 
    distributions for data (dots) and MC (histogram).
    The upper plots are for the \SDSS\ and lower plots are for \SNLS.
    The MC includes Poisson noise but no intrinsic brightness
    variation, and each MC distribution is scaled to have the 
    same sample size as the data. 
  }
  \label{fig:ovdatasim2_smearNONE}
\end{figure}

The data and MC dispersions are measured from the
following variables,
\begin{eqnarray}
   \SYMDELmu & \equiv & \DIFmu \nonumber \\
   \SYMDELc  & \equiv & \DIFc  \nonumber \\
   \SYMDELz  & \equiv & \DIFz  
  \label{eq:SYMDEL_def}
\end{eqnarray}
where the $(1+z)^{-1}$ factor is included to reduce the 
redshift-dependent variation from measurement \uncs.
These three quantities are shown in 
Figure~\ref{fig:ovdatasim2_smearNONE}
for the data and MC, and for both the \SDSS\ and \SNLS\ samples.
The MC includes only Poisson noise (no intrinsic variation),
and hence the data-MC difference in the width illustrates
the size of the intrinsic component that is needed.
The $\SYMDELmu$ comparison shows the most obvious discrepancy.
The $\SYMDELz$ and $\SYMDELc$ discrepancies are more subtle,
indicating that the effect of color variations is smaller
than the coherent variation.

To quantify the dispersions we compute the median,
$\MED \equiv {\rm median}\vert\Delta_x\vert$,
where $x = \mu,z,c$ indicates the variable type.
In particular, we compute the MC/data ratio of medians,
\begin{equation}
   \RATIO  =  { \MED^{\rm MC} } / {\MED^{\rm Data}} ~.
   \label{eq:define_RATIO}
\end{equation}
With the correct model of brightness variations
we expect $\RATIO=1$ for all three variables and
for both surveys.

The \unc\ on the median is calculated as follows.
For $N$ SNe, the statistical \unc\ on $N/2$ is 
$\sigHalf = \sqrt{N}/2$.
The median \unc\ ($\sigMED$) is defined such that 
$N-\sigHalf$ values of $\vert\Delta_x\vert$ lie below 
$M-\sigMEDminus$ and $N+\sigHalf$ values lie below $M+\sigMEDplus$.
For a rapidly falling distribution we 
typically find that $\sigMEDplus > \sigMEDminus$.
Here we define a symmetric \unc, 
$\sigMED \equiv (\sigMEDminus+\sigMEDplus)/2$.

% Test with [SURVEY]_NONE_SNRMAX_E4
As a numerical crosscheck we analyze \SDSS\
simulations in which the exposure time is adjusted 
for each SN so that the maximum S/N is $10^4$.
The resulting dispersions, 
defined simply as a Gaussian fitted $\sigma$, are
% ---------------------------------------
%           fitted sigma on
%        DELMU   DELZ     DELc
% SDSS:  2E-4    2E-3     8E-4
% SNLS:  3E-4    2E-3     7E-4
% ---------------------------------------
%
$0.0003$, $0.002$, and $0.001$~mag
for the three variables (Eq.~\ref{eq:SYMDEL_def}),
respectively; 
these dispersions are more than an order of magnitude 
smaller than the dispersions observed in the data.

% ##################################
 \section{Results}
 \label{sec:results}
% ##################################

The MC/data ratio of medians, 
$\RATIO$ (Eq.~\ref{eq:define_RATIO}),
is shown in Figure~\ref{fig:RATIOS} for all of the
models in Table~\ref{tb:models}, and for both surveys.
With no model of intrinsic scatter, $\RATIO$ is well below
unity in  all cases. Adding a coherent scatter (FUN-COH)
predicts the Hubble dispersion ($\SYMDELmu$), 
but has no impact on the color and \photoz\ dispersion.
The FUN-COLOR model almost predicts the Hubble dispersion,
but may overestimate the color precision.
FUN-MIX has been artificially tuned to predict the 
dispersion in all quantities, although the \photoz\
dispersion may still be underestimated.
The G10, C11\_0  and C11\_1 models provide decent predictions,
with a slight underestimate in the \photoz\ dispersion.
The C11\_2 model underestimates the Hubble dispersion.
Recall that the G10 model includes only positive correlations,
mainly from the coherent term $\sigCOH=0.09$,
while the C11 model includes both positive and negative correlations.
This G10 versus C11 comparison illustrates that there can be
significant degeneracies among models of intrinsic brightness variations.
% Breaking these degeneracies will require additional constraints,
% such as the redshift dependence of $\RATIO$ and \spec\ analyses.
The KRW09 models give a poorer description of the dispersion
because the Hubble dispersion is always underestimated.

\newcommand{\wid}{1.1in}
\begin{figure*}[h!]
\centering
\epsscale{1.}  % for preprint
   \includegraphics[width=\wid]{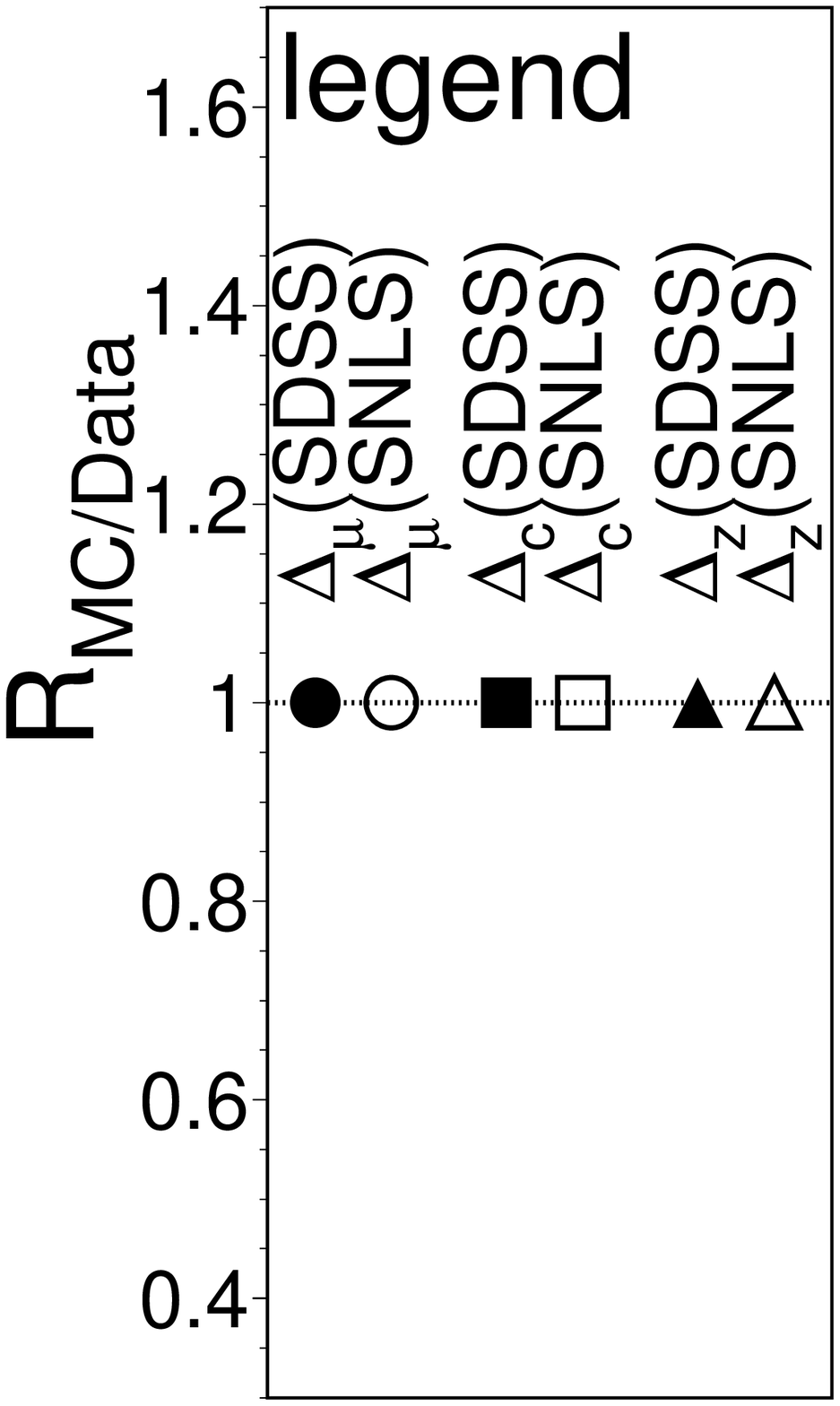}
   \includegraphics[width=\wid]{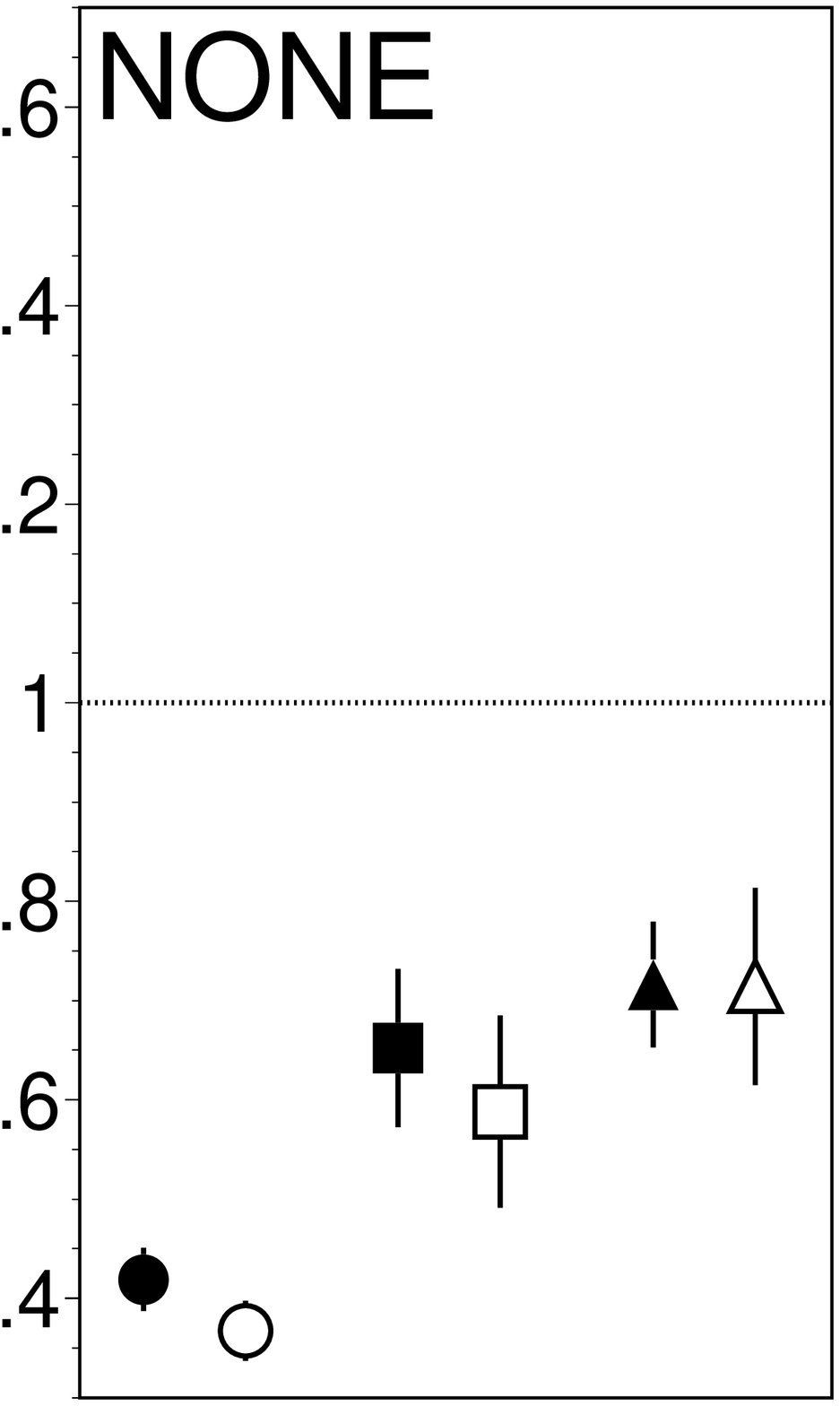}
   \includegraphics[width=\wid]{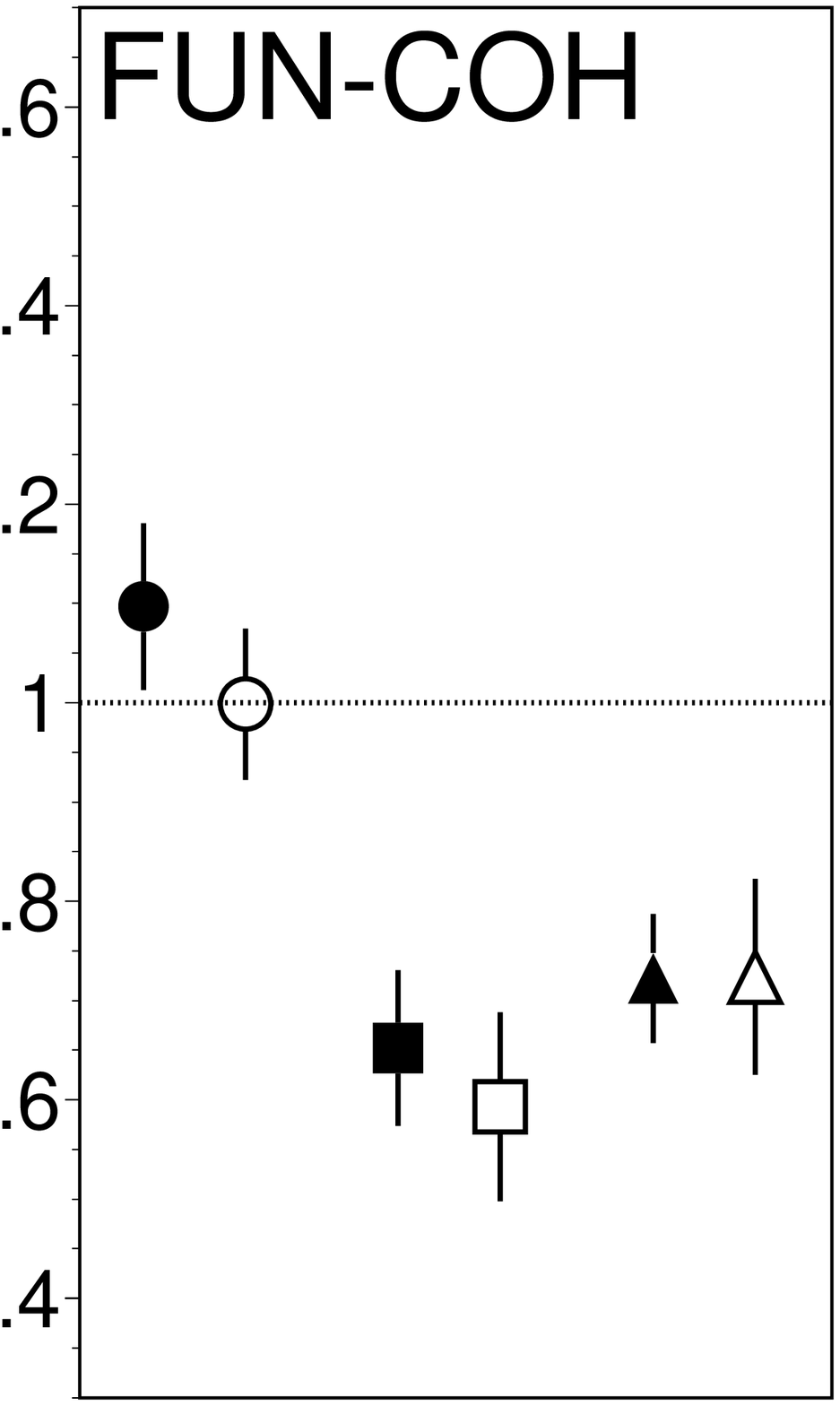}
   \includegraphics[width=\wid]{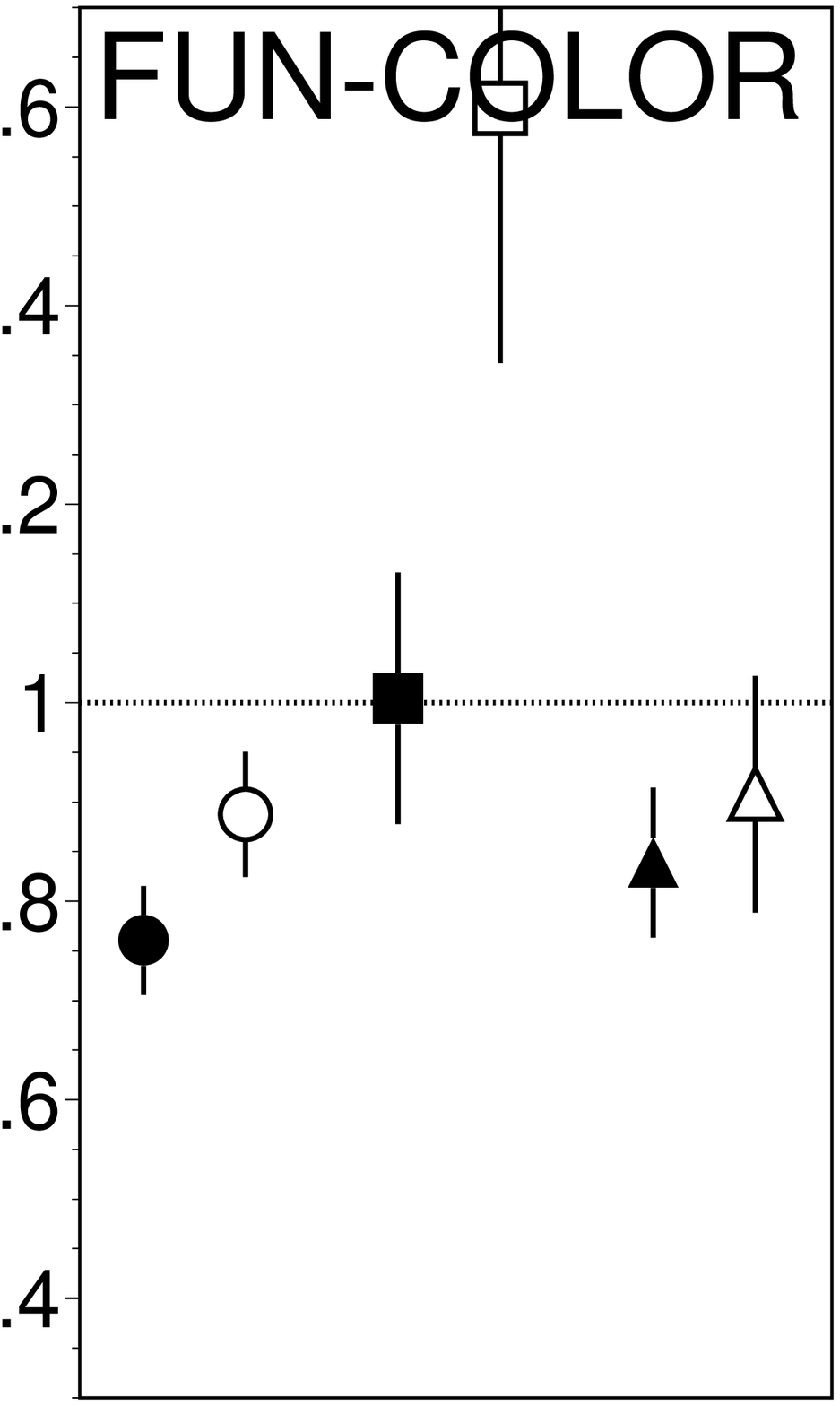}
   \includegraphics[width=\wid]{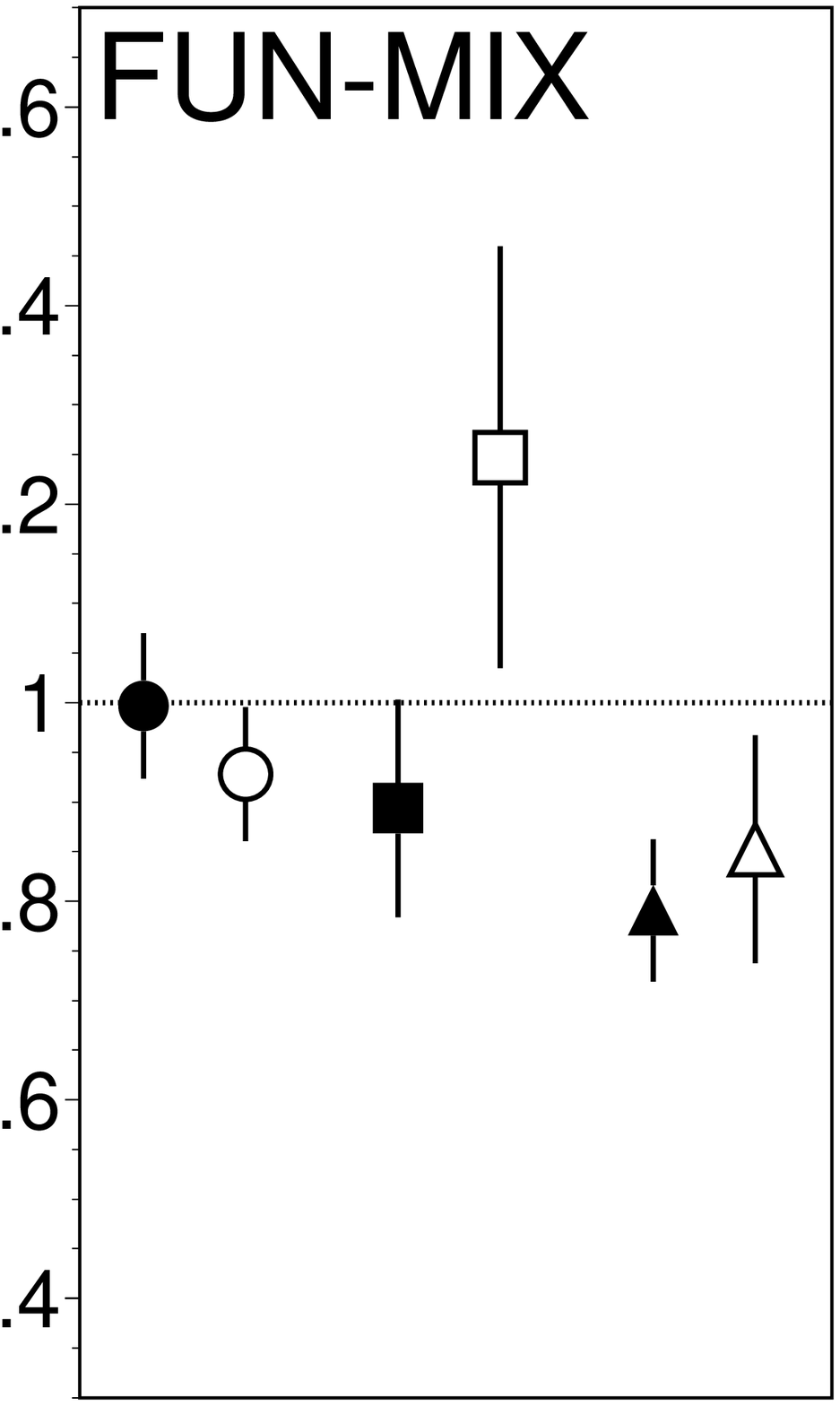}
   \includegraphics[width=\wid]{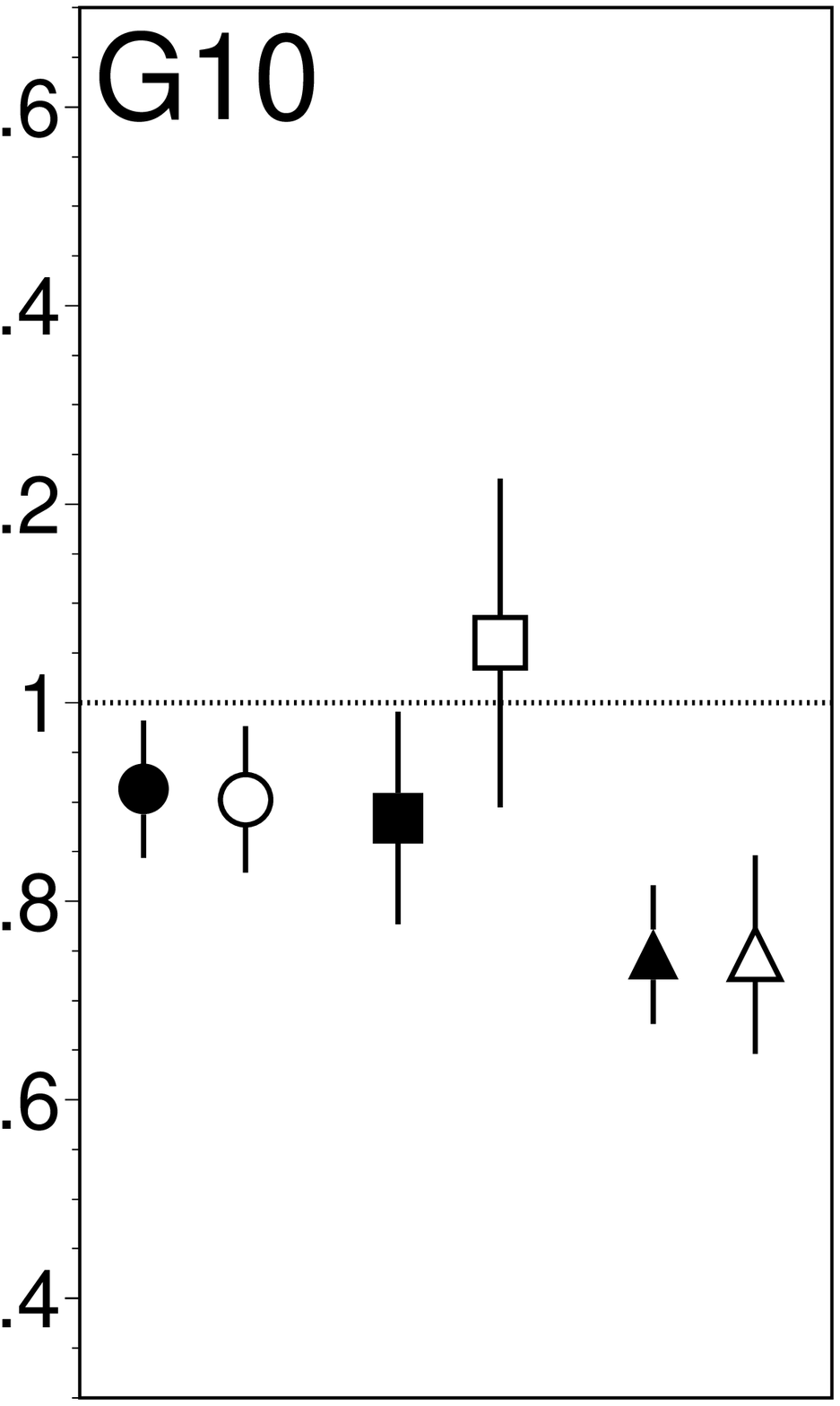}
   \includegraphics[width=\wid]{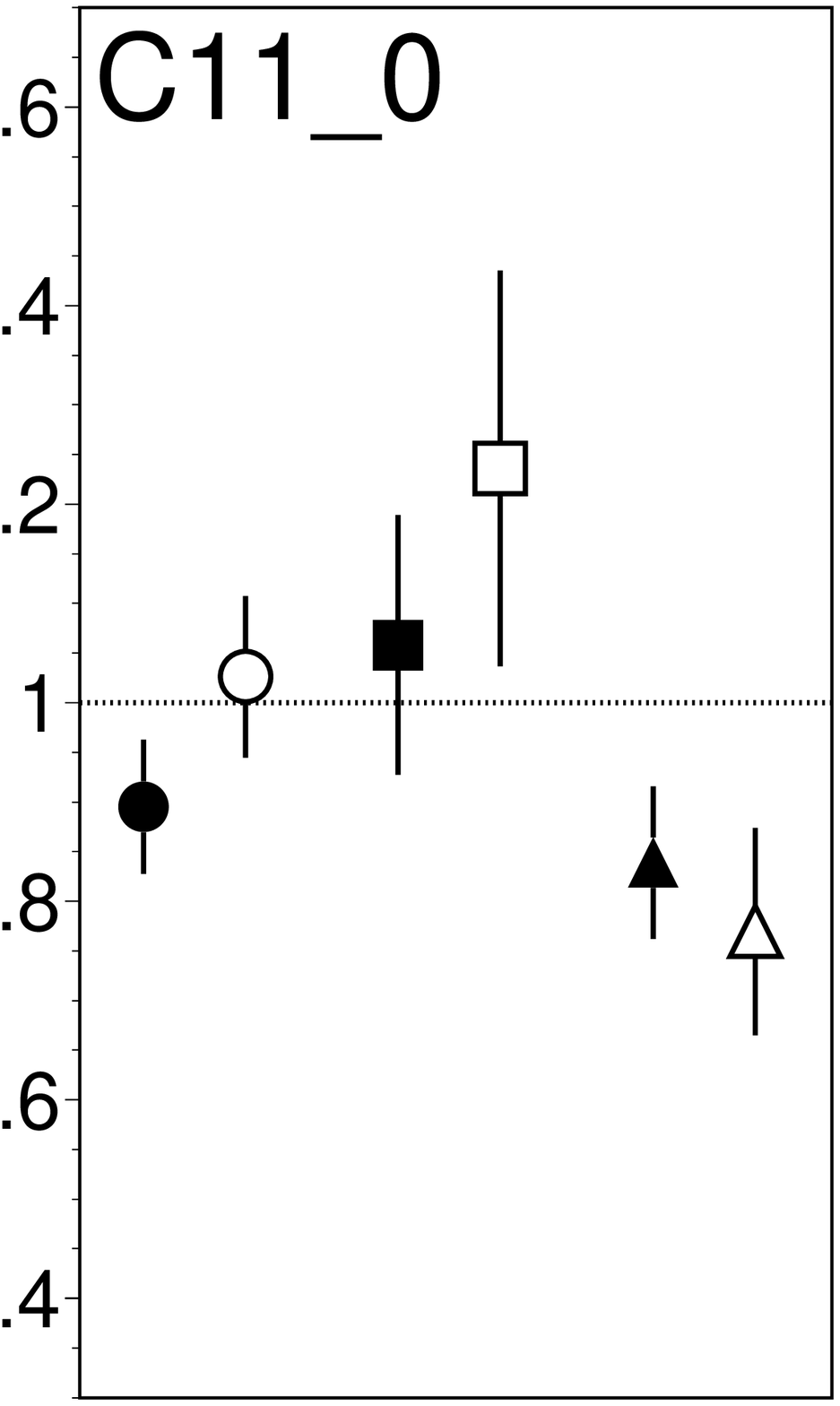}
   \includegraphics[width=\wid]{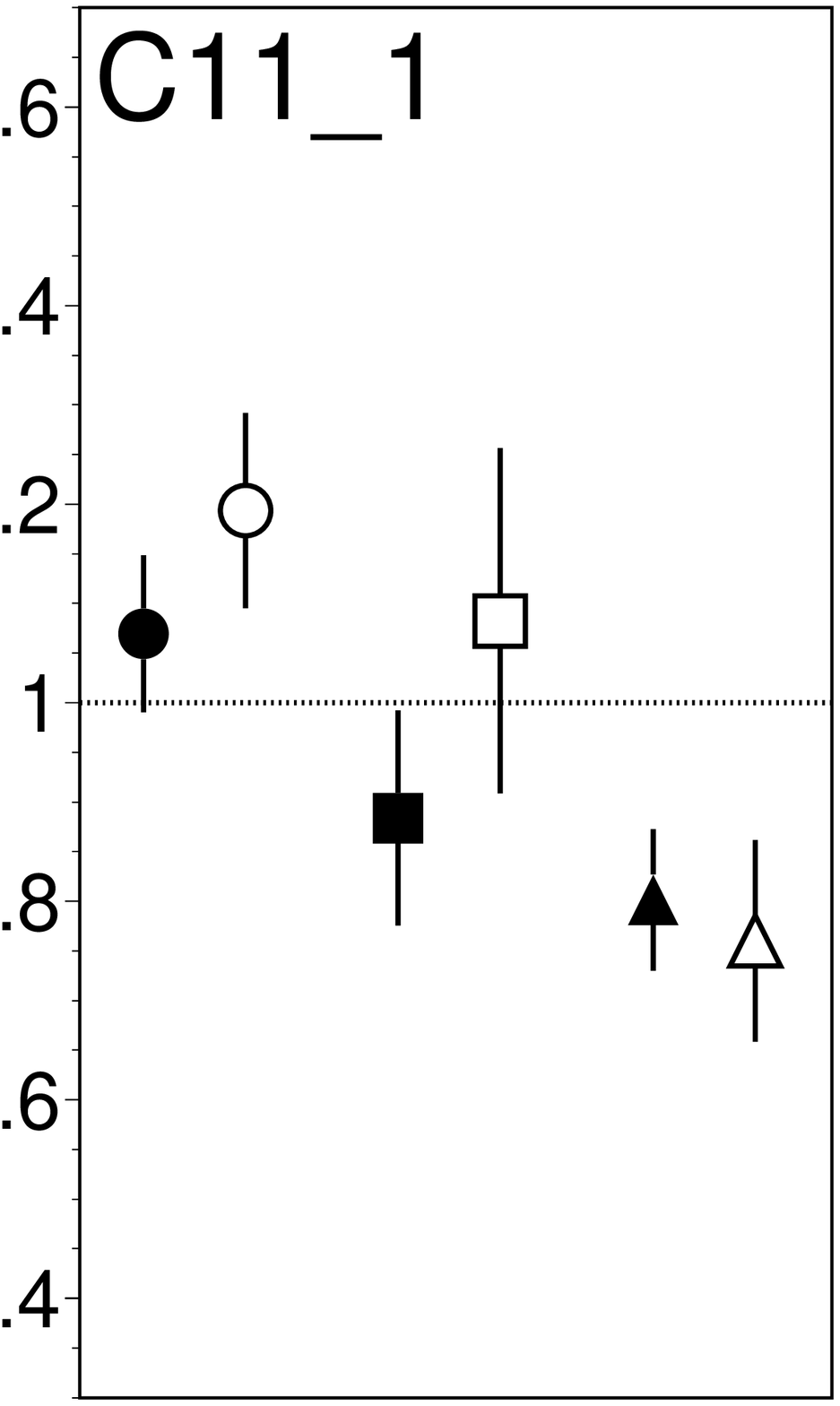}
   \includegraphics[width=\wid]{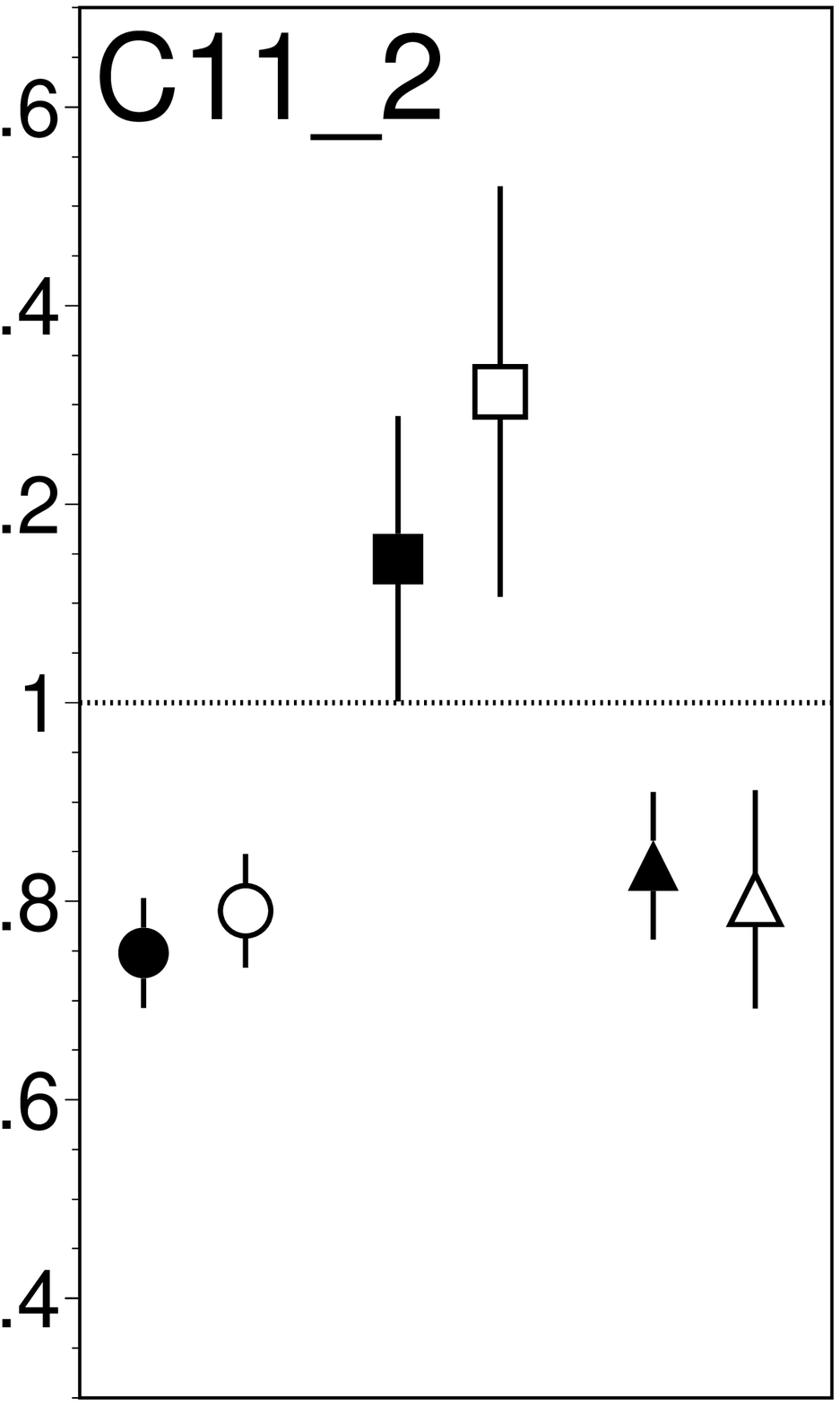}
   \includegraphics[width=\wid]{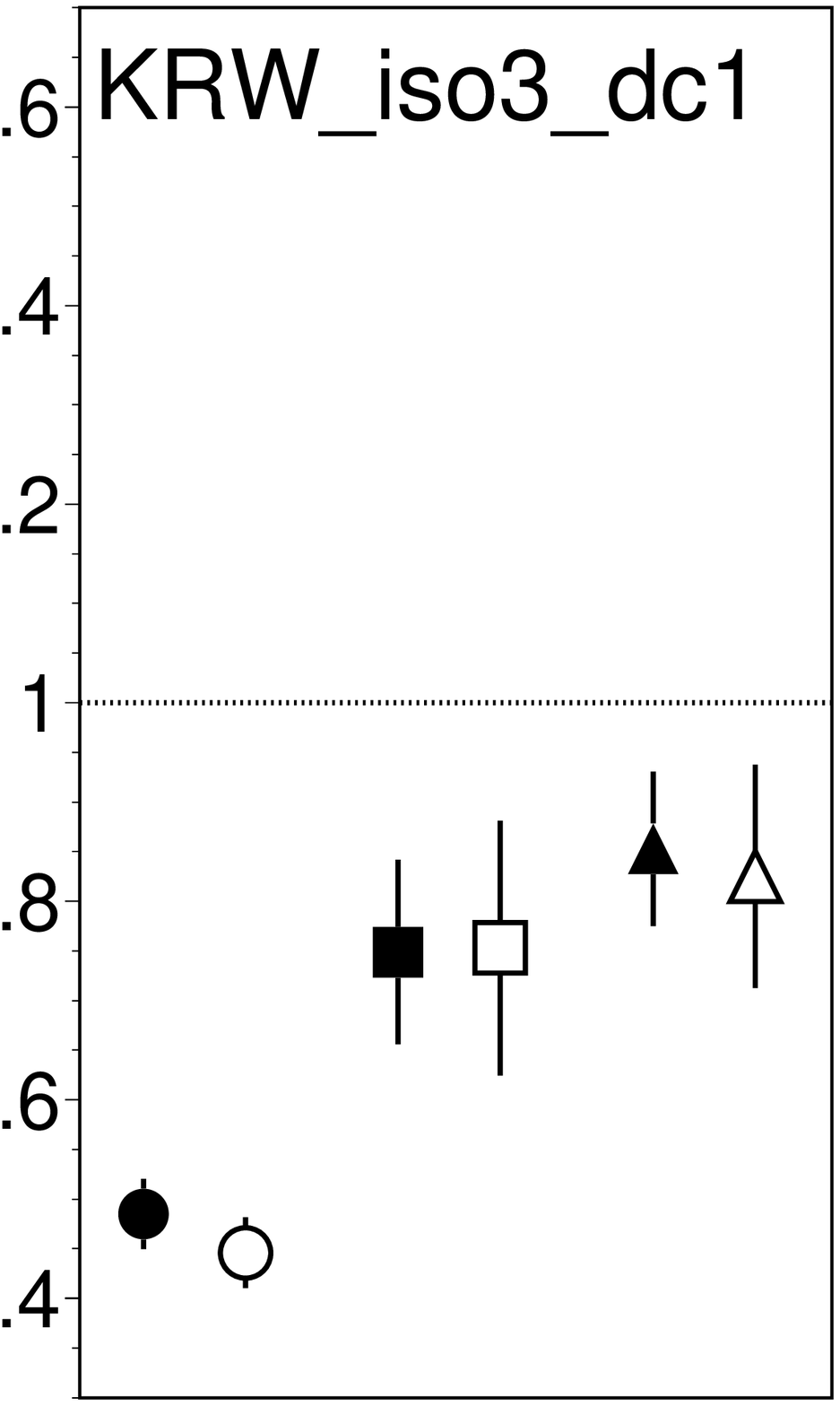}
   \includegraphics[width=\wid]{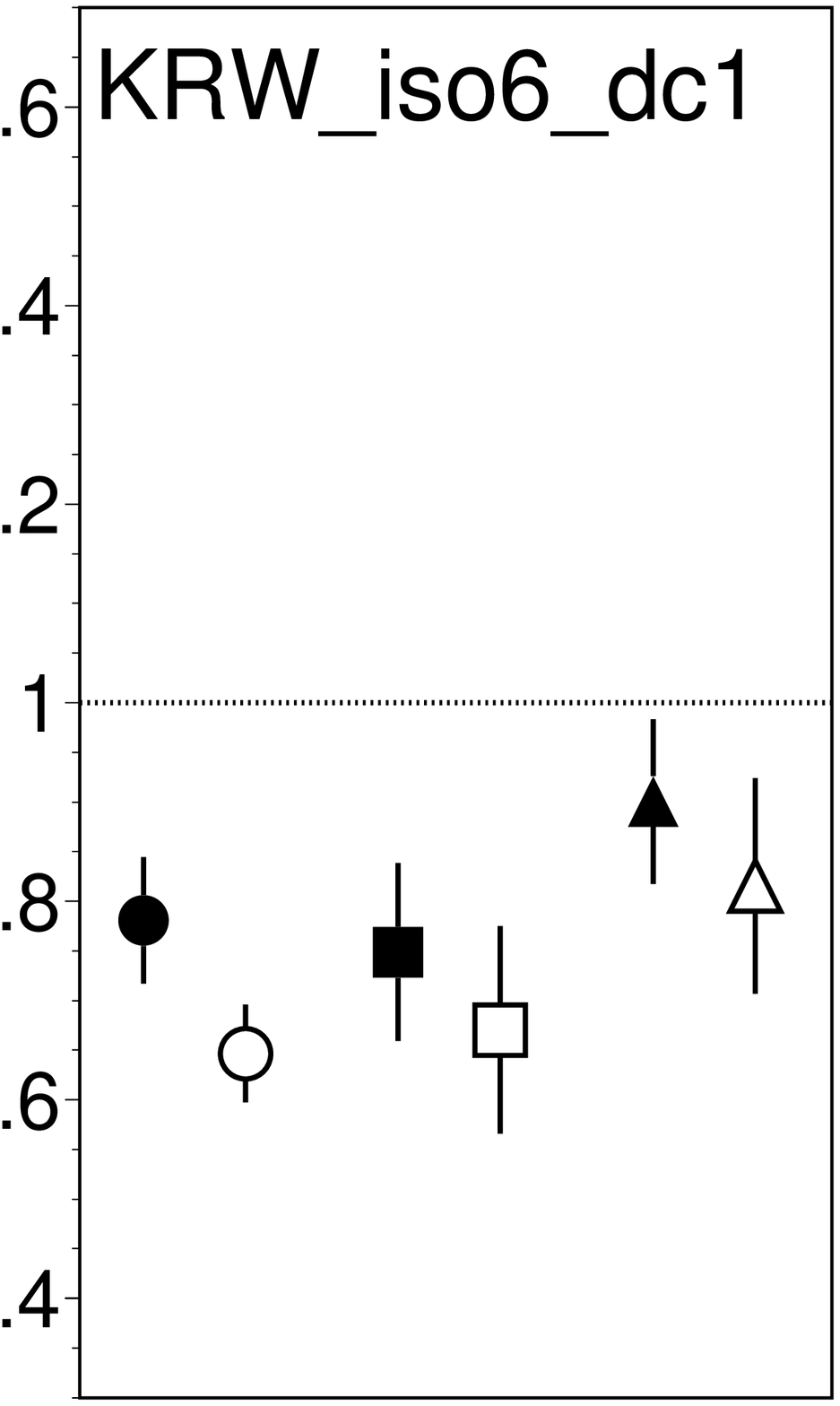}
   \includegraphics[width=\wid]{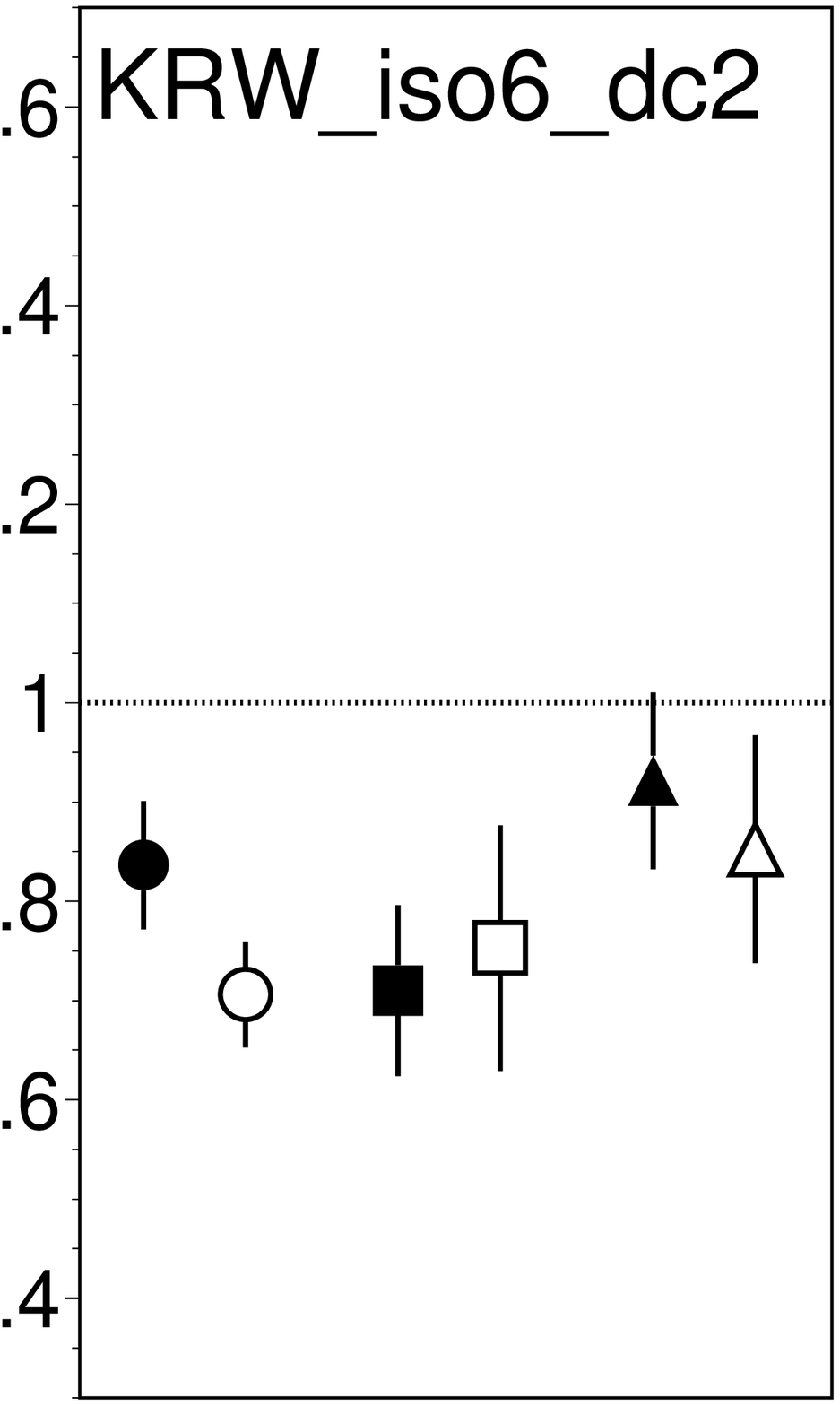}
   \includegraphics[width=\wid]{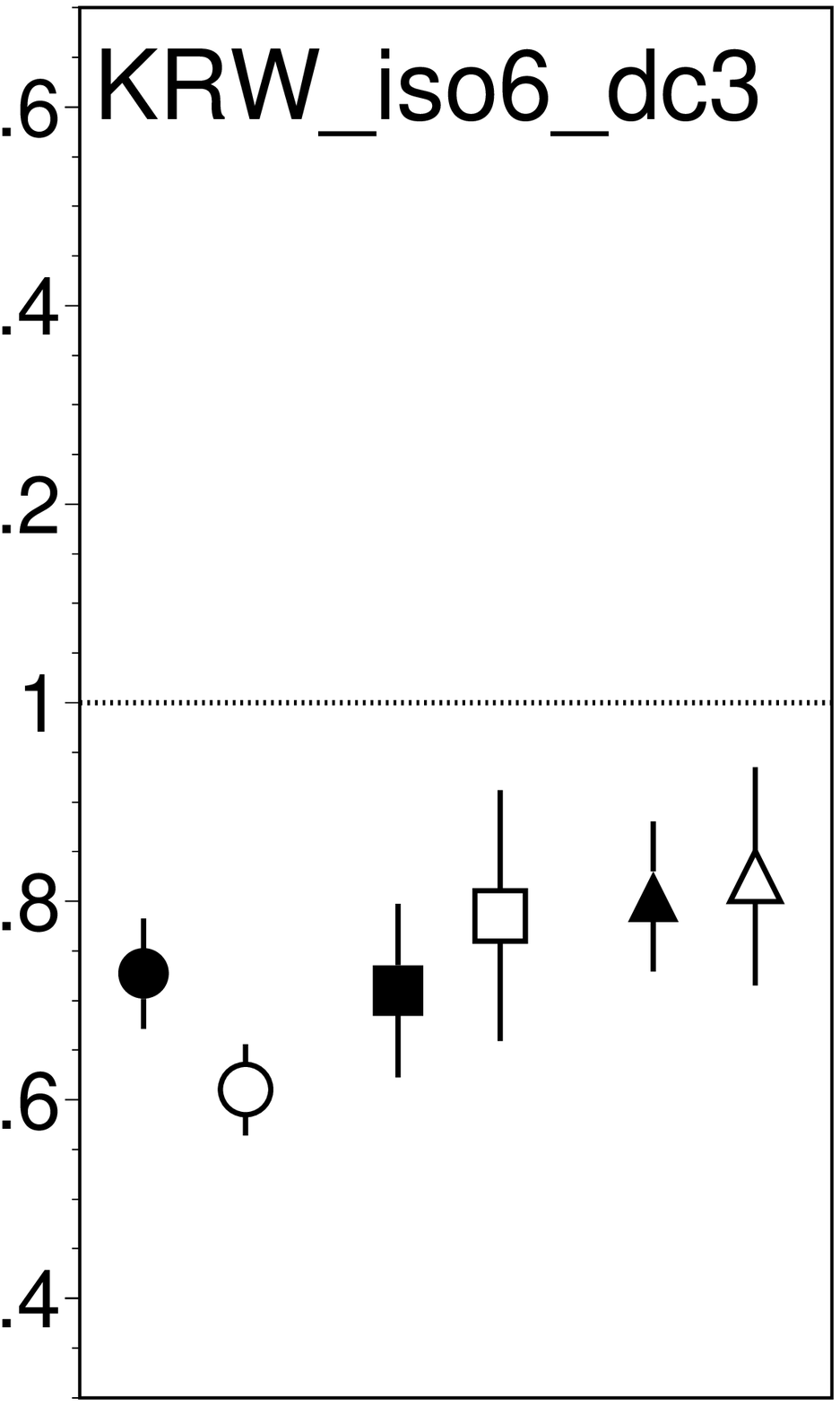}
   \includegraphics[width=\wid]{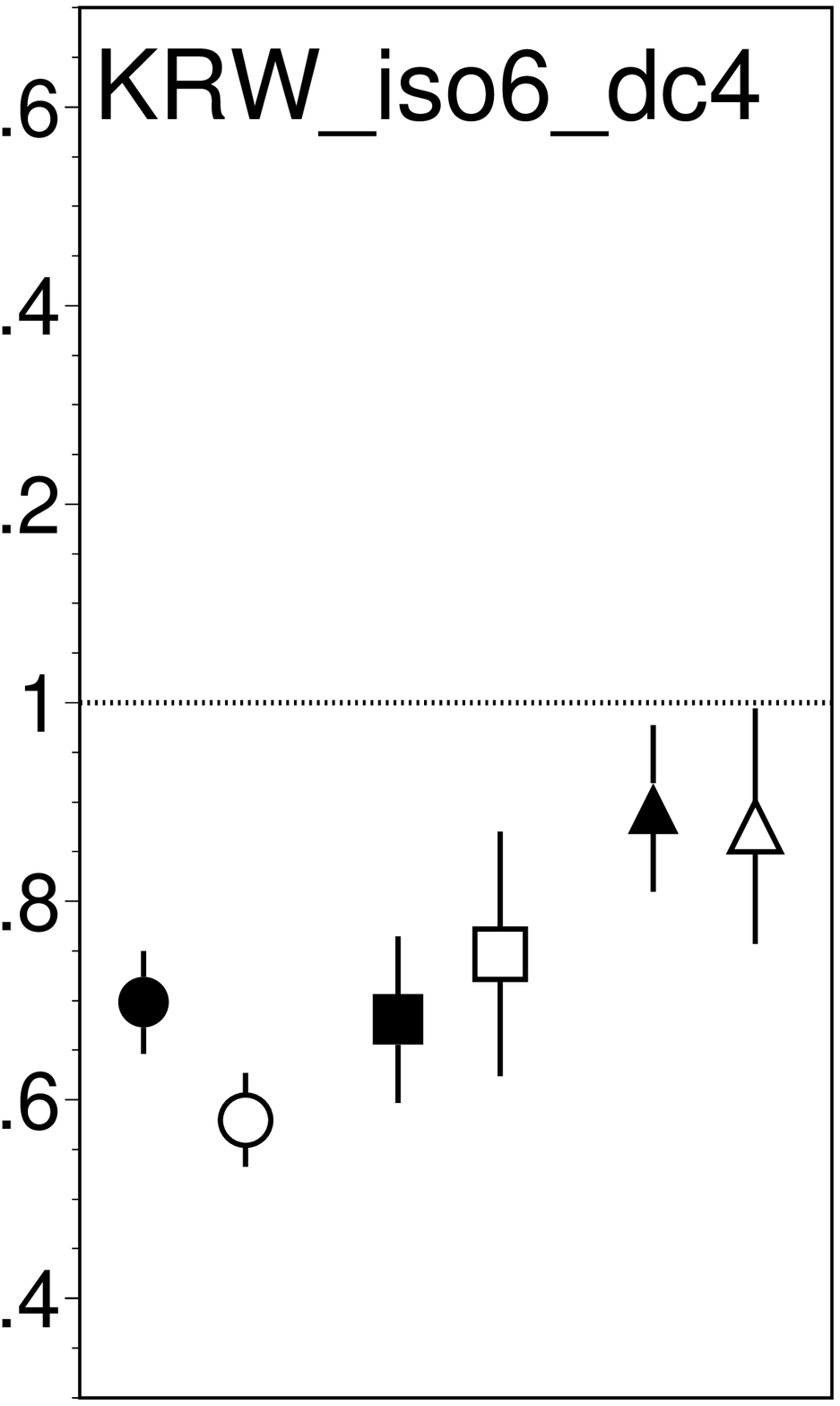}
   \includegraphics[width=\wid]{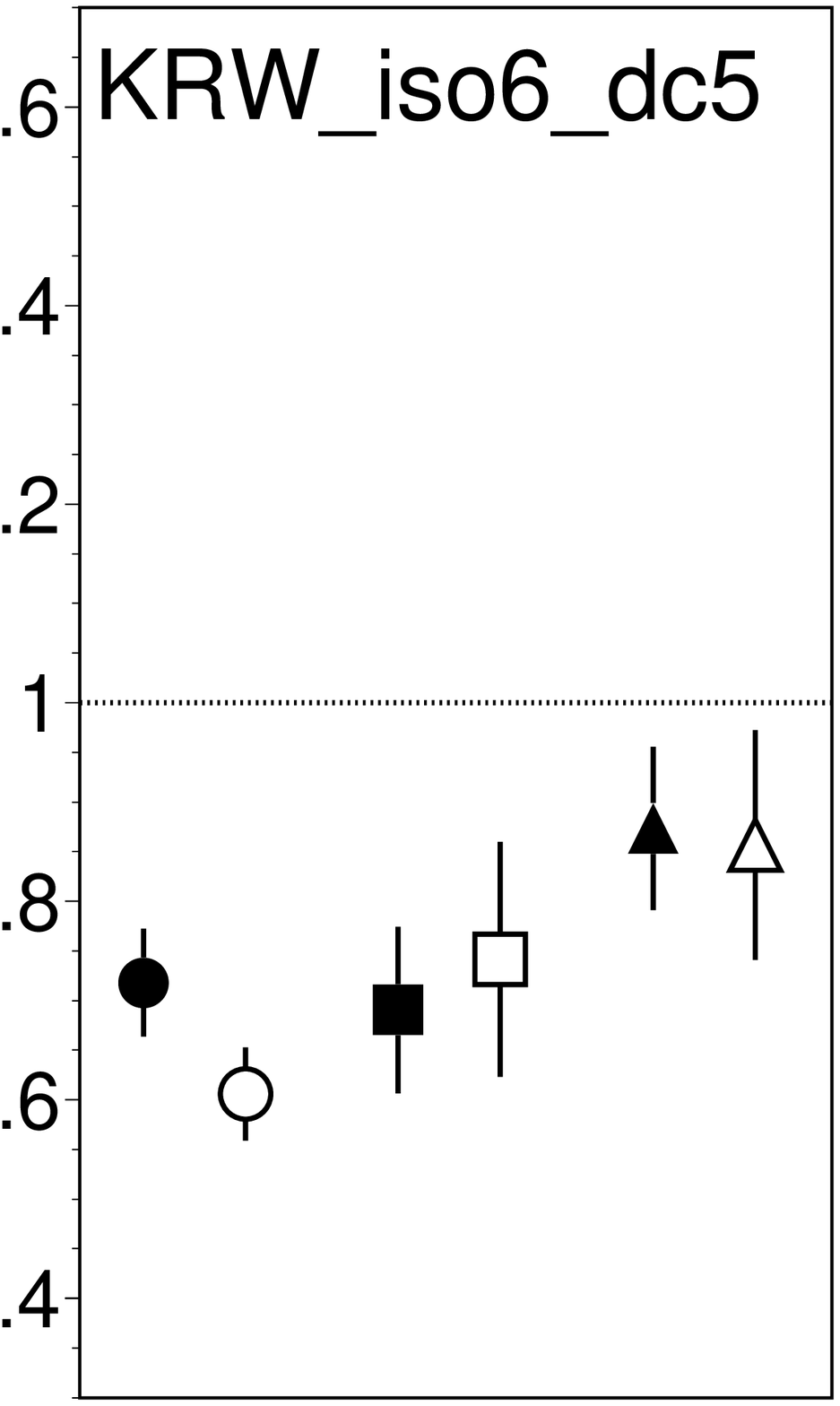}
   \includegraphics[width=\wid]{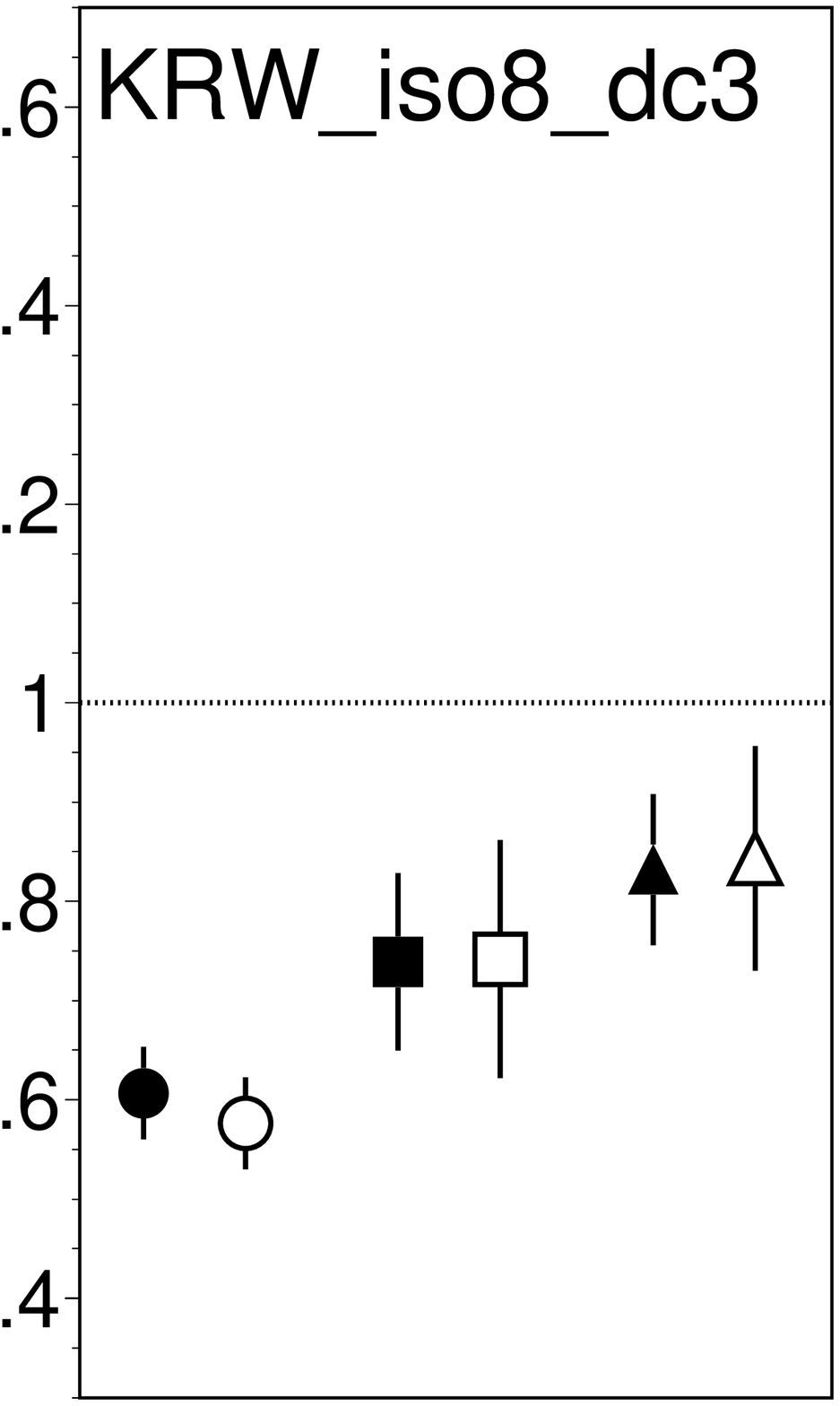}
   \includegraphics[width=\wid]{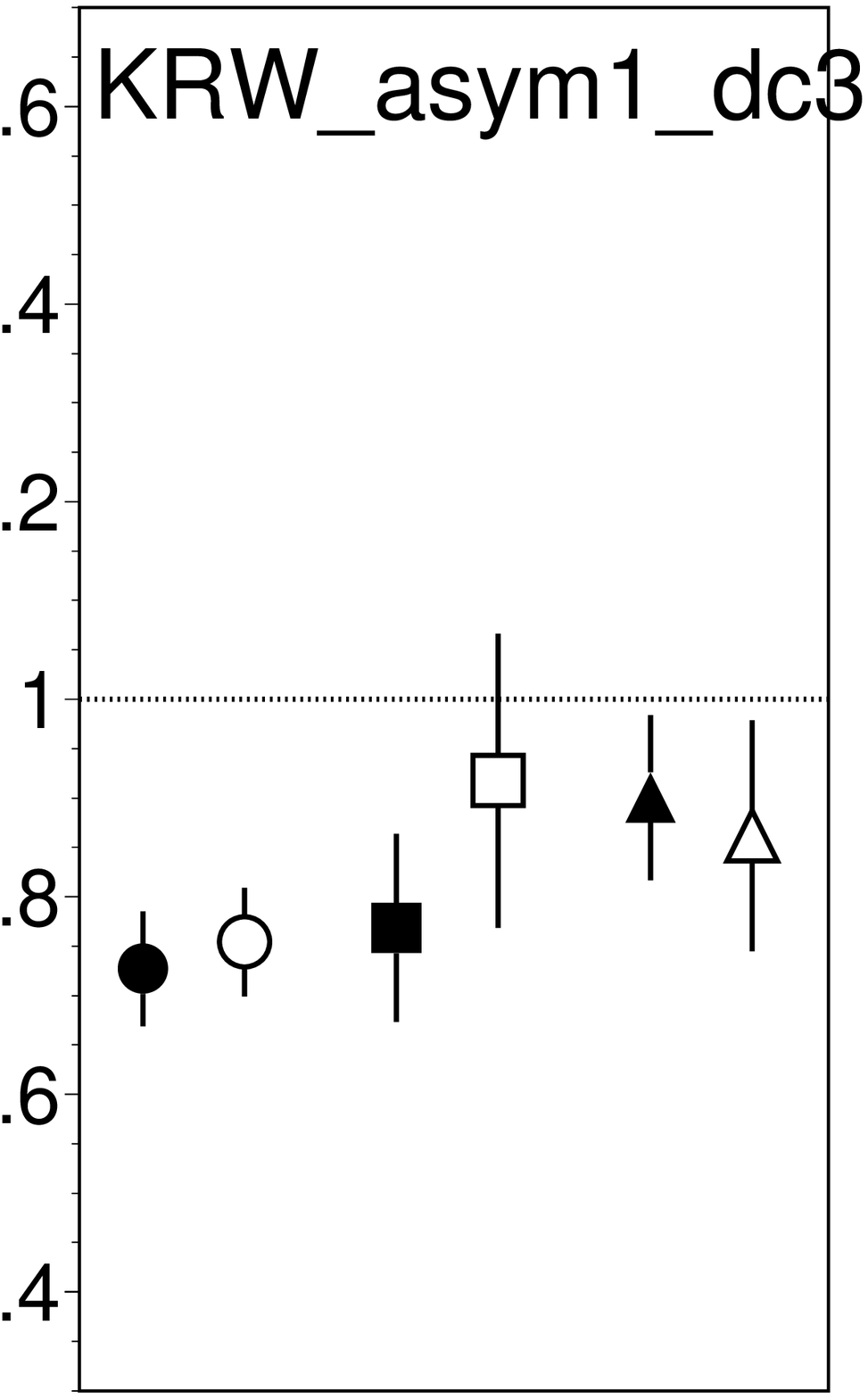}
  \caption{    
    $\RATIO$ for $\Delta_{\mu,z,c}$ as indicated in the legend panel.
    Solid-filled symbols are for the \SDSS\ and open symbols are for \SNLS.
    The intrinsic variation model for the simulation is indicated 
    at the top of each panel.
  }
  \label{fig:RATIOS}
\end{figure*}

% ------------------------------------------------
  \subsection{Systematics Tests}
  \label{subsec:results_syst}
% -----------------------------------------------

Here we describe some systematics tests to demonstrate 
the robustness of the results in Figure~\ref{fig:RATIOS}. 
We use the FUN-COH scatter model as the reference simulation for these 
tests which are summarized in Figure~\ref{fig:RATIO_FUNCOH_SYST}.
For each test a change is applied to the simulation
and then analyzed in exactly the same manner.

\newcommand{\widsyst}{0.9in}
\begin{figure}[h!]
\centering
\leavevmode
  \includegraphics[width=\widsyst]{median_ratios_legend.eps}
  \includegraphics[width=\widsyst]{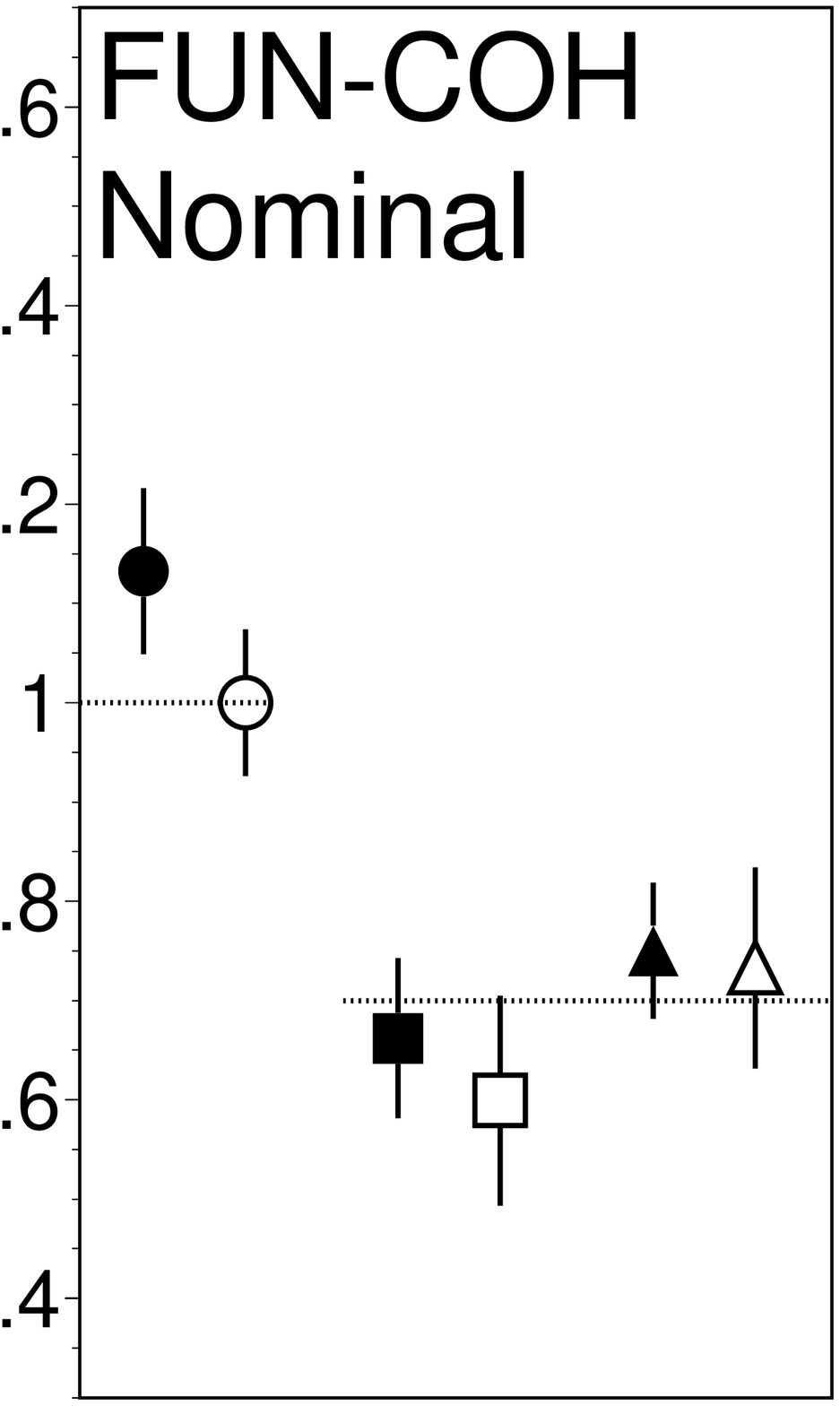}
  \includegraphics[width=\widsyst]{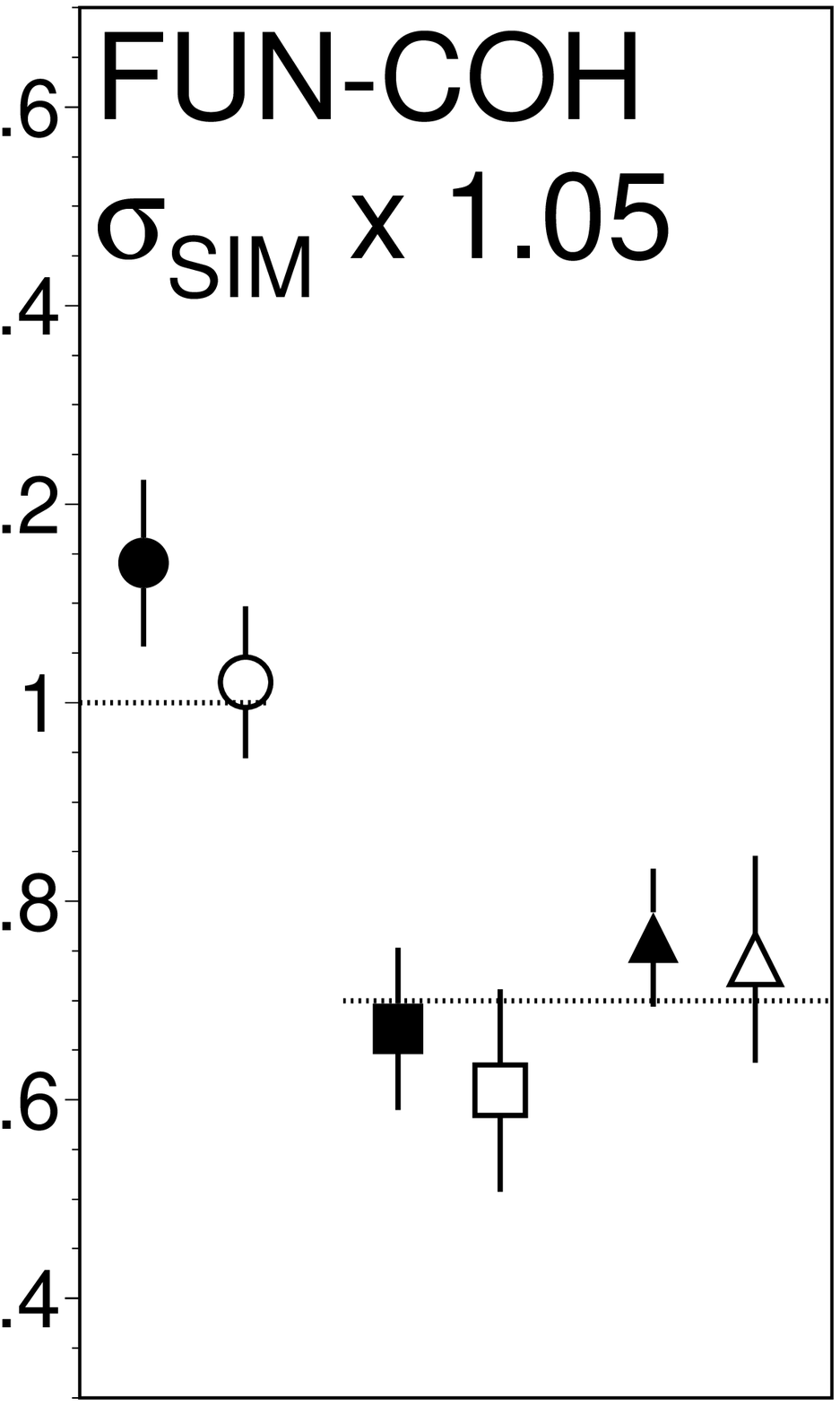}
  \includegraphics[width=\widsyst]{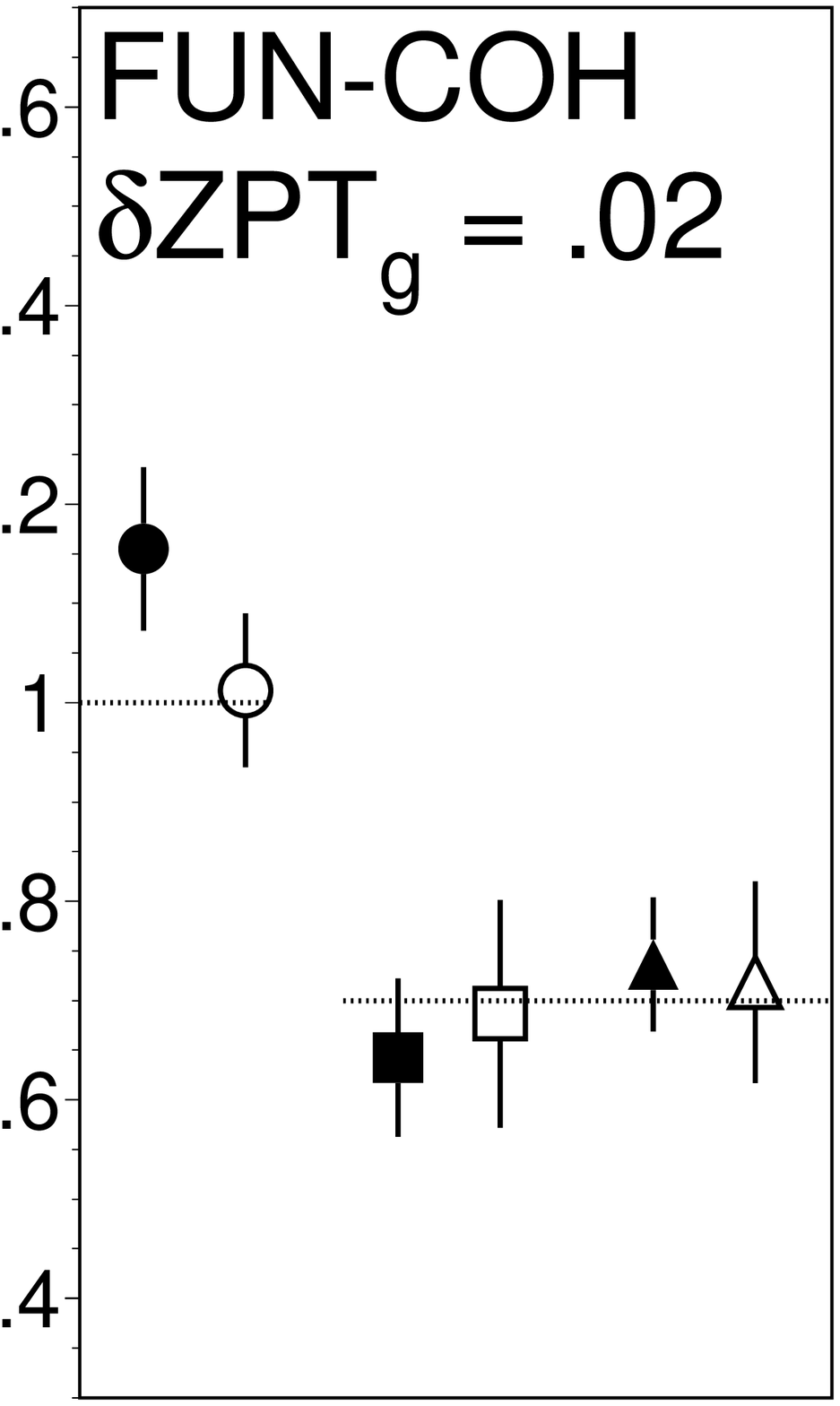}
  \includegraphics[width=\widsyst]{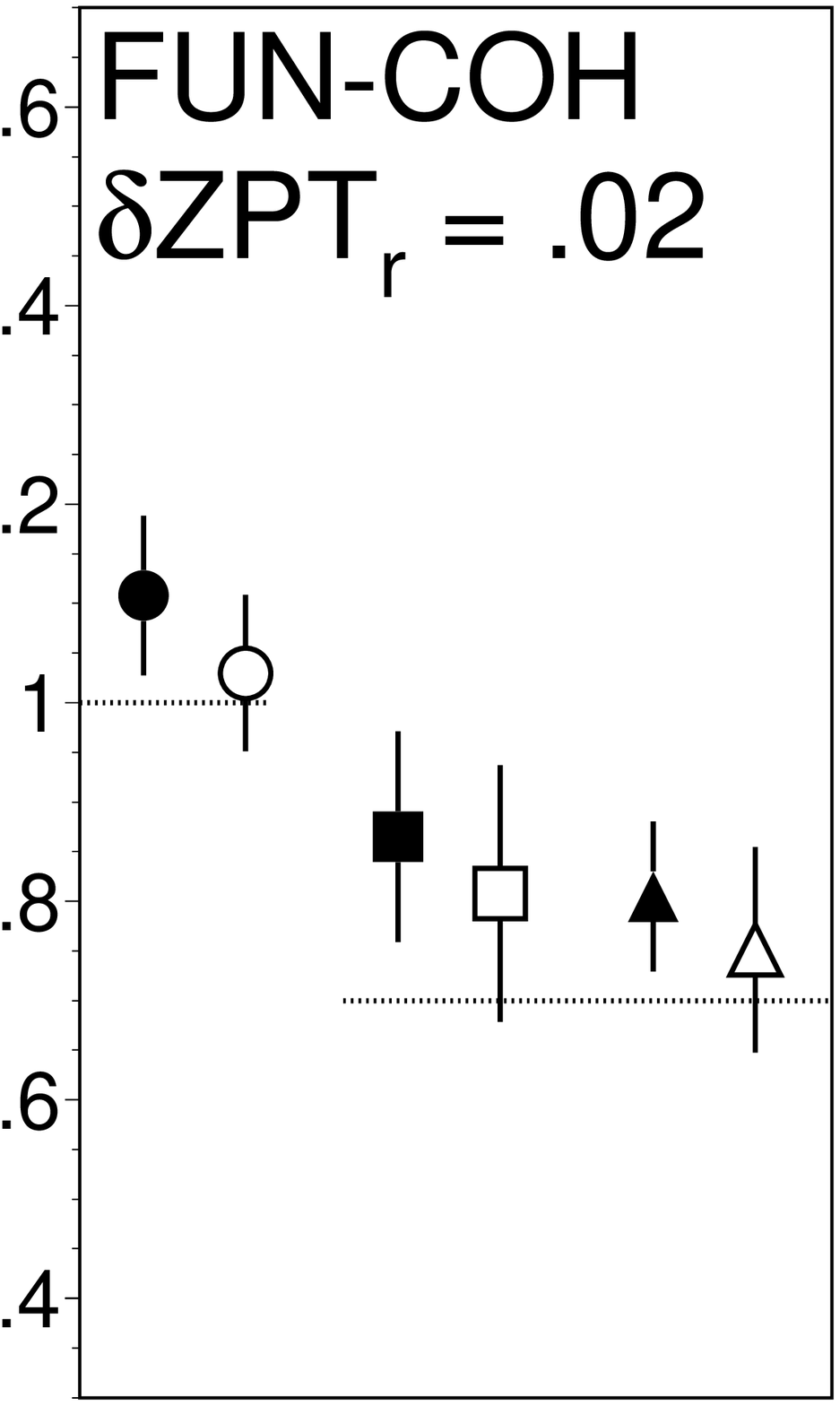}
  \includegraphics[width=\widsyst]{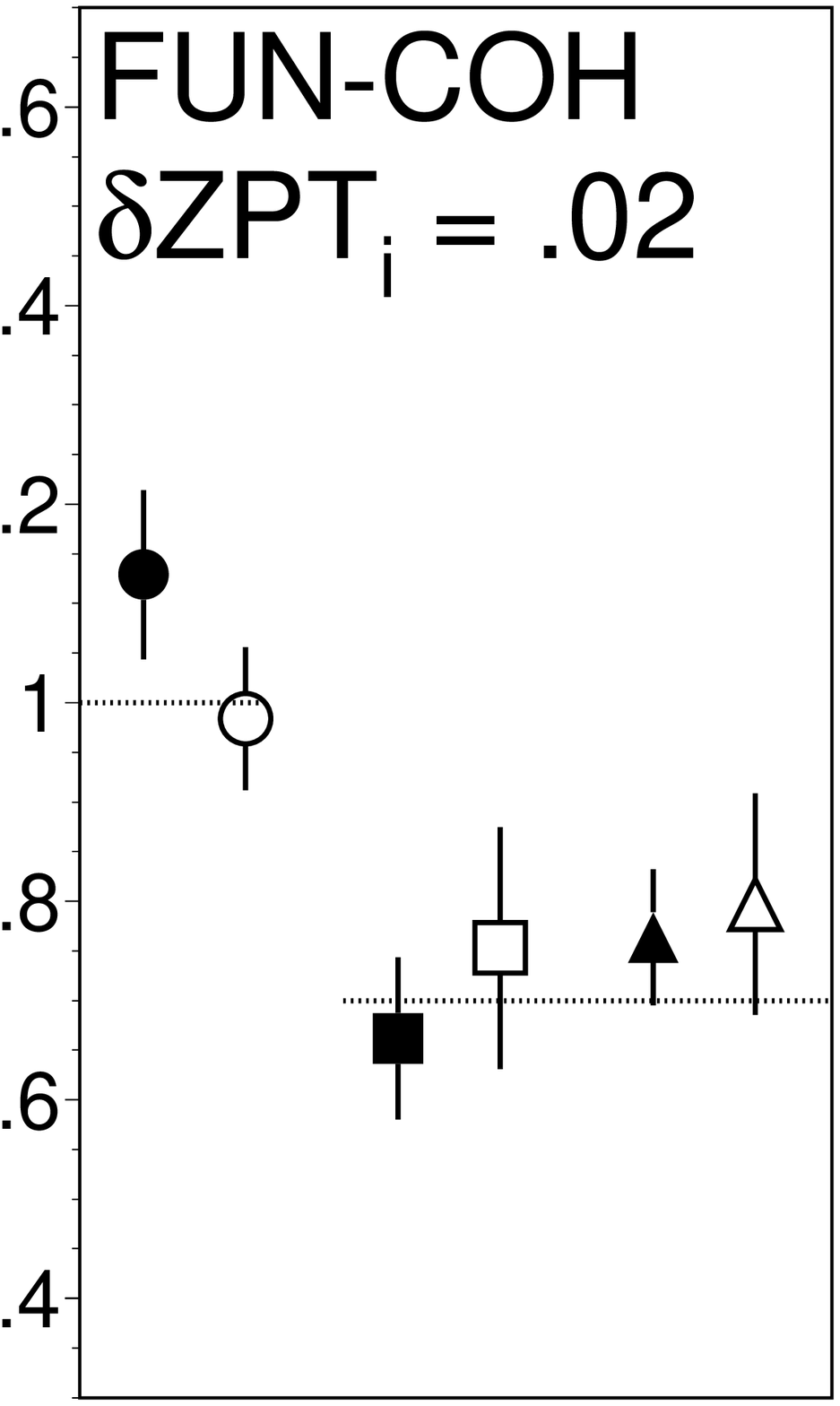}
  \includegraphics[width=\widsyst]{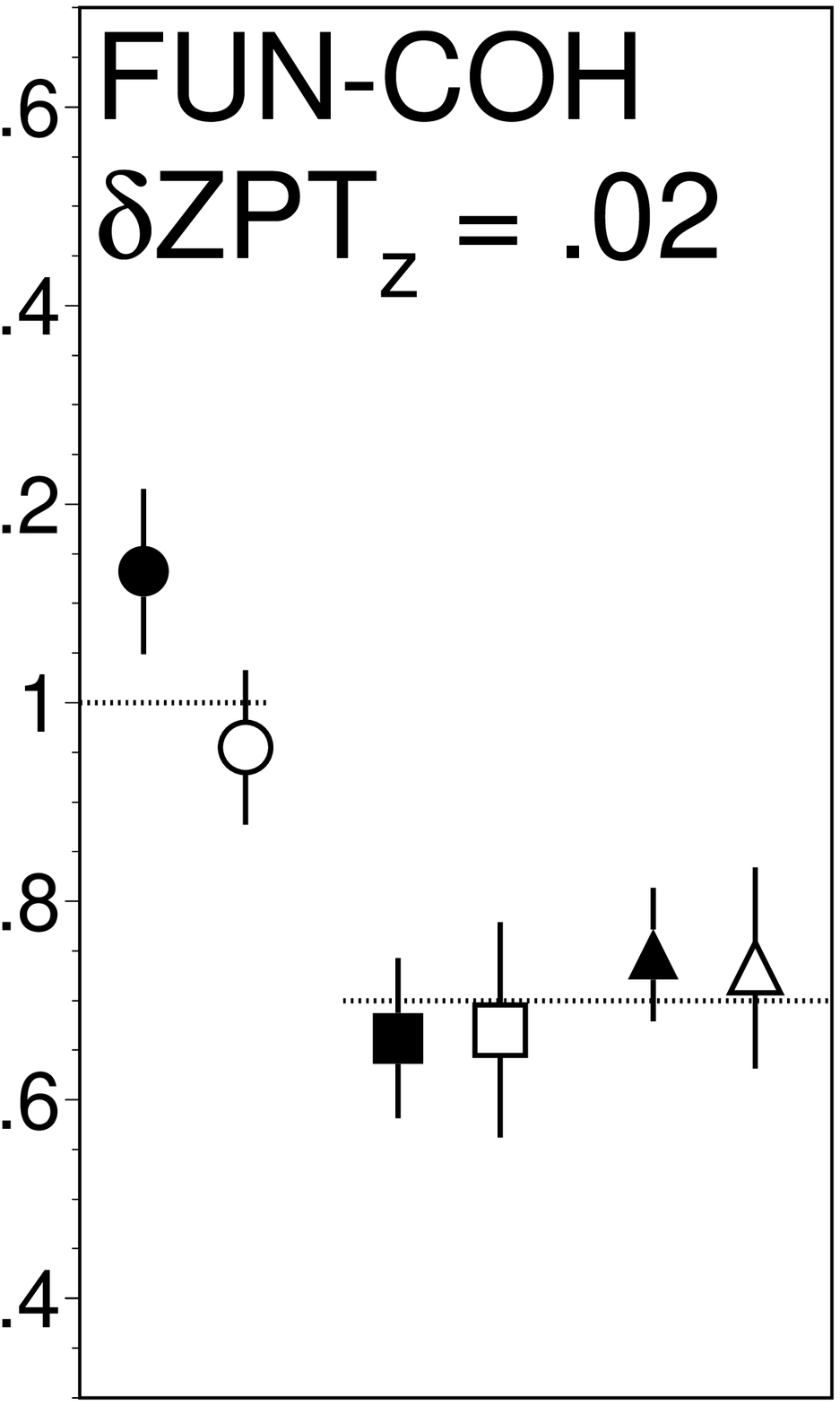}
  \includegraphics[width=\widsyst]{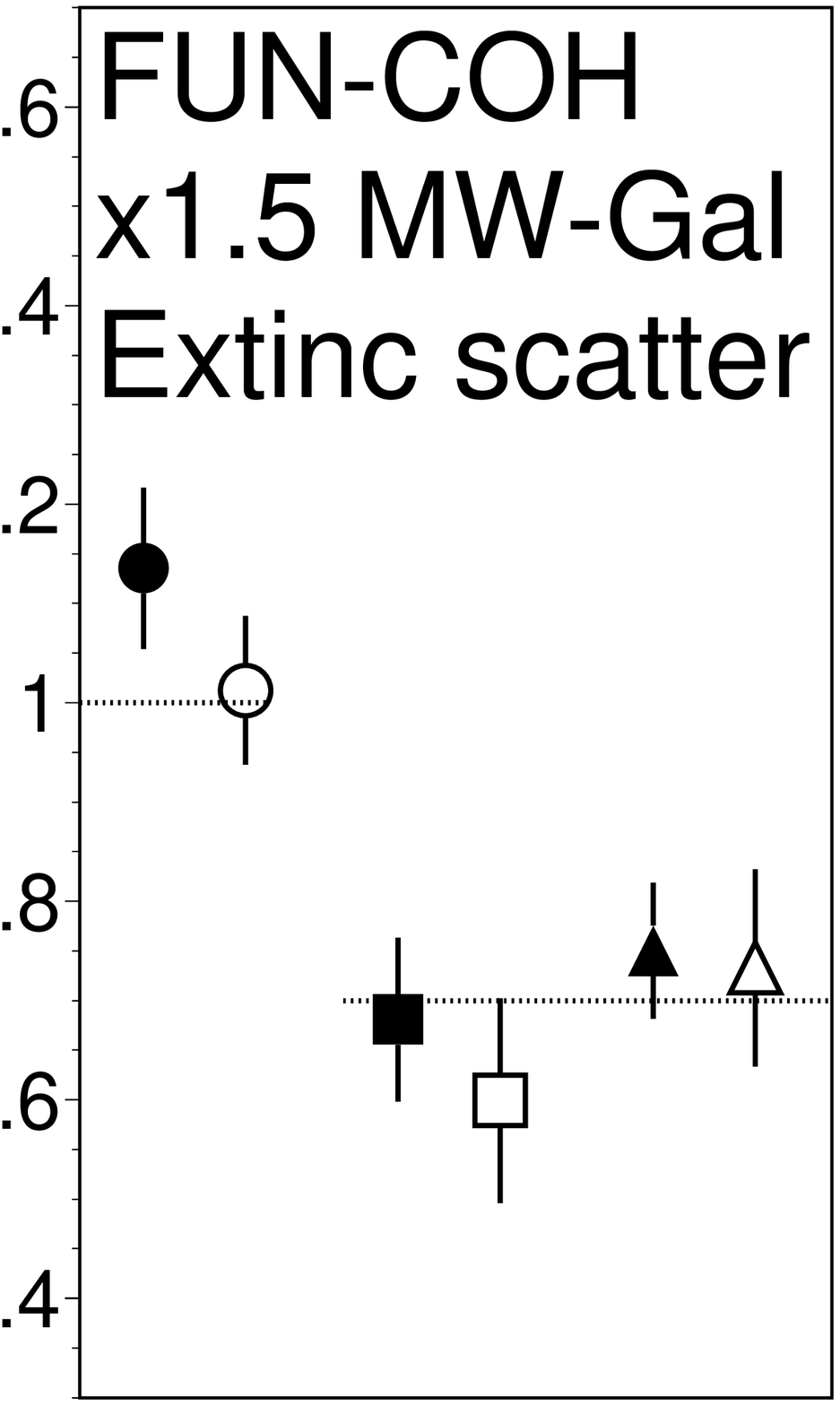}
  \includegraphics[width=\widsyst]{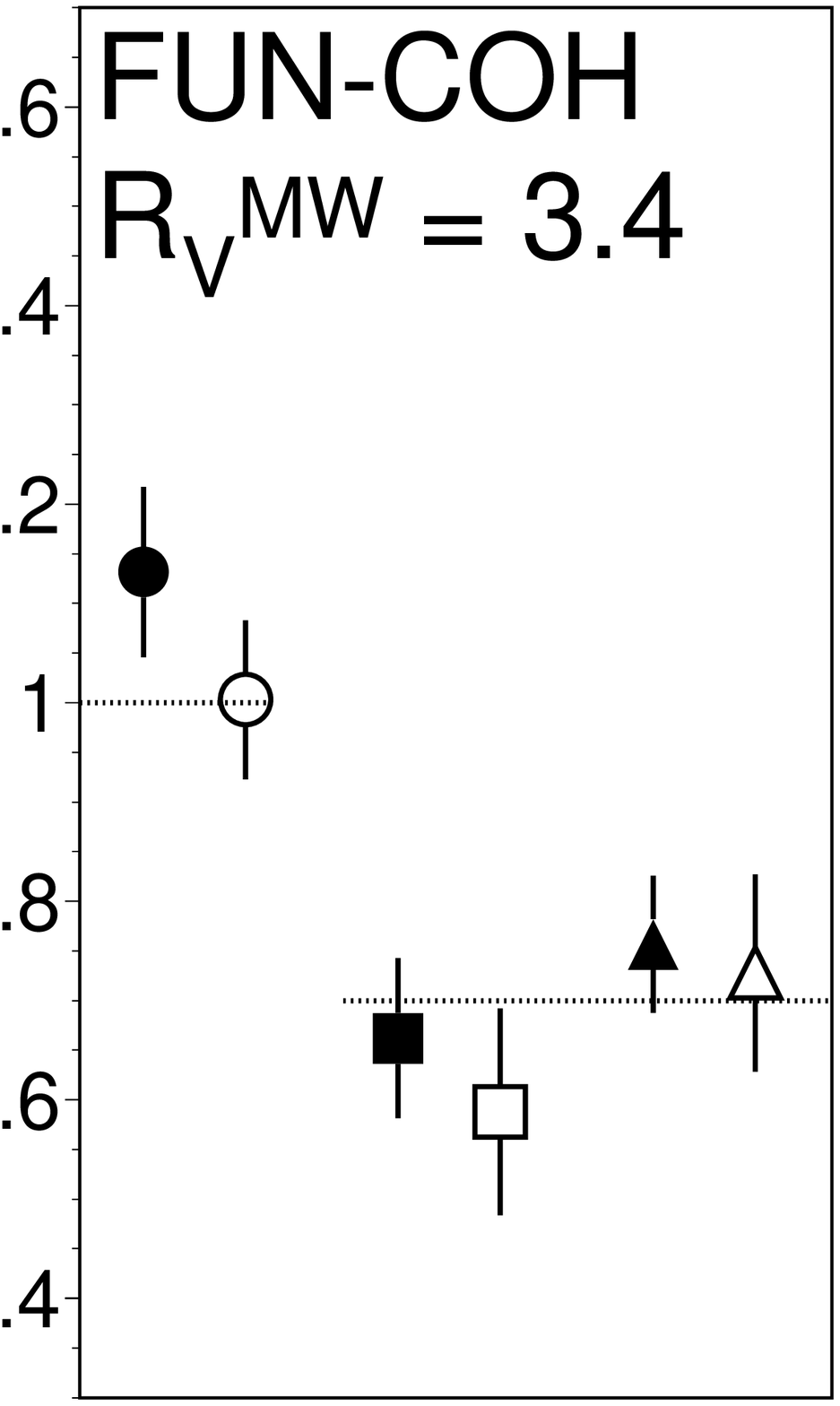}
  \includegraphics[width=\widsyst]{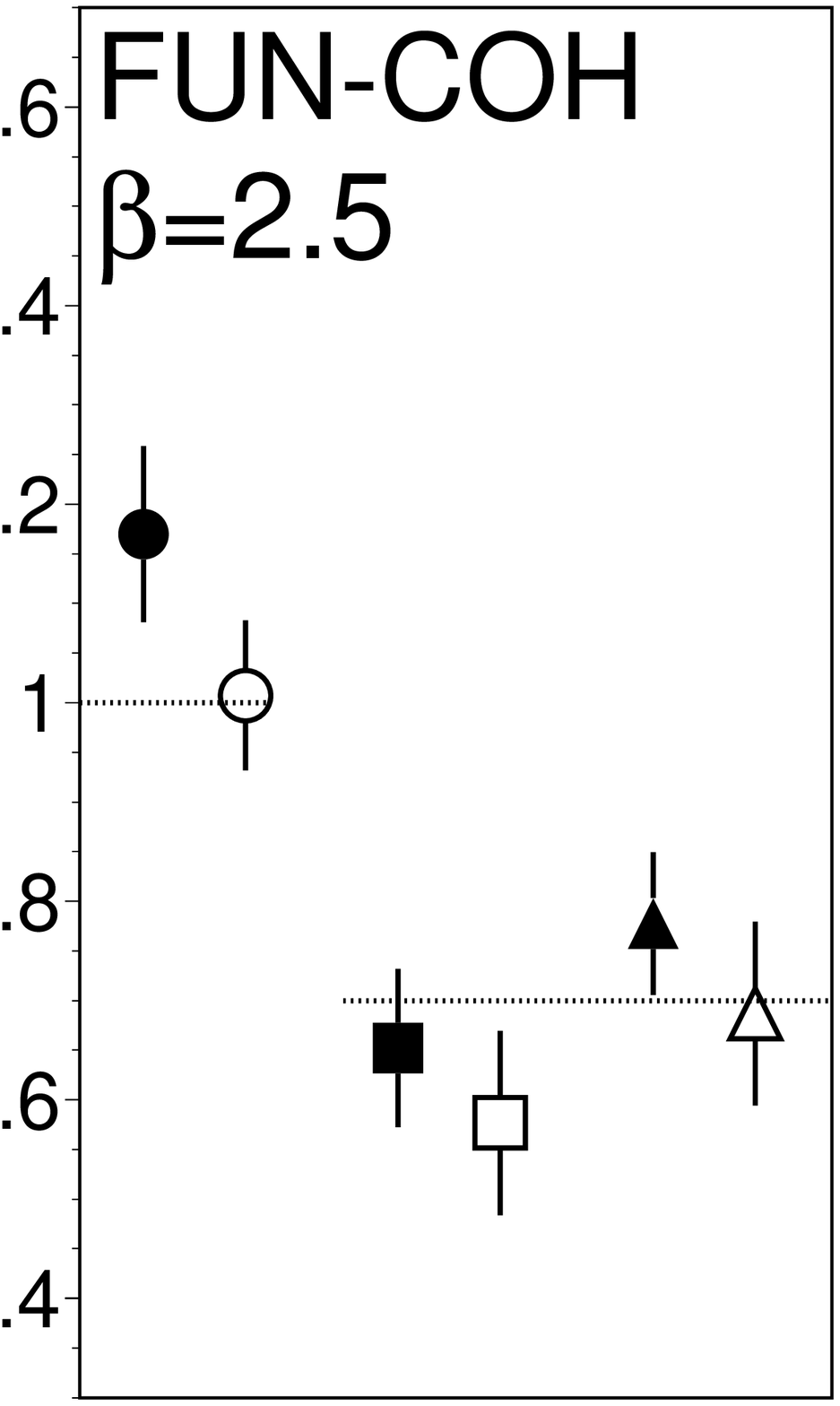}
  \includegraphics[width=\widsyst]{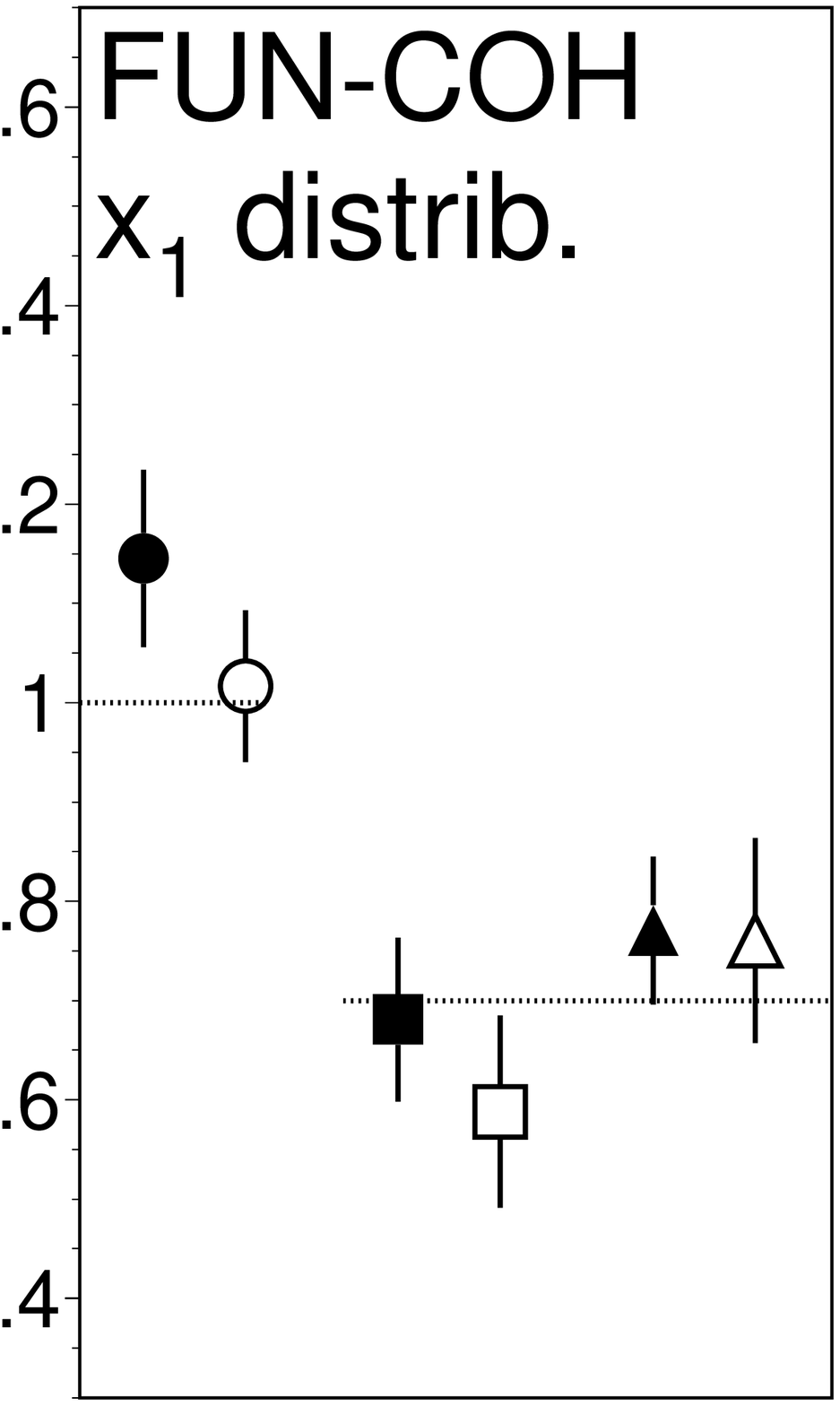}
  \includegraphics[width=\widsyst]{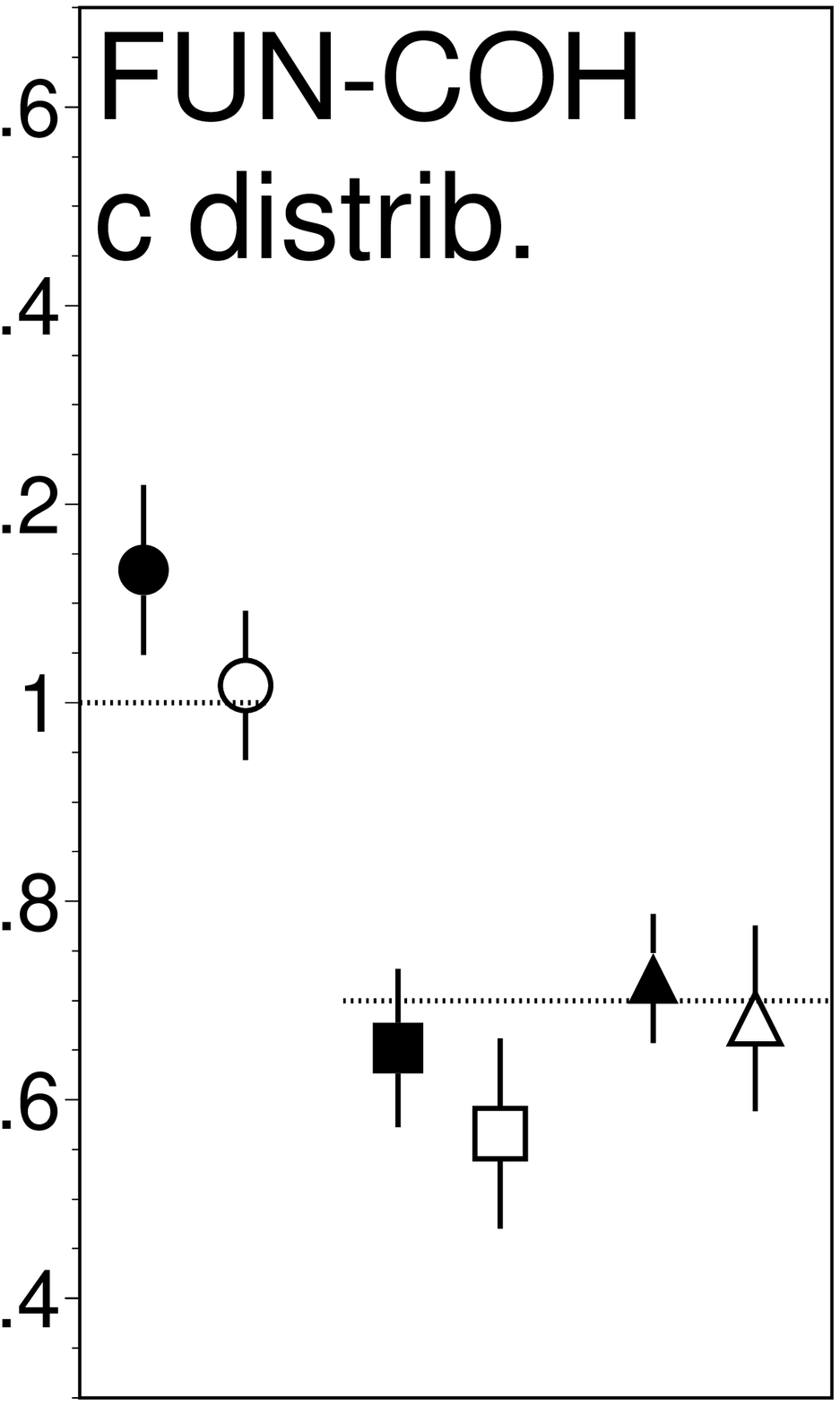}
  \caption{    
    $\RATIO$ for $\Delta_{\mu,z,c}$ as indicated in the legend panel
    from Figure~\ref{fig:RATIOS}. The first panel is for the FUN-COH
    scatter model using the nominal simulation.     
    The systematic test indicated above each panel is applied
    to the simulation using the same FUN-COH scatter model.
    The horizontal dashed lines are the same in each panel,
    and are intended to guide the eye.
  }
  \label{fig:RATIO_FUNCOH_SYST}
\end{figure}

The first test is based on the precision in the flux \uncs\
in the data. 
For pre-explosion epochs in which the true SN flux 
is known to be zero, 
examining the S/N distribution shows that the \uncs\
are accurate to within 5\%.  
The test labeled $\sigSIM\times 1.05$ corresponds
to a simulation with 5\% larger \uncs\ on all of the fluxes.

The next set of tests is based on a 0.02~mag zeropoint
change in each filter ($\delta{\rm ZPT}_{griz}$).
Note that this change is two times larger than the 
\unc\ reported by each survey team.

\Uncs\ on the Galactic extinction are examined by 
first increasing the estimated 16\% scatter to 24\%
($\times 1.5$ MW-Gal), and then increasing the
reddening parameter by 10\%, to $R_V=3.4$.

The next test is based on changing $\beta$ from 3.2 to 2.5,
a $5\sigma$ change from G10.

\begin{figure}[h!]
\centering
\epsscale{\xxScale}  % 1.15 for emulateapj 
\plotone{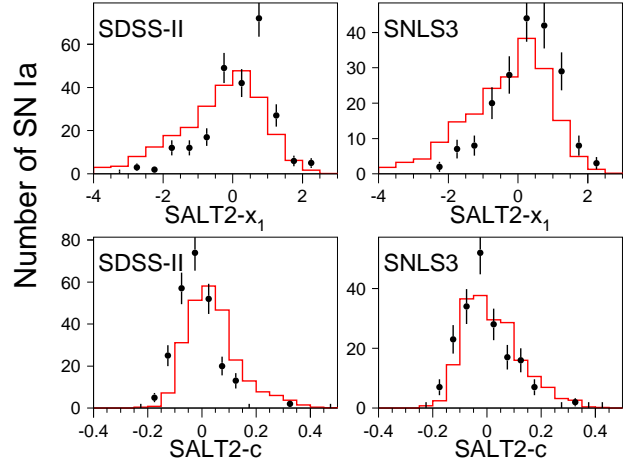}
  \caption{    
    Comparison of fitted stretch ($x_1$) and color ($c$)
    distributions for data (dots) and {\it modified} MC (histogram),
    where each MC distribution is scaled to have the same statistics 
    as the data.
    The simulated populations have been modified for
    systematic tests explained in the text.
  }
  \label{fig:ovsyst_x1c}
\end{figure}

The next two tests are based on changing the population
parameters for $x_1$ and $c$ (Table~\ref{tb:unfold_x1c}).
The simulated $x_1$ population is shifted toward faster-declining
light curves by setting $\sigPLUS = 0.5$ and $\sigMINUS=1.8$ 
(compare to nominal parameters in Table~\ref{tb:unfold_x1c}).
The simulated color population is shifted toward the red by
setting $\sigPLUS = 0.18$ and $\sigMINUS=0.05$.
For these systematic tests, the resulting data/MC comparisons 
for stretch and color are shown in Figure~\ref{fig:ovsyst_x1c};
the data and MC are clearly discrepant.

For all of these systematic tests, $\RATIO$ remains
significantly below unity for the $B-V-\cc$  and \photoz\ 
variables.
We also note that the independent \SDSS\ and \SNLS\ results
are consistent, showing consistency over different redshift
ranges.

% -------------------------------------------------
%  \clearpage
  \section{Impact of Intrinsic Scatter Model on the Hubble Diagram}
  \label{sec:results_Hubble}
% -------------------------------------------------

Here we investigate the potential Hubble diagram bias 
from using an incorrect model of intrinsic scatter.
We use four intrinsic scatter models that give reasonable 
data/MC agreement in Figure~\ref{fig:RATIOS}:
FUN-MIX, G10, C11\_0 and C11\_1.
Recall that data/MC agreement in these dispersion variables
means that the intrinsic scatter model cannot be ruled out,
but the agreement  does not ensure that the underlying model 
is correct.

The Hubble bias is determined from the difference between 
an ideal analysis using the correct intrinsic scatter matrix 
(Section~\ref{subsec:scatter_matrix}),
and a conventional analysis that adds a
wavelength-independent scatter to bring the
reduced $\chi^2$ to unity (Section~\ref{subsec:fit4mu}).
The ideal analysis is based on simulations with the correct 
model of intrinsic scatter, and thus does not reflect a 
realistic analysis that could be applied to data. 
The conventional analysis, however, reflects
a realistic analysis that has often been applied to data.
In Section~\ref{subsec:wbias} the Hubble bias is translated 
into a bias on the dark energy equation of state parameter $w$,
and in Section~\ref{subsec:evalMC} the biases are reevaluated
with Malmquist bias corrections.

% ----------------------------------------------------
 \subsection{Determining the Intrinsic-scatter Matrix}
 \label{subsec:scatter_matrix}
% ----------------------------------------------------

To evaluate the effect of intrinsic scatter in the analysis
of cosmological parameters, we first need to briefly 
summarize the concept of an intrinsic scatter matrix 
introduced in Section~2 of M11. Cosmology fitters in general 
minimize the function 
\begin{equation}
   \chi^2 = \sum_i {\DMU}_i / ({\sigstati}^2 + {\siginti}^2 )
   \label{eq:muchi2}
\end{equation}
where $\DMU_i$ is the difference between the fitted and calculated 
distance modulus (Eq.~\ref{eq:DMU}) for the $ith$ SN,
$\sigstati$ is the statistical (fitted) error on $\DMU_i$,
and $\siginti$ is an ad hoc parameter defined so that
$\chi^2/\Ndof = 1$. 
Since $\sigstati$ is computed from a statistical correlation
matrix between the \SALTII\ fit parameters ($m_B,x_1,c$),
M11 proposed an analogous ``intrinsic-scatter covariance matrix''
(denoted $\C$) to compute $\sigint$. 
Dropping the SN index $i$,
the ad-hoc error term is 
\begin{eqnarray}
  \sigint^2 & = & \C_{00} + \alpha^2\C_{11} + \beta^2\C_{cc} 
              \nonumber \\
            & + & 2\alpha\C_{01} - 2\beta\C_{0c} - 2\alpha\beta\C_{1c}~~,
  \label{eq:sigint}
\end{eqnarray}
%
%\begin{equation}
%  \sigint^2 =  \C_{00} + \alpha^2\C_{11} + \beta^2\C_{cc}               
%            +  2\alpha\C_{01} - 2\beta\C_{0c} - 2\alpha\beta\C_{1c}~~,
%  \label{eq:sigint}
%\end{equation}
%
where the subscript correspondence is  $0,1,c \to m_B,x_1,c$,
$\sqrt{\C_{00}} \equiv \sigintmB$, 
$\sqrt{\C_{11}} \equiv \sigintx$, and 
$\sqrt{\C_{cc}} \equiv \sigintc$.
All SNIa-cosmology analyses to date have used the
simplifying assumption that
$\sigint = \sigintmB = \sqrt{\C_{00}}$,
and ignored the other $\C$ terms in Eq.~\ref{eq:sigint}.
We refer to this method as the ``\WIS'' method,
while the $\C$-fit method refers to using additional terms
in Eq.~\ref{eq:sigint}.
The \WIS\ method is valid if the intrinsic scatter is 
independent of wavelength,
or if $\sigstati$ from the light curve fit includes the 
wavelength dependence of the scatter.
The FUN-COH panel in in Figure~\ref{fig:RATIOS}
clearly shows that the intrinsic scatter 
cannot be constant (i.e., wavelength independent).
M11 noted that using the \WIS\ method can lead to biased 
values of $\alpha$ and $\beta$.
Here we go a step further and examine
biases in simulated Hubble diagrams.

\begin{figure}[hb!]
\centering
\epsscale{\xScale}  % 1.1 for emulateapj 
\plotone{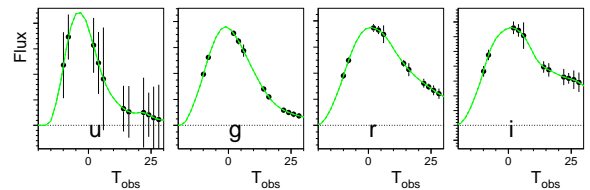}
  \caption{    
    Simulated \SDSS\ light curve (solid dots) with no Poisson
    noise and no intrinsic scatter. Each curve is the best-fit
    \SALTII\ model.
  }
  \label{fig:snlc_noPoisson}
\end{figure}

Since we do not have reliable methods for 
determining $\C$ from the data, 
we determine $\C$ from an artificial analysis using a 
simulation with the correct model of intrinsic scatter but
{\it without} Poisson fluctuations from the
calculated measurement {\uncs} (Eq.~\ref{eq:uncModel});
therefore the only source of scatter is from 
intrinsic variations. 
Although Poisson fluctuations are not applied,
the \uncs\ are included in the light curve fit-$\chi^2$
calculations
so that the correct filter-dependent weights are used.
For example, the SDSS $u$ band has relatively 
poor S/N compared to the other bands and therefore this
passband has less weight in determining $\C$.
We refer to these simulations as ``{\intonly}'' to distinguish
them from the ``nominal'' simulations that include Poisson 
fluctuations.
This \intonly\ simulation is illustrated in 
Figure~\ref{fig:snlc_noPoisson}
for the special case with no intrinsic scatter;
the simulated fluxes lie exactly on the best-fit \SALTII\
model and they have the correct \uncs\ corresponding
to real \obss.

To better compare the resulting bias to the \unc\ reported in 
\cite{Sullivan11}, the simulations have been adjusted
to better match the data sample  used in this SNLS3 analysis.
First, the \SDSS\ sample size is reduced by a factor
of three to correspond to the first-season sample used
in the SNLS3 analysis. The next change is that the
S/N requirement in the three passbands
(see end of Section~\ref{sec:data}) is relaxed from 8 to 5.
Finally, we have included a simulated nearby ($z<0.1$) 
sample as explained in the Appendix.
To measure biases with good precision,
the MC sample sizes correspond to $\MCDATAratiowbias$ 
times the data statistics.

After performing \SALTII\ light curve fits on the
simulated intrinsic-only sample, we define
$\SIMDELmB \equiv \mBfit - \mBtrue$ and
$\SIMDELc \equiv \cfit - \ctrue$,
where ``true'' indicates the true value from the simulation
and ``fit'' indicates the result from a light curve fit.
The true values are defined by the underlying \SALTII\ 
model before the intrinsic smearing model is applied.
% For other models such as those from KRW09
% there is no clear definition of $\mBtrue$ and $\ctrue$,
% and thus it is not always obvious how to define the intrinsic
% scatter matrix within the \SALTII\ framework.
The covariance terms with the stretch parameter $x_1$ are 
negligible because the intrinsic scatter models are 
epoch-independent and thus do not change the light curve shape;
the $x_1$-terms in $\C$ are therefore ignored.

The $2\times 2$ intrinsic scatter matrix is defined to be
\begin{eqnarray}
  \sqrt{{\C}_{00}} & = & \sigintmB = {\rm rms}|\SIMDELmB|  
              \nonumber \\
  \sqrt{{\C}_{cc}} & = & \sigintc  = {\rm rms}|\SIMDELc| 
              \nonumber \\
  {\C}_{0c} & = & \langle \SIMDELmB \SIMDELc \rangle ~~,
  \label{eq:covdef}
\end{eqnarray}
where $\langle\rangle$ indicates the mean value of the enclosed quantity.
Another caveat is that $\C$ depends on the redshift
and on which filters are included in the light curve fit.
This dependence is linked to the \SALTII\ color parameter ($c$)
that is evaluated by extrapolating a color law to the
central wavelengths of the $B$ and $V$ passbands.
To address this caveat, $\C$ is evaluated as a function
of redshift and sample as  shown in Figure~\ref{fig:intcov}.
A second-order polynomial function of redshift is adequate
to describe the components of $\C$.
For the FUN-MIX and G10 models, $\sigintmB$
and $\beta\sigintc$ give a comparable contribution 
($\sim 0.1$) to $\sigint$.
For the C11 models, $\sigintmB \sim 0.03$ is much smaller
than the contribution from $\beta\sigintc$.

\newcommand{\wwid}{3.3in}
\begin{figure}[hb!]
\centering
\epsscale{\xxScale}  % 1.15 for emulateapj 
   \includegraphics[width=\wwid]{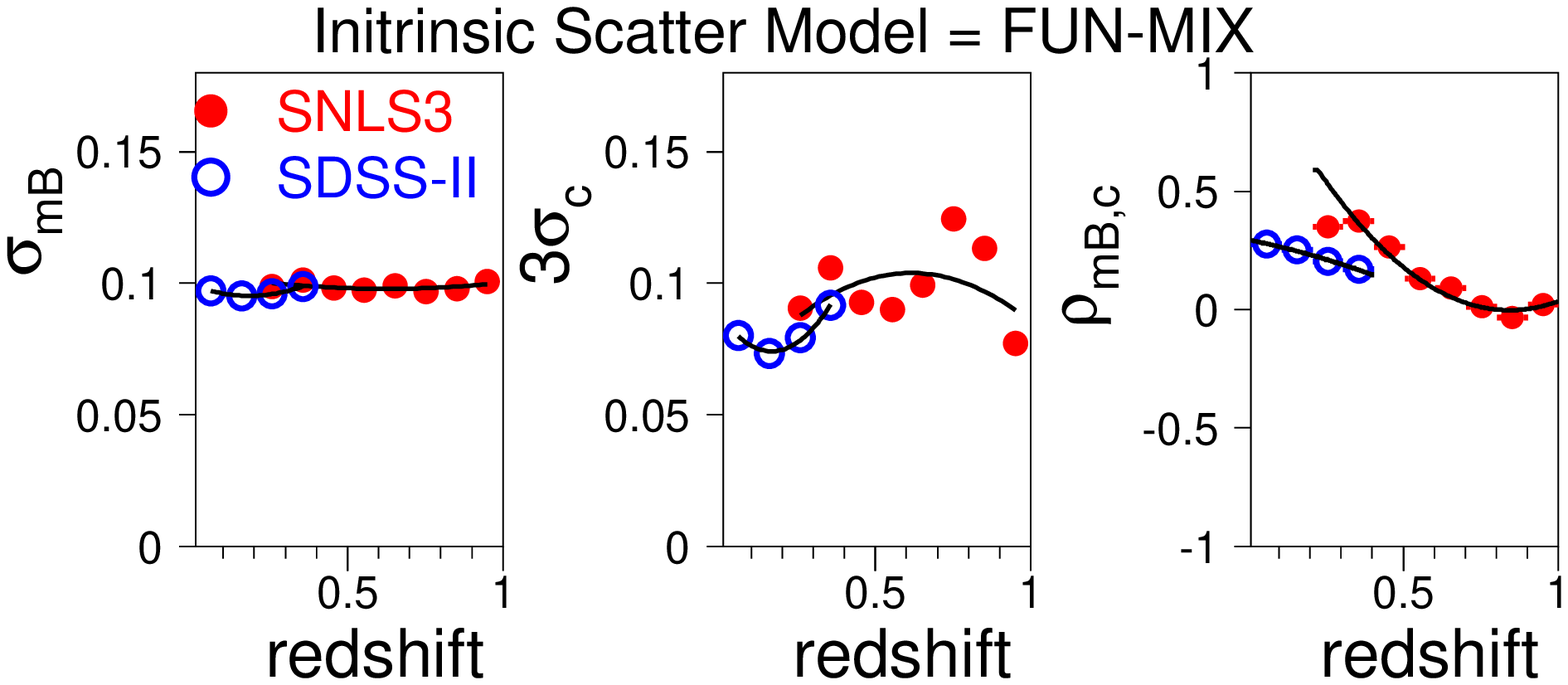}
   \includegraphics[width=\wwid]{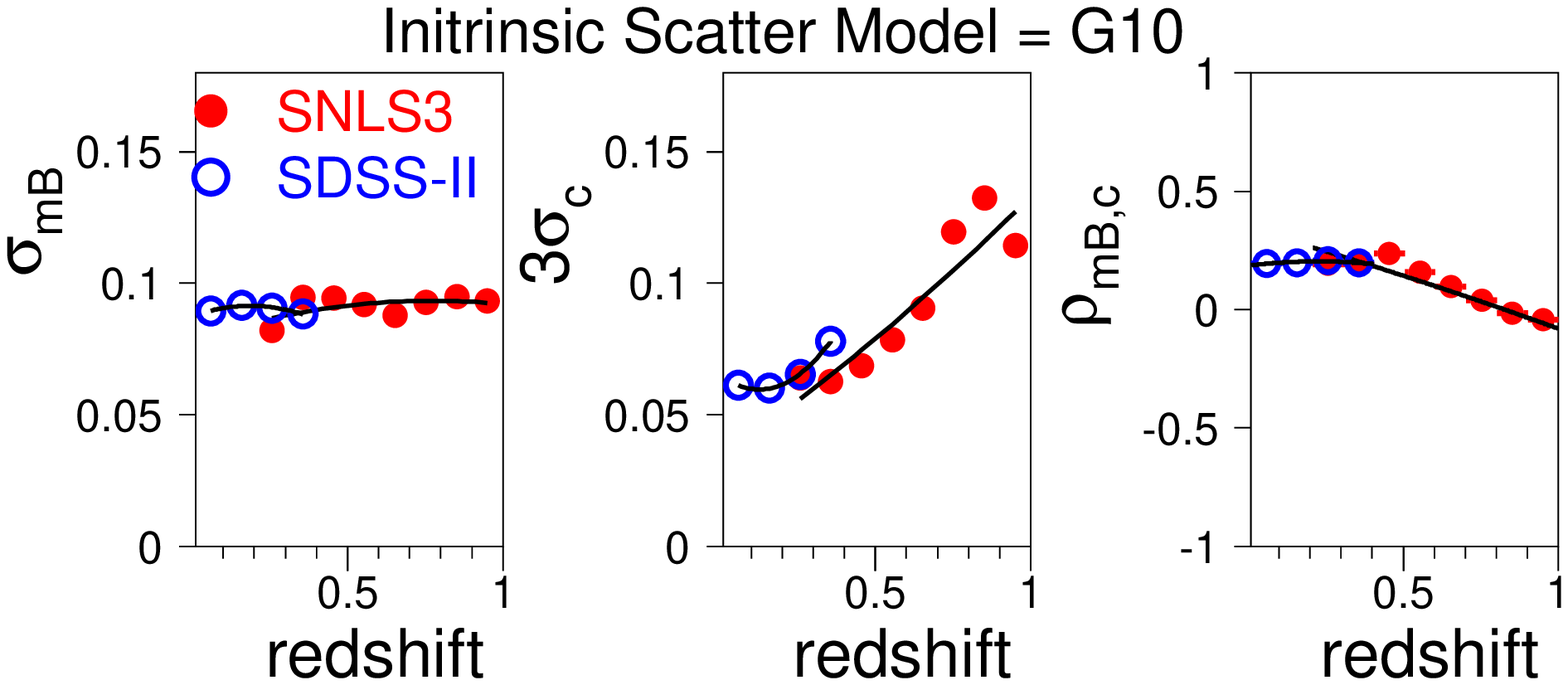}
   \includegraphics[width=\wwid]{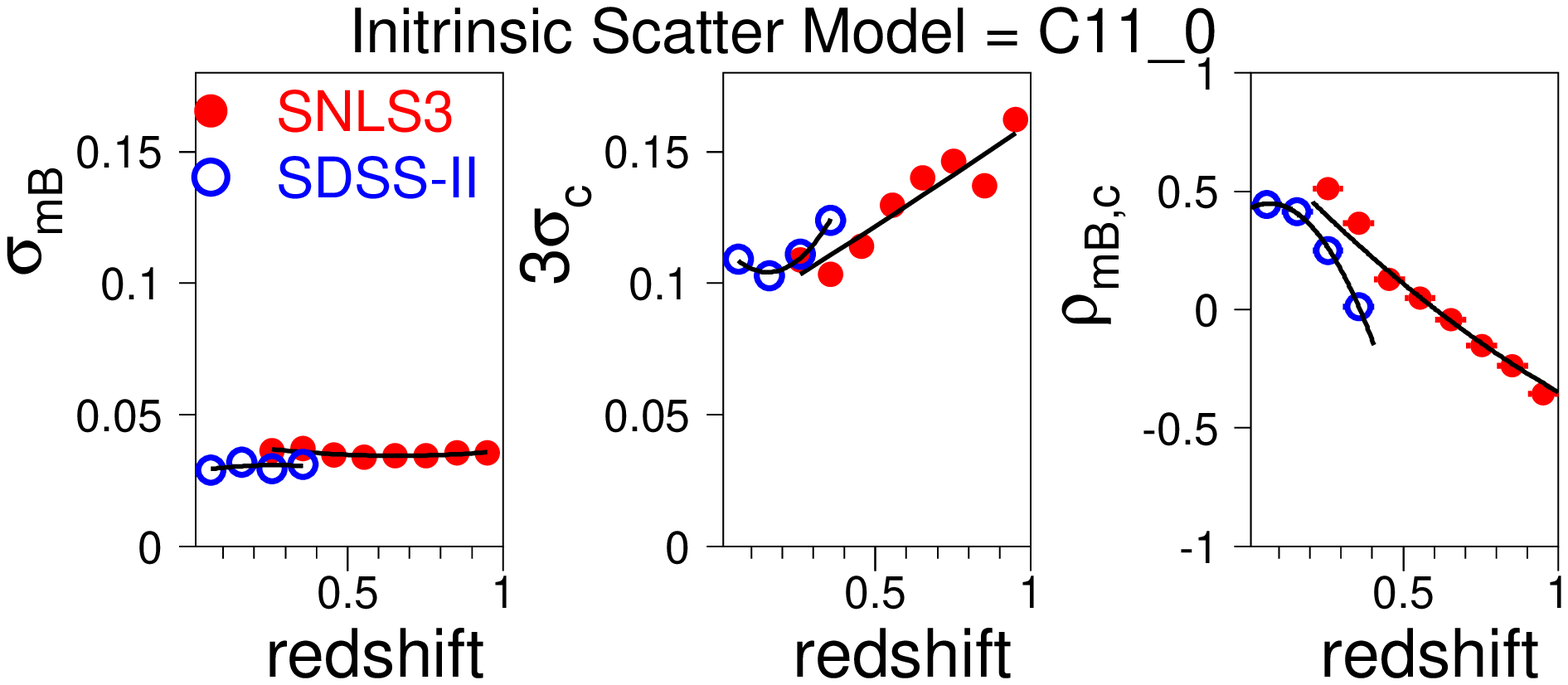}
   \includegraphics[width=\wwid]{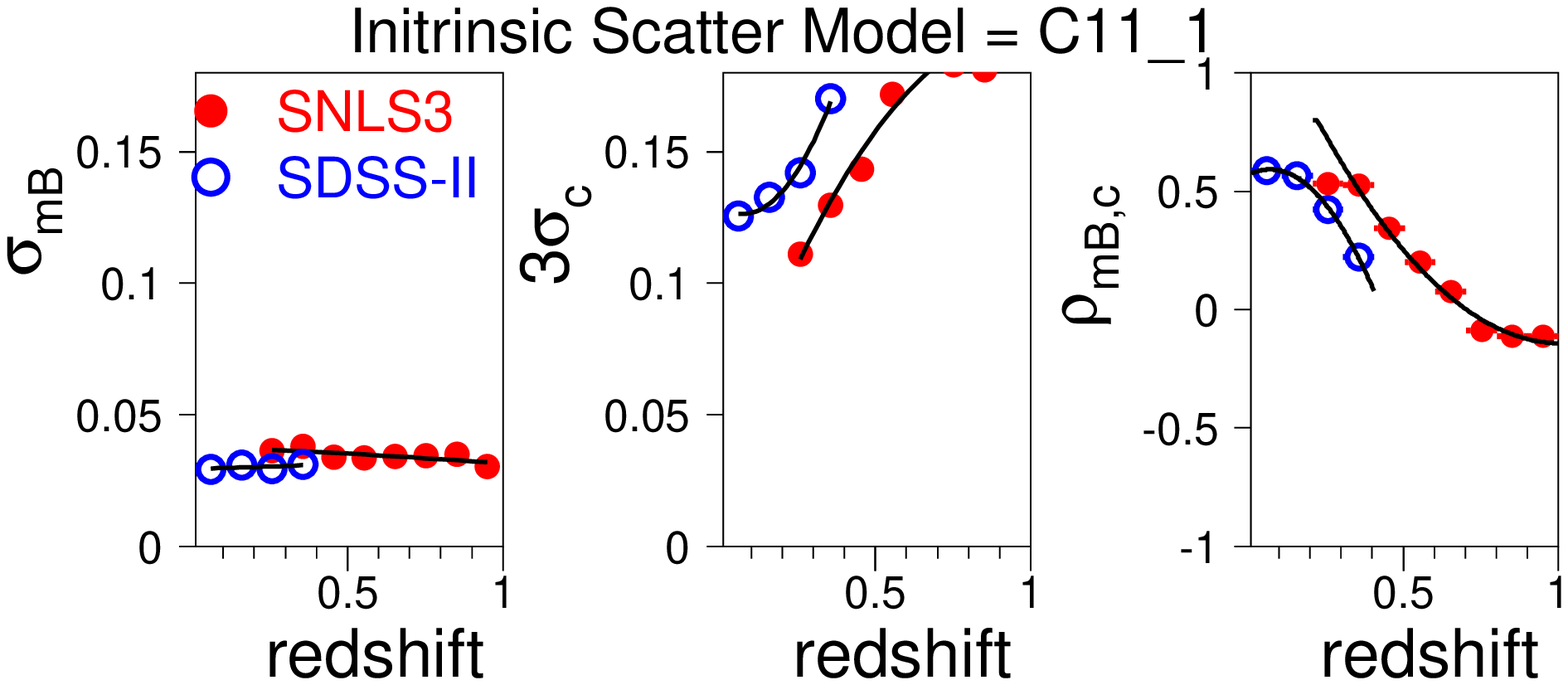}
  \caption{    
    Elements of intrinsic scatter matrix ($\C$ ) 
    for the simulated models indicated above each set of plots.
    The panels show the following parameters vs. redshift:
    $\sigmB = \sqrt{\C_{00}}$ (left),
    $3\sigc  = 3\sqrt{\C_{cc}}$ (middle), and
    $\rho_{m_B,c} = \C_{0c}/(\sigmB\sigc)$ (right).
    The ``int'' superscripts have been dropped for clarity.
    Note that $\sigmB$ and $3\sigc$ are in roughly the
    same units as the scatter in the distance modulus residual.
    The simulated \SNLS\ ({\SDSS}) samples are indicated 
    by solid (open) circles. Each solid curves represents
    a second-order polynomial function of redshift used
    to describe $\C$.
  }
  \label{fig:intcov}
\end{figure}

% --------------------------------------------------
  \subsection{Fitting for $\alpha$, $\beta$, and the Distance Moduli}
  \label{subsec:fit4mu}
% --------------------------------------------------

Using the fitted {\SALTII} parameters and \uncs\ from the
nominal MC, we use the {\SALTtomu} program (M11) to minimize 
Eq.~\ref{eq:muchi2}. This minimization gives the best-fit values
of $\alpha$ and $\beta$, along with an independent offset ($M_z$)
in each 0.1-wide redshift bin (Eq.~\ref{eq:muSALT2}). 
The separate offsets are used to eliminate the $\mu$-dependence 
on cosmological parameters. 
These fitted parameters are used to compute a distance modulus
for each SN, and the resulting redshift+distance pairs form
a Hubble diagram that can be fit for cosmological parameters.
To properly evaluate the fit $\chi^2$ using
the scatter matrix $\C$, 
the light curve fits have been done with the {\SALTII} 
model \uncs\ set to zero (hence reducing $\sigstati$) 
because they are now included in the {\SALTtomu} fit as
part of the $\sigint$ calculation. 
% Without a model \unc\ in the fit, 
% the $\Pfit$ cut has also been removed to avoid rejecting
% high S/N light curves with significant variations from the
% intrinsic scatter model.
This modification only affects the \SALTtomu\ $\chi^2$ 
and has a negligible impact on the fitted parameters.

The \WIS\ fit is based on the traditional method of tuning 
$\sigintmB$ such that $\chi^2/\Ndof=1$,
while setting all of the other covariance terms to zero.
The light curve fits include the \SALTII\ model errors,
and hence the intrinsic \uncs\ corresponding to the G10 model
are included.
We note that there is an improved statistical treatment
in \cite{March_11}, using a Bayesian hierarchical model.
While they obtain an \unc\ on $\sigintmB$, their scatter 
model is fundamentally the same as our \WIS\ model because 
they do not include additional covariance terms.

The {\SALTtomu} fit results for both intrinsic scatter methods
(\WIS\ and $\C$)
are shown in Table~\ref{tb:bias}.
For both methods the fitted values of $\alpha$ agree well 
with the simulated input, $\alphaSIM = 0.11$. 
For the FUN-MIX and G10 scatter models,
the fitted values of $\beta$ agree well with
the simulated input ($\betaSIM=3.2$)
for both methods.
For the C11 models, however, the situation is quite different. 
The \WIS\ fitted $\beta$ values are too low by 0.3--0.6,
and the significance of this bias ranges from
6 to 15 standard deviations.
The $\C$-fit $\beta$ values are consistent with $\betaSIM$.

\begin{table*}[hb!]
\caption{
  $\alpha$ and $\beta$ Fit Results from Simulations 
  with Different Intrinsic-scatter Models,
  and $w$-bias Results\tablenotemark{a}  from Cosmology Fits
  } % end caption
\begin{center}
{\tiny % \footnotesize  % for preprint
\begin{tabular}{ l cccc | ccc }
\tableline\tableline
 Simulated     & Intrinsic  & & & & & &  Malmquist- \\                 
 Scatter       & Scatter &
                 & Fitted                    
                 & Fitted  & & Uncorrected & 
                      Corrected\tablenotemark{d} \\
 Model        &  Matrix  
                 & $\chi^2/\Ndof$ 
                 &  $\alpha$\tablenotemark{b} 
                 &  $\beta$\tablenotemark{c} 
                 &  $\beta$-bias
                 &  $w$-bias
                 &  $w$-bias
         \\
\tableline\tableline
% --------------------------------------------------------------------
% use auto-generated table copied from $SNDP62/magSmear_paper/SALT2mu
  &  \multicolumn{4}{c}{SDSS-II} \\   
  FUN-MIX &  $\sigintmB=0.087$  
		 & 1.004   % chi2/dof  
		 & $0.114 \pm 0.003$   % fit alpha  
		 & $3.25 \pm 0.04$   % fit beta  
 		 &  $ -0.01  \pm  0.01 $     % Beta bias   
		 &  $ +0.007  \pm 0.006 $      % raw w bias      
		 &  $ -0.022  \pm 0.005 $   % w bias with Malm cor   
		 \\   
      G10 &  $\sigintmB=0.085$  
		 & 0.993   % chi2/dof  
		 & $0.109 \pm 0.003$   % fit alpha  
		 & $3.20 \pm 0.03$   % fit beta  
 		 &  $ +0.04  \pm  0.01 $     % Beta bias   
		 &  $ -0.002  \pm 0.006 $      % raw w bias      
		 &  $ +0.011  \pm 0.007 $   % w bias with Malm cor   
		 \\   
   C11\_0 &  $\sigintmB=0.082$  
		 & 0.980   % chi2/dof  
		 & $0.117 \pm 0.003$   % fit alpha  
		 & $2.94 \pm 0.03$   % fit beta  
 		 &  $ -0.27  \pm  0.02 $     % Beta bias   
		 &  $ +0.070  \pm 0.009 $      % raw w bias      
		 &  $ -0.023  \pm 0.008 $   % w bias with Malm cor   
		 \\   
   C11\_1 &  $\sigintmB=0.106$  
		 & 0.985   % chi2/dof  
		 & $0.117 \pm 0.003$   % fit alpha  
		 & $2.72 \pm 0.03$   % fit beta  
 		 &  $ -0.50  \pm  0.03 $     % Beta bias   
		 &  $ +0.109  \pm 0.013 $      % raw w bias      
		 &  $ +0.011  \pm 0.011 $   % w bias with Malm cor   
		 \\   
 \\  % ---------------  
  FUN-MIX &  $\C$  
		 & 1.038   % chi2/dof  
		 & $0.109 \pm 0.003$   % fit alpha  
		 & $3.25 \pm 0.04$   % fit beta  
 		 &  ---     % Beta bias   
		 &  ---     % raw w bias      
		 &  ---  % w bias with Malm cor   
		 \\   
      G10 &  $\C$  
		 & 1.085   % chi2/dof  
		 & $0.106 \pm 0.003$   % fit alpha  
		 & $3.17 \pm 0.03$   % fit beta  
 		 &  ---     % Beta bias   
		 &  ---     % raw w bias      
		 &  ---  % w bias with Malm cor   
		 \\   
   C11\_0 &  $\C$  
		 & 1.076   % chi2/dof  
		 & $0.112 \pm 0.003$   % fit alpha  
		 & $3.21 \pm 0.03$   % fit beta  
 		 &  ---     % Beta bias   
		 &  ---     % raw w bias      
		 &  ---  % w bias with Malm cor   
		 \\   
   C11\_1 &  $\C$  
		 & 1.009   % chi2/dof  
		 & $0.110 \pm 0.004$   % fit alpha  
		 & $3.20 \pm 0.04$   % fit beta  
 		 &  ---     % Beta bias   
		 &  ---     % raw w bias      
		 &  ---  % w bias with Malm cor   
		 \\   
 \hline  % ---------------  
  &  \multicolumn{4}{c}{SNLS3} \\   
  FUN-MIX &  $\sigintmB=0.087$  
		 & 0.980   % chi2/dof  
		 & $0.114 \pm 0.002$   % fit alpha  
		 & $3.17 \pm 0.02$   % fit beta  
 		 &  $ -0.04  \pm  0.02 $     % Beta bias   
		 &  $ -0.011  \pm 0.008 $      % raw w bias      
		 &  $ +0.012  \pm 0.008 $   % w bias with Malm cor   
		 \\   
      G10 &  $\sigintmB=0.069$  
		 & 1.009   % chi2/dof  
		 & $0.112 \pm 0.002$   % fit alpha  
		 & $3.22 \pm 0.02$   % fit beta  
 		 &  $ +0.11  \pm  0.01 $     % Beta bias   
		 &  $ -0.017  \pm 0.011 $      % raw w bias      
		 &  $ +0.002  \pm 0.011 $   % w bias with Malm cor   
		 \\   
   C11\_0 &  $\sigintmB=0.094$  
		 & 0.998   % chi2/dof  
		 & $0.109 \pm 0.002$   % fit alpha  
		 & $2.86 \pm 0.02$   % fit beta  
 		 &  $ -0.31  \pm  0.01 $     % Beta bias   
		 &  $ +0.072  \pm 0.009 $      % raw w bias      
		 &  $ +0.001  \pm 0.009 $   % w bias with Malm cor   
		 \\   
   C11\_1 &  $\sigintmB=0.127$  
		 & 0.989   % chi2/dof  
		 & $0.108 \pm 0.003$   % fit alpha  
		 & $2.55 \pm 0.02$   % fit beta  
 		 &  $ -0.65  \pm  0.02 $     % Beta bias   
		 &  $ +0.127  \pm 0.012 $      % raw w bias      
		 &  $ +0.060  \pm 0.011 $   % w bias with Malm cor   
		 \\   
 \\  % ---------------  
  FUN-MIX &  $\C$  
		 & 1.143   % chi2/dof  
		 & $0.104 \pm 0.002$   % fit alpha  
		 & $3.20 \pm 0.02$   % fit beta  
 		 &  ---     % Beta bias   
		 &  ---     % raw w bias      
		 &  ---  % w bias with Malm cor   
		 \\   
      G10 &  $\C$  
		 & 1.207   % chi2/dof  
		 & $0.103 \pm 0.002$   % fit alpha  
		 & $3.12 \pm 0.02$   % fit beta  
 		 &  ---     % Beta bias   
		 &  ---     % raw w bias      
		 &  ---  % w bias with Malm cor   
		 \\   
   C11\_0 &  $\C$  
		 & 1.302   % chi2/dof  
		 & $0.104 \pm 0.002$   % fit alpha  
		 & $3.16 \pm 0.02$   % fit beta  
 		 &  ---     % Beta bias   
		 &  ---     % raw w bias      
		 &  ---  % w bias with Malm cor   
		 \\   
   C11\_1 &  $\C$  
		 & 1.297   % chi2/dof  
		 & $0.105 \pm 0.003$   % fit alpha  
		 & $3.20 \pm 0.03$   % fit beta  
 		 &  ---     % Beta bias   
		 &  ---     % raw w bias      
		 &  ---  % w bias with Malm cor   
		 \\   
 \hline  % ---------------  
  &  \multicolumn{4}{c}{SDSS-II + SNLS3} \\   
  FUN-MIX &  $\sigintmB=0.086$  
		 & 0.991   % chi2/dof  
		 & $0.114 \pm 0.002$   % fit alpha  
		 & $3.20 \pm 0.02$   % fit beta  
 		 &  $ -0.02  \pm  0.01 $     % Beta bias   
		 &  $ -0.019  \pm 0.006 $      % raw w bias      
		 &  $ +0.010  \pm 0.006 $   % w bias with Malm cor   
		 \\   
      G10 &  $\sigintmB=0.075$  
		 & 0.997   % chi2/dof  
		 & $0.112 \pm 0.002$   % fit alpha  
		 & $3.22 \pm 0.02$   % fit beta  
 		 &  $ +0.07  \pm  0.01 $     % Beta bias   
		 &  $ -0.003  \pm 0.005 $      % raw w bias      
		 &  $ -0.001  \pm 0.005 $   % w bias with Malm cor   
		 \\   
   C11\_0 &  $\sigintmB=0.088$  
		 & 1.007   % chi2/dof  
		 & $0.112 \pm 0.002$   % fit alpha  
		 & $2.89 \pm 0.02$   % fit beta  
 		 &  $ -0.26  \pm  0.01 $     % Beta bias   
		 &  $ +0.035  \pm 0.004 $      % raw w bias      
		 &  $ +0.018  \pm 0.004 $   % w bias with Malm cor   
		 \\   
   C11\_1 &  $\sigintmB=0.119$  
		 & 0.998   % chi2/dof  
		 & $0.111 \pm 0.002$   % fit alpha  
		 & $2.63 \pm 0.02$   % fit beta  
 		 &  $ -0.56  \pm  0.02 $     % Beta bias   
		 &  $ +0.069  \pm 0.010 $      % raw w bias      
		 &  $ +0.040  \pm 0.011 $   % w bias with Malm cor   
		 \\   
 \\  % ---------------  
  FUN-MIX &  $\C$  
		 & 1.130   % chi2/dof  
		 & $0.105 \pm 0.002$   % fit alpha  
		 & $3.22 \pm 0.02$   % fit beta  
 		 &  ---     % Beta bias   
		 &  ---     % raw w bias      
		 &  ---  % w bias with Malm cor   
		 \\   
      G10 &  $\C$  
		 & 1.177   % chi2/dof  
		 & $0.104 \pm 0.002$   % fit alpha  
		 & $3.15 \pm 0.02$   % fit beta  
 		 &  ---     % Beta bias   
		 &  ---     % raw w bias      
		 &  ---  % w bias with Malm cor   
		 \\   
   C11\_0 &  $\C$  
		 & 1.241   % chi2/dof  
		 & $0.107 \pm 0.002$   % fit alpha  
		 & $3.15 \pm 0.02$   % fit beta  
 		 &  ---     % Beta bias   
		 &  ---     % raw w bias      
		 &  ---  % w bias with Malm cor   
		 \\   
   C11\_1 &  $\C$  
		 & 1.267   % chi2/dof  
		 & $0.108 \pm 0.002$   % fit alpha  
		 & $3.20 \pm 0.02$   % fit beta  
 		 &  ---     % Beta bias   
		 &  ---     % raw w bias      
		 &  ---  % w bias with Malm cor   
		 \\   
 \hline  % ---------------  
  &  \multicolumn{4}{c}{Nearby + SDSS-II + SNLS3} \\   
  FUN-MIX &  $\sigintmB=0.085$  
		 & 1.003   % chi2/dof  
		 & $0.115 \pm 0.002$   % fit alpha  
		 & $3.18 \pm 0.02$   % fit beta  
 		 &  $ -0.06  \pm  0.01 $     % Beta bias   
		 &  $ -0.013  \pm 0.003 $      % raw w bias      
		 &  $ -0.002  \pm 0.003 $   % w bias with Malm cor   
		 \\   
      G10 &  $\sigintmB=0.072$  
		 & 1.010   % chi2/dof  
		 & $0.112 \pm 0.001$   % fit alpha  
		 & $3.23 \pm 0.02$   % fit beta  
 		 &  $ +0.07  \pm  0.01 $     % Beta bias   
		 &  $ -0.001  \pm 0.003 $      % raw w bias      
		 &  $ +0.001  \pm 0.003 $   % w bias with Malm cor   
		 \\   
   C11\_0 &  $\sigintmB=0.086$  
		 & 0.990   % chi2/dof  
		 & $0.113 \pm 0.001$   % fit alpha  
		 & $2.86 \pm 0.01$   % fit beta  
 		 &  $ -0.28  \pm  0.01 $     % Beta bias   
		 &  $ +0.013  \pm 0.003 $      % raw w bias      
		 &  $ -0.017  \pm 0.003 $   % w bias with Malm cor   
		 \\   
   C11\_1 &  $\sigintmB=0.108$  
		 & 0.998   % chi2/dof  
		 & $0.111 \pm 0.002$   % fit alpha  
		 & $2.68 \pm 0.02$   % fit beta  
 		 &  $ -0.48  \pm  0.01 $     % Beta bias   
		 &  $ +0.020  \pm 0.003 $      % raw w bias      
		 &  $ -0.006  \pm 0.003 $   % w bias with Malm cor   
		 \\   
 \\  % ---------------  
  FUN-MIX &  $\C$  
		 & 1.077   % chi2/dof  
		 & $0.108 \pm 0.001$   % fit alpha  
		 & $3.23 \pm 0.02$   % fit beta  
 		 &  ---     % Beta bias   
		 &  ---     % raw w bias      
		 &  ---  % w bias with Malm cor   
		 \\   
      G10 &  $\C$  
		 & 1.104   % chi2/dof  
		 & $0.105 \pm 0.001$   % fit alpha  
		 & $3.16 \pm 0.01$   % fit beta  
 		 &  ---     % Beta bias   
		 &  ---     % raw w bias      
		 &  ---  % w bias with Malm cor   
		 \\   
   C11\_0 &  $\C$  
		 & 1.127   % chi2/dof  
		 & $0.109 \pm 0.001$   % fit alpha  
		 & $3.15 \pm 0.02$   % fit beta  
 		 &  ---     % Beta bias   
		 &  ---     % raw w bias      
		 &  ---  % w bias with Malm cor   
		 \\   
   C11\_1 &  $\C$  
		 & 1.145   % chi2/dof  
		 & $0.108 \pm 0.002$   % fit alpha  
		 & $3.17 \pm 0.02$   % fit beta  
 		 &  ---     % Beta bias   
		 &  ---     % raw w bias      
		 &  ---  % w bias with Malm cor   
		 \\   
 \hline  % ---------------  
 
\tableline  % ------------------------------------
\end{tabular}
  \tablenotetext{1}{Bias \uncs\ are MC statistical \uncs. }
  \tablenotetext{2}{Simulated $\alphaSIM = 0.11$.}
  \tablenotetext{3}{Simulated $\betaSIM = 3.20$.}
  \tablenotetext{4}{
                   For \WIS\ method, 
                   correction is from simulation using fitted
                    $\beta$ value and G10 intrinsic scatter model.
              } % ende tablenote{3}
}  % end small env
\end{center}
  \label{tb:bias}
\end{table*}

The reduced $\chi^2$ values are close to
unity for the \WIS\ method
because of the explicit $\sigintmB$ tuning. 
However, there is no such tuning for the $\C$-fit method and the 
reduced $\chi^2$ are within about 10\%--20\%  of unity.
In summary, using the correct intrinsic scatter matrix $\C$ 
in the \SALTtomu\ fit results in unbiased $\beta$
values, and $\chi^2/\Ndof \simeq 1$.
Using the simplistic \WIS\ method results in a significant
bias on $\beta$ for the C11 models.

\newcommand{\cmean}{\langle c^{\rm fit}\rangle}

\begin{figure}[h!]
\centering
\epsscale{\xxScale}  % 1.15 for emulateapj 
\plotone{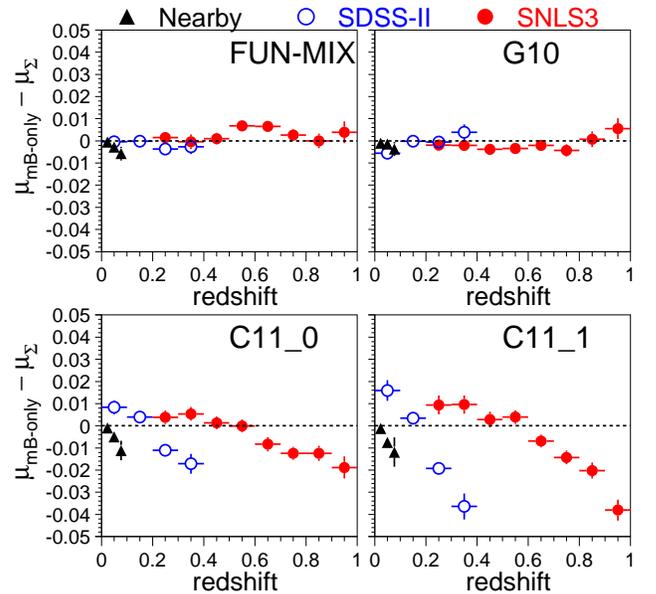}
  \caption{    
    Average difference between the distance modulus from the 
    \WIS\ fit and that from the $\C$-fit, vs. redshift.
    The simulated (true) scatter model is indicated in each panel.
    The simulated samples are 
    nearby (solid triangles),
    \SDSS\ (open circles), and
    \SNLS\ (solid circles).
  }
  \label{fig:bias_mudif}
\end{figure}

\begin{figure}[h!]
\centering
\epsscale{\xxScale}  % 1.15 for emulateapj 
\plotone{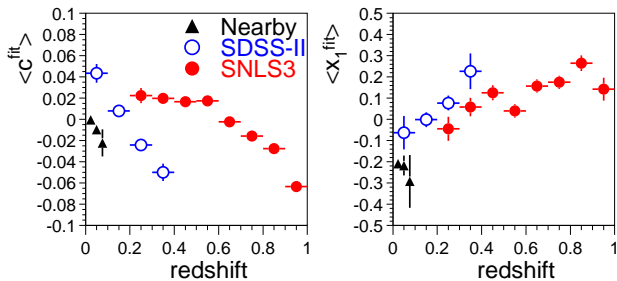}
  \caption{    
    Mean fitted \SALTII\ color vs. redshift for the 
    nearby (solid triangles),
    \SDSS\ (open blue circles) and
    \SNLS\ (solid red circles) simulation.
    Right panel shows mean fitted \SALTII\ 
    stretch parameters ($x_1$) vs. redshift.   
    Although this simulation uses the G10 intrinsic scatter model,
    there is little difference for the other scatter models.
    Also note that the analogous plots in the bottom panels of
    Figures~\ref{fig:ov1_SDSS} and \ref{fig:ov1_SNLS}
    show a larger variation because of the more strict 
    selection requirement of three passbands with S/N$>8$.
  } 
  \label{fig:fitpar_zbias}
\end{figure}

Since the $\C$-fit method results in unbiased
$\beta$ values we define the $\mu$-bias to be 
$\DMU = \mu_{mB{\rm -only}} - \mu_{\C}$:
$\mu_{\C}$ is the distance modulus from the $\C$-fit method and
$\mu_{mB{\rm -only}}$ is the distance from the \WIS\ method.
Figure~\ref{fig:bias_mudif}  displays
$\DMU$ versus redshift for each scatter model.
For the FUN-MIX and G10 scatter models, 
the maximum $\DMU$ variation is only $\sim 0.01$~mag
over the redshift range of each survey.
For the C11 models $\DMU$ varies by several hundredths
over the redshift range of each survey.
This $\mu$-bias is caused by the bias in $\beta$ combined with 
Malmquist bias that results in bluer SN with 
increasing redshift.
Figure~\ref{fig:fitpar_zbias} shows the mean fitted
color value versus redshift, 
and the corresponding $\mu$-bias is 
$\DMU \simeq \Delta\beta \times \cmean$  
where $\Delta\beta$ is the bias in $\beta$
and $\cmean$ is the mean \SALTII\ color.
The redshift dependence on $\DMU$ is therefore directly
proportional to the redshift dependence of $\cmean$.
Figure~\ref{fig:fitpar_zbias} also shows the mean fitted
stretch parameter ($x_1$) versus redshift. However,
since the corresponding bias on $\alpha$ is less than $0.01$,
the resulting $\mu$-bias is less than $0.01\times 0.3 \sim 0.003$,
and is thus much smaller than the bias from the color term.

To summarize,
these bias tests show that the G10 model is internally 
consistent and that the wavelength dependence of the 
intrinsic scatter can be described with both methods.
In the first method the scatter is described by the
\uncs\ in the \SALTII\ light curve fitting model,
and the \WIS\ scatter matrix is used in the
\SALTtomu\ stage. 
In the second $\C$-fit method the light curve fitting
model \uncs\ are set to zero and the full intrinsic scatter 
matrix is used in the \SALTtomu\ fitting stage.
The difference in results between these two methods,
as applied to simulations based on the G10 model,
is negligible.
However, if the simulated scatter model includes large 
anti-correlations such as those suggested by C11, 
then these two methods are not consistent and the
traditional \WIS\ method results in a 
significant redshift-dependent bias
in the distance moduli (Figure~\ref{fig:bias_mudif}).
More specifically, the bias is a result of incorrectly 
using the \SALTII\ (i.e., G10) \uncs\ in the light curve 
fits of the samples generated with the C11 model.

% -------------------------------------------------
  \subsection{Estimating the $w$-bias}
  \label{subsec:wbias}
% -------------------------------------------------

To estimate the bias on the dark energy parameter $w$,
we fit the simulated Hubble diagram in a manner similar
to that described in Section~8 of K09.
The effect of Malmquist corrections is discussed later in 
Section~\ref{subsec:evalMC}.
The $w$-bias is defined to be the difference between
$w$ obtained from the \WIS\ method and the $\C$-fit method.
Priors are included from measurements of 
baryon acoustic oscillations from the SDSS Luminous
Red Galaxy sample \citep{Eisen05} and from 
the cosmic microwave background temperature anisotropy 
measured from the
{\it Wilkinson Microwave Anistropy Probe}
\citep{Komatsu08}.

Since the simulated SN sample consists of 
$\MCDATAratiowbias$ times the data statistics,
the \uncs\ on the priors have been reduced by
a factor of $\sqrt{\MCDATAratiowbias}$.
We checked this procedure by splitting the
simulated sample into $\MCDATAratiowbias$ independent
sub-samples and fitting each sub-sample with the
nominal priors; the average of the $\MCDATAratiowbias$
fitted $w$-values is in good agreement with that from 
using the full simulation and priors with reduced \uncs.
The $\beta$-bias and 
Malmquist-uncorrected
$w$-bias values are defined as the difference between the 
\WIS\ and $\C$-fit methods,
and they are shown in Table~\ref{tb:bias}. 
The \unc\ on the $w$-bias is given by 
${\rm rms}/\sqrt{\MCDATAratiowbias}$,
where rms is the dispersion among the
$\MCDATAratiowbias$ independent  $w$-bias measurements.

For the FUN-MIX and G10 models the $w$-bias is small 
($<0.02$) for all sample combinations. This small bias
is expected since the fitted $\beta$ and distance moduli
show a very small bias.
For the C11 models, the $w$-bias is larger.
Fitting the \SDSS\ or \SNLS\ sample alone, 
the $w$-bias is  
$\sim 0.07$ for the C11\_0 model and
$\sim 0.12$ for the C11\_1 model.
Combining the {\SDSS} $+$ \SNLS\ samples, 
the $w$-bias is reduced by a factor of $\sim 2$ 
for each of the C11 models.
Including the nearby sample (nearby+{\SDSS}+{\SNLS})
the $w$-bias is further reduced:
$\wBiasCzero$ and $\wBiasCone$ for the C11\_0 and C11\_1
models, respectively.
The reduction in $w$-bias as more samples are combined
is a result of the fortuitous circumstance that the
mean \SALTII\ color for each sample is very close:
$-0.005$, $-0.015$, and $-0.005$ for the 
nearby, \SDSS, and \SNLS, respectively. 
To illustrate this point we have redone the analysis 
using a simulated nearby sample that has no Malmquist 
bias and a mean \SALTII\ color that is 0.06~mag redder 
than the data; the resulting $w$-bias on the combined
nearby+{\SDSS}+{\SNLS} sample increases to 0.05.

% -------------------------------------------------
  \subsection{Monte Carlo Correction for Malmquist Bias}
  \label{subsec:evalMC}
% -------------------------------------------------

\newcommand{\MqSIM}{MqSIM}
\newcommand{\mufit}{\mu_{\rm fit}}
\newcommand{\musim}{\mu_{\rm true}}

In Section~\ref{subsec:wbias} the $w$-bias is determined
without accounting for differences in the 
Malmquist bias correction. 
While the true Malmquist bias should not depend on the
analysis method (\WIS\ versus $\C$-method),
here we show that the evaluated Malmquist bias,
using the fitted value of $\beta$, does indeed
depend on the  analysis method. In addition, using the
evaluated Malmquist bias reduces the $\mu$-bias shown in 
Figure~\ref{fig:bias_mudif}. 
In the discussion below, ``{\MqSIM}''
refers to the simulation used to evaluate
the Malmquist bias correction.

For the $\C$-fit method the \MqSIM\ uses the correct model of intrinsic  
scatter and the correct $\alpha$ and $\beta$ parameters. In principle 
the intrinsic-scatter matrix would have to be translated into a 
wavelength-dependent scatter model for the simulation, 
but we have not preformed this step.
For the \WIS\ method the \MqSIM\ for each sample uses 
the G10 model and the fitted (biased) value of $\bfit$ from 
Table~\ref{tb:bias}. This procedure is used because it closely
reflects the procedure in previous analyses.
The distance-modulus corrections are from a second-order 
polynomial fit to $\mufit - \musim$ versus redshift:
$\musim$ is the true distance modulus in the \MqSIM,
and $\mufit$ is the distance computed from the 
SALT2-fitted parameters (color and stretch) 
and the simulated values of $\alpha$ and $\beta$.

\begin{figure}[h!]
\centering
\epsscale{1.} 
\plotone{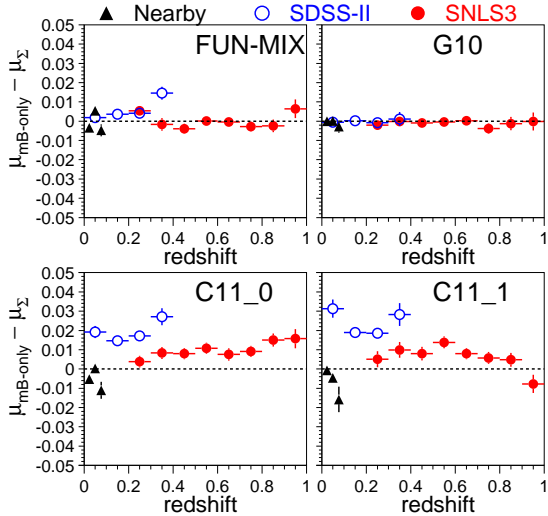}
  \caption{    
      Same as Figure~\ref{fig:bias_mudif},
      but after applying Malmquist bias correction.
  } % end caption
  \label{fig:biascor_mudif}
\end{figure}

Figure~\ref{fig:biascor_mudif} shows the $\mu$-bias versus 
redshift with the Malmquist correction applied. 
Overall the redshift-dependent bias is smaller than for the 
uncorrected distances in Figure~\ref{fig:bias_mudif},
but the bias is still significant for the C11 models.
The bias is zero for the G10 model because the fitted
light curve model (G10) corresponds to the correct model
of intrinsic scatter.
The Malmquist-corrected $w$-bias results are shown in the 
last column of Table~\ref{tb:bias}. 
Compared to the bias from the uncorrected Hubble diagrams,
the $w$-bias is typically smaller when the Malmquist correction 
is applied.

While using an incorrect model of intrinsic scatter 
can lead to a significant bias in the Hubble diagram,
this bias is somewhat reduced by simply applying a 
simulated Malmquist bias correction using the fitted 
value of $\beta$ in the simulation. 
The bias reduction depends on the intrinsic scatter model, 
and we cannot rule out increased sensitivity on other
systematic effects.

% ##################################
 \section{Conclusions}
 \label{sec:conclude}
% ##################################

We have used high quality \SDSS\ and \SNLS\ data and 
simulations to show that SN~Ia intrinsic brightness variations 
include wavelength dependent variations resulting in a color
dispersion of $\sim 0.02$~mag. 
A broad range of simulated intrinsic-scatter models 
(Table~\ref{tb:models})
is roughly consistent with the following photometric observables:
Hubble scatter, dispersion in $B-V-\cc$, and \photoz\ residuals.
These models include the G10 model that is dominated by a 
coherent term and has only positive wavelength correlations,
and the C11 model that has a small coherent term
and large anti-correlations.

We have used these intrinsic scatter models in 
high-statistics simulations
to test the standard procedure of adding a constant 
distance-modulus \unc\ ($\sigint$) to the measured \uncs\ 
so that $\chi^2/\Ndof=1$ for cosmology fits to the 
SN~Ia Hubble-diagram. 
The constant $\sigint$ assumption is valid if the
light curve fits include model \uncs\ with the correct
wavelength dependence of the scatter.
If these model \uncs\ are not correct, such as using
the \SALTII\ \uncs\ to fit a simulated sample with large
anti-correlations in the scatter (C11 model),
then significant biases can appear in the Hubble diagram.
For the specific simulation tests reported here
the distance-modulus bias varies by up to {\MAXMUBIAS}~mag
over the redshift range of the each survey,
although this bias is roughly halved after applying
Malmquist bias corrections.
% The resulting $w$-bias is large ($\sim 0.1$)
% from cosmology fits to either survey by itself.
For the combined nearby+{\SDSS}+{\SNLS} 
simulated sample, which corresponds closely to the 
472 SNe~Ia analyzed in \cite{Sullivan11},
the $w$-bias is only $\sim 0.02$, in part because
the mean color for each sample is very nearly the same.
While this $w$-bias is well below the total systematic 
\unc\ reported in \cite{Sullivan11} ($\dwsyst = 0.06$),
there is no assurance that future analyses on larger
data sets will benefit from the fortuitous cancellations.
For example, replacing the biased nearby sample with
an unbiased sample increases the $w$-bias to $0.05$,
comparable to the current systematic \unc.

We have also shown that applying a simulated Malmquist bias correction,
based on the fitted (biased) $\beta$ value, may reduce the bias from 
using an incorrect model of intrinsic scatter. However, we urge
caution in relying on this apparent ``free lunch'' because we have 
not fully explored if this naive correction increases the sensitivity 
to other systematic effects.

It is important not to interpret our results as proof
that such biases exist in current SNIa-cosmology results;
we have shown, for example, that this intrinsic scatter bias 
is negligible if the G10 model is correct.
However, until the intrinsic scatter correlations can
be better constrained, 
our results suggest that an additional 
systematic \unc\ should be included.

Although the intrinsic-scatter models considered here 
depend only on wavelength (except for the KRW09 models),
the true behavior could include
a dependence on epoch, color, stretch, and redshift.
The impact of the scatter models on the \SALTII\
training is under investigation and will be published
later. To obtain more robust systematic
constraints on cosmological parameters,
we encourage additional studies to measure or constrain 
the nature of intrinsic scatter.

%\bigskip
J.F. and R.K. are grateful for the support of 
National Science Foundation grant 1009457,
a grant from ``France and Chicago Collaborating in the Sciences'' (FACCTS),
and support from the 
Kavli Institute for Cosmological Physics at the University of Chicago.

The SDSS is managed by the Astrophysical Research Consortium
for the Participating Institutions.  The Participating Institutions are
the American Museum of Natural History,
Astrophysical Institute Potsdam,
University of Basel,
Cambridge University,
Case Western Reserve University,
University of Chicago,
Drexel University,
Fermilab,
the Institute for Advanced Study,
the Japan Participation Group,
Johns Hopkins University,
the Joint Institute for Nuclear Astrophysics,
the Kavli Institute for Particle Astrophysics and Cosmology,
the Korean Scientist Group,
the Chinese Academy of Sciences (LAMOST),
Los Alamos National Laboratory,
the Max-Planck-Institute for Astronomy (MPA),
the Max-Planck-Institute for Astrophysics (MPiA), 
New Mexico State University, 
Ohio State University,
University of Pittsburgh,
University of Portsmouth,
Princeton University,
the United States Naval Observatory,
and the University of Washington.

This work is based in part on observations made at the 
following telescopes.
The Hobby-Eberly Telescope (HET) is a joint project of the 
University of Texas at Austin,
the Pennsylvania State University,  Stanford University,
Ludwig-Maximillians-Universit\"at M\"unchen, and 
Georg-August-Universit\"at G\"ottingen.  
The HET is named in honor of its principal benefactors,
William P. Hobby and Robert E. Eberly.  The Marcario Low-Resolution
Spectrograph is named for Mike Marcario of High Lonesome Optics, who
fabricated several optical elements 
for the instrument but died before its completion;
it is a joint project of the Hobby-Eberly Telescope partnership and the
Instituto de Astronom\'{\i}a de la Universidad Nacional Aut\'onoma de M\'exico.
The Apache Point Observatory 3.5 m telescope is owned and operated by 
the Astrophysical Research Consortium. We thank the observatory 
director, Suzanne Hawley, and site manager, Bruce Gillespie, for 
their support of this project. The Subaru Telescope is operated by the 
National Astronomical Observatory of Japan. The William Herschel 
Telescope is operated by the Isaac Newton Group on the island of 
La Palma in the Spanish Observatorio del Roque 
de los Muchachos of the Instituto de Astrofisica de 
Canarias. The W.M. Keck Observatory is operated as a scientific partnership 
among the California Institute of Technology, the University of 
California, and the National Aeronautics and Space Administration. 
The Observatory was made possible by the generous financial support 
of the W. M. Keck Foundation.

% =========================
  \appendix
  \section{Simulation of Nearby SN~Ia Sample}
  \label{app:simLOWZ}
% =========================

Here we describe the simulation of the nearby ($z<0.1$) SN~Ia sample
corresponding to the 123 nearby SNe~Ia used in \cite{Conley11}.
While the nearby SN data are from several surveys and filter sets,
we simplify the simulation by considering only the
CFA3-Keplercam light curves \citep{Hicken09} using
the $UBVr$ filters. 
The $i$ band is dropped because it is outside the valid 
wavelength range of the \SALTII\ model.
Since the CFA3-Keplercam simulation is a proxy for the entire
nearby SN sample, we simulate the correct nearby-SN statistics 
corresponding to about half that of the \SNLS\ sample.

Since we do not have the observing conditions 
(mainly PSF and sky noise)
needed for the \SNANA\ simulation, we adopt a different strategy.
The basic idea is to use each observed SN to define an
observational sequence. The observed redshift and time
of peak brightness are used along with the cadence.
A random \SALTII\ stretch and color are chosen for each SN,
and the simulated S/N for each epoch is essentially scaled 
from the observed S/N. 
More technically, we artificially fixed the PSF to be 
$0.8^{\prime\prime}$
and then for each epoch compute the sky noise needed to 
simulate the observed \unc; 
this strategy allows generating nearby SNe in exactly the 
same manner as for the \SDSS\ and \SNLS.
With little knowledge of the \spec\ selection criteria
we cannot simulate the selection bias from first principles. 
We therefore modified the population parameters in 
Table~\ref{tb:unfold_x1c} so that the simulated 
color and stretch distributions
match those of the nearby sample.
For the generated color distribution the peak-probability 
value depends  on redshift, $-0.02 - z/2$, 
and the Gaussian width parameters are 
$\sigMINUS = 0.06$ and $\sigPLUS=0.10$.
For the generated stretch distribution,
the peak-probability value is reduced to 0.2.

The intrinsic-scatter matrix is determined for the nearby 
sample with the same procedure used on the \SDSS\ and \SNLS\
simulations, and using \SALTtomu\ we have checked that the 
fitted $\beta$ values are in good agreement with the simulated input,
$\betaSIM=3.2$. 
We have also examined the analogous data/MC comparisons in
Figures~\ref{fig:ov1_SDSS} and \ref{fig:ov1_SNLS}, and find
equally good agreement.

% ==============================================================
% BIBLIOGRAPHY
\bibliographystyle{apj}
\bibliography{intrinsic_magSmear}

% ####################################
  \end{document}